\documentclass{lmcs}
\pdfoutput=1

\usepackage{lastpage}
\lmcsdoi{20}{1}{9}
\lmcsheading{}{\pageref{LastPage}}{}{}%
{Nov.~01,~2022}{Jan.~26,~2024}{}


\usepackage[utf8]{inputenc}
\usepackage{stmaryrd}
\usepackage{xspace}
\usepackage{hyperref}
\usepackage{float}

\newcommand{\Xtup}[1][x]{\Xs[#1]}
\newcommand{\Ytup}[1][y]{\Ys[#1]}

\newcommand{\Xs}[1][x]{\mathbf{#1}}
\newcommand{\Ys}[1][y]{\mathbf{#1}}

\newcommand{\Xset}[1][x]{\Set{\Xtup[#1]}}
\newcommand{\Yset}[1][y]{\Set{\Ytup[#1]}}
\newcommand{\Set}[1]{\mathit{set}(#1)}

\newcommand{\fhtw}[1] {\mathsf{fhtw}(#1)}

\newcommand{\dom} {D}
\newcommand{\db} {\mathbb{D}}

\newcommand{\varx} {\mathcal{X}}

\newcommand{\R}{\mathbb{R}}
\newcommand{\Rp}{\mathbb{R}^+}
\newcommand{\N}{\mathbb{N}}

\newcommand{\Consts}{\mathcal{C}}

\newcommand{\fv}{\mathit{fv}}

\newcommand{\MAXIMIZE}{\textbf{maximize }}
\newcommand{\MINIMIZE}{\textbf{minimize }}
\newcommand{\SUBJECTTO}{\textbf{subject  to }}

\newcommand{\FORALL}[3]{\forall \XwhereQ{#1}{#2}.\ #3}
\newcommand{\XwhereQ}[2]{#1\ST#2}
\newcommand{\ST}{{:}}

\newcommand{\weightq}[3]{\mathbf{weight}_{#1 \ST #3}(#2)}
\newcommand{\weight}[2]{\mathbf{weight}_{#2}(#1)}

\newcommand\num{\mathbf{num}}
\newcommand{\true}{\mathit{true}}

\newcommand{\real}{r}

\newcommand{\CQ}{\mathit{CQ}_\Sigma}
\newcommand{\LP}{\mathit{L}\textsc{p}(\CS)}
\newcommand{\SUM}{\mathit{Sum}(\CS)}
\newcommand{\EXP}{\mathit{Exp}(\CS)}
\newcommand{\NUM}{\mathit{Num}(\CS)}
\newcommand{\LC}{\LCSig{\CS}}

\newcommand{\lcqs}[1]{\mathsf{Queries_w}{(#1)}}
\newcommand{\lcqfo}[1]{\mathsf{Queries}_\forall{(#1)}}
\newcommand{\lcqsum}[1]{\mathsf{Queries}_\Sigma{(#1)}}

\newcommand{\agm}{\mathsf{AGM}}
\newcommand{\agmw}[1]{\agm_{\mathsf{w}}{(#1)}}
\newcommand{\agmfo}[1]{\agm_\forall{(#1)}}
\newcommand{\agmsum}[1]{\agm_\Sigma{(#1)}}

\newcommand{\LSym}[1]{\stackrel{.}{#1}}
\newcommand{\LEq}{{\LSym{=}}}

\newcommand\subs[1]{\mathit{subs}_{#1}}
\newcommand{\gammabar}{\tilde{\gamma}}

\newcommand{\tup}[2]{\theta^{#2}_{#1}}

\newcommand{\sem}[1]{\llbracket #1 \rrbracket}
\newcommand{\opsem}[1]{opt(#1)}

\newcommand{\closure}[3]{\closurebase{#1}^{\db, #3}}

\newcommand{\closurebase}[1]{close(#1)}
\newcommand\closureno[3]{\closurebase{#1}^{#3}}

\newcommand{\interpret}[1]{\langle#1\rangle}

\newcommand{\weightbs}[1][{L}]{\Xi_{#1,T}^\db}
\newcommand{\weightB}[3]{\xi_{#1, #2}^{#3}}

\newcommand{\localsound}[1]{\rho_{C}^{T,\db}(#1)}
\newcommand{\localsoundc}[3]{\mathit{E}^{#2,\db}_{#3}({#1})}

\newcommand{\CS}{\CQ} 
\newcommand\constraints{\Constraints}
\newcommand\Constraints{\CQ}
\newcommand\C{\Query}

\newcommand{\Sol}[1]{\mathit{sol}^{#1}}

\newcommand{\Soldb}[1]{\sem{#1}^\db}

\newcommand{\projt}[2] {{#1}_{\mid {#2}}}
\newcommand{\projs}[2] {{#1}_{\mid {#2}}}
\newcommand{\projsb}[2] {{#1}_{\mid {\Bag(#2)}}}
\newcommand{\projtb}[2]{\projt{#1}{\Bag(#2)}}

\newcommand{\Figure}[1]{Figure~\ref{#1}\xspace}

\newcommand\ignore[1]{}

\newcommand{\sP}{\#\mathsf{P}}
\newcommand{\NP}{\mathsf{NP}}
\newcommand{\coNP}{\mathsf{coNP}}

\newcommand{\Wset}{\mathsf{W}}
\newcommand{\Fset}{\mathsf{F}}
\newcommand{\Oset}{\mathsf{O}}
\newcommand{\Bset}{\mathsf{B}}

\newcommand{\torder}{\mathit{order}}
\newcommand{\tfactory}{\mathit{prod}}
\newcommand{\troute}{\mathit{route}}
\newcommand{\twarehouse}{\mathit{store}}

\newcommand\deliver{\mathit{dlr}}

\newcommand\CLOSED{_{clos}}
\newcommand{\LSclosed}{\mathit{LS}\CLOSED(\CQ)}
\newcommand{\LCclosed}{\mathit{LC}\CLOSED(\CQ)}
\newcommand{\LPclosed}{\ensuremath{LP\CLOSED(\CQ)}\xspace}

\renewcommand{\LP}{\mathit{LP}}
\newcommand{\LS}{\mathit{LS}}
\renewcommand{\LC}{\mathit{LC}}
\newcommand\REPL{\mathit{repl}}
\newcommand\Rho[1][\nu]{\REPL^{#1}}
\newcommand\LPCQ{\ensuremath{\mathit{LP(\CQ)}}\xspace}

\newcommand\Xwheredelta[2]{#1\ST#2}

\renewcommand\SUM{\LS(\CS)}

\newcommand\EQU{{\LSym{=}}}
\newcommand{\Bool}{\mathbb{B}}
\newcommand\Reals{\R}
\newcommand\Vars{\mathcal{X}}
\newcommand{\domain}{\mathsf{D}}

\newcommand\SLP{\mathit{L}\textsc{p}}
\newcommand\SLC{\mathit{L}\textsc{c}}
\newcommand{\lpvarset}{\Xi}
\newcommand\vlp{\xi}

\newcommand\theSUBJECTTO{\textbf{subject  to }}
\newcommand\SSUBJECTTO{\text{ \theSUBJECTTO}}

\newcommand\eval[2]{eval_{#1}(#2)}

\renewcommand\Vert{\mathcal{V}}
\newcommand\TLong{T=(\Vert, \Edg, \Lab)}

\newcommand\Edg{\mathcal{E}}
\newcommand{\Nat}{\mathbb{N}}
\newcommand\Rels{\mathcal{R}}
\newcommand\RelsN[1][n]{\Rels^{(#1)}}
\newcommand\DB{\mathit{db}_\Sigma}

\newcommand{\inExpr}{}
\newcommand{\Soldbv}[2]{\sem{#2}_{#1}^\db}
\newcommand{\SoldbX}[1]{\Soldbv{X}{#1}}
\newcommand\Query{Q}
\newcommand{\thetaQpar}[1]{\tup{Q}{(#1)}}

\newcommand{\Bag}{\mathcal{B}}

\newcommand{\addnode}{extend\xspace}
\newcommand{\forgnode}{project\xspace}
\newcommand{\joinnode}{join\xspace}

\newcommand{\size}[1]{|#1|}

\newcommand\repl[1][\nu]{\mathit{wc}^{#1}}

\newcommand\LPCQfrag{\ensuremath{\mathit{LP}^{\mathit{proj}}(\CQ)} \xspace}

\newcommand\wC[1][]{\repl(#1)}

\newcommand{\trees}[1]{\mathcal{T}_{#1}}

\newcommand{\rewet}[1]{\rho_{W}^{T,\db}({#1})}
\newcommand{\reweq}[1]{\rho_W^{\trees{Q}, \db}({#1})}
\newcommand{\rewts}[1]{\rho_W^{\trees{}, \db}({#1})}
\newcommand{\rewe}[1]{\rewbase^{\trees{}, \db}(#1)}

\newcommand{\rewei}{\rewbase^{\trees{}}}
\newcommand{\rewew}{\rewbase_W^{\trees{}}}
\newcommand{\rewec}{\rewbase_C^{\trees{}}}

\newcommand{\localsoundcal}[1]{\rho_{C}^{\trees{#1},\db}(#1)}
\newcommand{\localsoundcalt}[1]{\rho_{C}^{\trees{},\db}(#1)}

\newcommand\rewbase{\rho}

\newcommand{\emptyassign}{[]}

\newcommand{\tups}[1][\queryset]{\Theta_{#1}^\db}

\newcommand{\qfp}[1]{\mathit{qf(#1)}}
\newcommand\Priv{\mathit{Priv}}
\newcommand\Test{\mathit{Test}}
\newcommand\Sens{\mathit{Sens}}
\newcommand\reladb[1]{#1^\db}
\newcommand\rela[1]{#1}
\newcommand\InSt{\mathit{InStudy}}
\newcommand\Para[1]{\ensuremath{(#1)}}
\newcommand\w{w}
\newcommand{\wset}[1][]{\mathbf{W}_{#1}}

\renewcommand\Vert{\mathcal{V}}
\renewcommand\TLong{T=(\Vert, \Edg, \Bag)}

\newcommand\W{\Omega}
\newcommand{\Wfam}{\W}
\newcommand{\Wfamlong}{(\Wfam_v)_{v \in \Vert}}

\newcommand{\omusym}[1]{\omega_{#1}}
\newcommand{\omu}[2]{\omusym{#1}(#2)}

\newcommand{\var}[1]{\Bag(#1)}

\newcommand{\vart}[1]{\var{\descorequal{#1}}}

\newcommand{\descorequal}[1]{{{\downarrow}{#1}}} 

\newcommand{\varc}[1]{\var{\context{#1}}} 
\newcommand\context[1]{{\uparrow} #1}

\newcommand\decompositiontree{decomposition tree\xspace}

\newcommand{\projw}[2]{\pi_{#2}(#1)}
\newcommand{\projwb}[2]{\projw{#1}{\Bag(#2)}}
\newcommand{\projwi}[3]{\projw{#1}{\interset{#2}{#3}}}

\newcommand\conjunctive{conjunctively decomposed\xspace}
\newcommand\conjdecomp{conjunctive decomposition\xspace}

\newcommand{\projtv}[2]{\projt{#1}{\vart{#2}}}
\newcommand{\projsv}[2]{\projs{#1}{\vart{#2}}}

\newcommand{\projsc}[2]{\projs{#1}{\varc{#2}}}

\newcommand{\extset}[2]{#2[#1]}

\newcommand{\extseta}[2]{\extset{#1}{\projs{A}{#2}}}
\newcommand{\extsetav}[2]{\extset{#1}{\projsv{A}{#2}}}
\newcommand{\extsetab}[2]{\extset{#1}{\projsb{A}{#2}}}

\newcommand\WC{weighting collection\xspace}
\newcommand{\TWD}[1][T]{\WC on \ensuremath{#1}\xspace}

\newcommand{\ltri}{the inclusion from the left to the right\xspace}
\newcommand{\rtli}{the inclusion from the right to the left\xspace}

\newcommand{\decomptree}{decomposition tree\xspace}

\newcommand{\interset}[2]{{\Bag^{#1#2}}}

\newcommand{\assignVal}[2]{\mathit{eval}^{#2,#1}}

\newcommand{\mapf}{\mu}

\newcommand{\om}[1][] {\omega_{#1}}

\newcommand{\sxd}[2]{S_{{#1} = {#2}}}

\newcommand\SSUM{\mathit{L}\textsc{e}}

\begin{document}

\title[Linear Programs with Conjunctive Queries]{Linear Programs with Conjunctive Database Queries}

\author[F. Capelli]{Florent Capelli\lmcsorcid{0000-0002-2842-8223}}[a,b]
\author[N. Crosetti]{Nicolas Crosetti}[b]
\author[J. Niehren]{Joachim Niehren\lmcsorcid{0000-0002-2611-8950}}[b]
\author[J. Ramon]{Jan Ramon\lmcsorcid{0000-0002-0558-7176}}[b]

\address{{Universit\'{e} de Lille}
  }
\email{florent.capelli@univ-lille.fr}

\address{{Inria, Lille}
}
\email{nicolas.crosetti@inria.fr, joachim.niehren@inria.fr, jan.ramon@inria.fr}

\thanks{
This work was partially supported by the French Agence Nationale de la Recherche, AGGREG project reference ANR-14-CE25-0017-01, Headwork project reference ANR-16-CE23-0015, KCODA project ANR-20-CE48-0004 and by a grant of the Conseil Régional Hauts-de-France. The project DATA, Ministère de l’Enseignement Supérieur et de la Recherche, Région Nord-Pas de Calais and European Regional Development Fund (FEDER) are acknowledged for supporting and funding this work. We also thank Sylvain Salvati, Sophie Tison and Yuyi Wang for fruitful discussions and anonymous reviewers of a previous version of this paper for their helpful comments. 
}

\begin{abstract}
In this paper, we study the problem of optimizing a linear program
whose variables are the answers to a conjunctive query. For this we
propose the language LP(CQ) for specifying linear programs whose
constraints and objective functions depend on the answer sets of
conjunctive queries. We contribute an efficient algorithm for solving
programs in a fragment of LP(CQ). The natural approach constructs a
linear program having as many variables as there are elements in the answer set
of the queries. Our approach constructs a linear program having the
same optimal value but fewer variables. This is done by exploiting the
structure of the conjunctive queries using generalized hypertree decompositions of
small width to factorize elements of the answer set together.  We
illustrate the various applications of LP(CQ) programs on three
examples: optimizing deliveries of resources, minimizing noise for differential privacy,
and computing the s-measure of patterns in graphs as needed for data
mining.

\end{abstract}

\maketitle

\tableofcontents  

\section{Introduction}

\ignore{Renaming
$$\begin{array}{l@{\Rightarrow}l@{\qquad}l}
    \LSym{\omega} & w            & \text{weighting on $\theta$'s} \\
    \LSym{W}           & W            & \text{weighting on $\xi$'s}  \\
    \omega              &  \omega  & \text{weighting on answer sets}  \\
    W                       & \Omega  & \text{weighting collection on answer sets}
  \end{array}
  $$}
When modeling optimization problems it often seems
natural to separate the logical constraints from the relational data. This
  holds for linear programming with AMPL~\cite{ampl} and 
for constraint programming in MiniZinc~\cite{minizinc}. It was
also noticed in the context of database research,  when using
integer linear programming for finding optimal 
database repairs as proposed by Kolaitis, Pema 
and Tan \cite{kolaitis_efficient_2013}, or when 
using linear optimization to explain the result 
of a database query to the user as proposed by
Meliou and Suciu~\cite{tiresias}.  
Moreover, tools like 
\texttt{SolveDB}~\cite{vsikvsnys16} have been developed 
to better integrate mixed integer programming and thus
linear programming into relational
databases.

We also find it natural to define the relational data of linear optimization problems by database queries. For this reason, we propose the language of linear programs with conjunctive queries \LPCQ in the present paper.  An \LPCQ program is a linear program with constructs allowing to express linear constraints and linear sums over the weightings of an answer set of database queries. It hence allows us to express an optimization problem with a linear objective function subject to linear constraints that are parameterized by conjunctive queries. To do so, we define the \emph{natural interpretation} $\interpret{L}^\db$ of an \LPCQ $L$ over a database $\db$ that is a linear program whose variables are in correspondence with the answer set of the queries of $L$, hence, $\interpret{L}^\db$ expresses an optimization problem whose solutions are weightings of the answer sets of conjunctive queries. The optimal weightings of \LPCQ programs can be computed in a natural manner, by first answering the database queries, then generating the interpretation of $L$ over $\db$ and solving it by calling a linear solver. We then approach the  question -- to our knowledge for the first time -- of whether this can be done with lower complexity for subclasses of conjunctive queries such as the class of acyclic conjunctive queries.

As our main contribution we present a more efficient algorithm for computing the optimal value of a program in \LPCQ programs that is able to exploit hypertree decomposition of the queries to speed up the computation. Our algorithm operates in two phases: first, it unfolds universal quantifiers present in \LPCQ programs to generate a program in a more restrictive language that we call \LPclosed. Then, the algorithm exploits a hypertree decomposition to construct an alternate interpretation of an \LPclosed program over a database that we call the factorized interpretation. The factorized interpretation is a linear program having the same optimal value as the linear program resulting in the natural interpretation of \LPCQ while being more succinct. It uses different linear program variables that intuitively represent sums of the linear program variables in the natural interpretation. The number of linear program variables in the factorized interpretation depends only on the fractional hypertree width of hypertree decompositions of the queries provided in the input, rather than on the number of query variables. In this manner, our more efficient algorithm can decrease the data complexity, i.e., the degree of the polynomial in the upper bound of the run time of the naive algorithm based on computing the natural interpretation and solving it with a linear program solver. With respect to the combined complexity, even solving \LPclosed programs is $\NP$-hard and $\coNP$-hard in general, but our approach shows that some cases are tractable. 

We prove the correctness of the factorized interpretation with respect to the natural interpretation -- that is, the fact that factorized and natural interpretation generates linear program with the same optimal value -- by exhibiting a correspondence between weightings of answer sets on the natural interpretation, and weightings of answer sets on the factorized interpretation. This correspondence can be seen as an independent contribution as it shows that one can reconstruct a relevant weighting of the answer set of a quantifier free conjunctive query by only knowing the value of the projected weighting on the bags of a tree decomposition.  Conjunctive queries with existential quantifiers are dealt with by showing that one can find an equivalent projecting \LPCQ program with quantifier free conjunctive queries only.

\subsection{A Concrete Example}
\label{sec:introexample}

We start by illustrating the language \LPCQ on an example.
\newcommand{\theUNIT}{unit}
\newcommand{\UNITs}{\theUNIT s}
\newcommand{\object}{object}
\newcommand{\objects}{\object s}

\medskip\noindent\textbf{Resource Delivery Optimization.}
We consider a situation in logistics where a company received orders
for specific quantities of resource \objects{}.
The \objects{} must be produced at a factory, then transported to a warehouse before being delivered to the buyer.
The objective is to fulfill every order while minimizing the overall delivery costs and respecting the production
capacities of the factories as well as the storing capacities of the warehouses.

Let $\Fset$ be the set of factories, $\Oset$ the set of \objects{}, $\Wset$ the set of warehouses and
$\Bset$ the set of buyers.
We consider a database $\db$ with elements in the domain $D = \Fset \uplus \Oset \uplus \Wset \uplus \Bset \uplus \Rp$.
The database $\db$ has four tables. 
The first table $\tfactory^\db \subseteq \Fset \times \Oset \times
\Rp$ 
contains
triples $(f, o, q)$ stating that the factory $f$ can produce up to $q$ \UNITs{} of \object{} o.
The second table $\torder^\db : \Bset \times \Oset \times \Rp$ contains triples $(b, o, q)$ 
stating that the buyer $b$ orders $q$ \UNITs{} of object $o$.
The third table $\twarehouse^\db \subseteq \Wset \times \Rp$ contains pairs $(w, l)$ 
stating that the warehouse $w$ has a storing limit of $l$.
The fourth table $\troute^\db : (\Fset \times \Wset\times\Rp) \cup
  (\Wset \times \Bset \times \Rp)$  
contains triples $(f, w, c)$ stating that the transport from factory
$f$ to warehouse $w$ costs $c$, and triples $(w, b, c)$ stating 
that the transport from warehouse $w$ to buyer $b$ costs $c$.
The query:
\[
    \deliver(f, w, b, o) = \exists q.\exists q_2.\exists c. \exists c_2.\ \tfactory(f, o, q) \wedge \torder(b, o, q_2)
    \wedge \troute(f, w, c) \wedge \troute(w, b, c_2)
  \]
selects from the database $\db$ all tuples $(f, w, b, o)$ such that the factory $f$ 
can produce some \objects{} o to be delivered to buyer $b$ through the
warehouse $w$. Let $Q=\deliver(f',w',b',o')$ and let $\Sol\db(Q)$ be the answers of $Q$ on database $\db$. 
The goal is to determine for each of these possible deliveries
the quantity of the \object{} that should actually be sent.
These quantities are modeled by the unknown
  weights $\theta^\alpha_Q$ of the query answers
  $\alpha\in\Sol\db(Q)$. 
For any factory $f$ and warehouse $w$
the sum 
$\sum_{\alpha\in\Sol\db(Q\wedge w'\EQU w\wedge f'\EQU f)}\theta^\alpha_Q$
is described by the expression $
\weight{Q}{f' \EQU f  \wedge w' \EQU w} 
$ when interpreted over $\db$.

We use the \LPCQ  program in \Figure{fig:lpex2} to describe the optimal weights that minimize the overall delivery costs.

\begin{figure}[h]

  \[
\begin{array}{ll}
\MINIMIZE  & \sum_{(f, w, c): \troute(f, w, c)} \num(c)\  \weight{Q}{\Xtup: f' \EQU f
  \wedge w' \EQU w} + \\
           & \sum_{(w, b, c): \troute(w, b, c)} \num(c) \ \weight{Q}{\Xtup: w' \EQU w \wedge b' \EQU
     b}
\\
\SUBJECTTO & \\
           & \FORALL{(f, o, q)}{\tfactory(f, o, q)}{\weight{Q}{\Xtup: f' \EQU f \wedge o' \EQU o} \leq \num(q)}\ \wedge\\
           & \FORALL{(b, o, q)}{\torder(b, o, q)}{\weight{Q}{\Xtup: b' \EQU b \wedge o' \EQU o}   \geq \num(q)}\ \wedge\\
           & \FORALL{(w, l)}{\twarehouse(w, l)}{\weight{Q}{\Xtup: w' \EQU w} \leq \num(l)}
\end{array}
\]    

\caption{A \LPCQ program  for the resource delivery
optimization where $Q=\deliver(\Xtup)$ and $\Xtup = (f', w', b', o')$.
\label{fig:lpex2}}
\end{figure}

The weights depend on the interpretation of the program over the database, since $\db$ specifies the production capacities of the factories, the stocking limits of the warehouses, etc.
The program has the following constraints:
\begin{itemize}
    \item[-]  for each $(f, o, q) \in \tfactory^\db$ the overall quantity of \object{} $o$ produced 
        by  $f$ is at most $q$\footnote{In this constraint, we use the construct $\num(q)$ to explicitly specify that the domain of $q$ is $\Rp$ and that, when evaluating the \LPCQ program on a database $\db$, $q$ will be decoded back as an element of $\Rp$.}.
    \item[-] for each $(b, o, q)\in \torder^\db$ the overall quantity of \objects{} $o$ delivered to $b$ is at least $q$.
    \item[-] for each $(w, l) \in \twarehouse^\db$ the overall quantity of \objects{} stored in $w$ is at most $l$.
\end{itemize}

By answering the query $Q$ on the database $\db$ and  introducing a linear program variable $\theta^\alpha_Q$ for each of the query answer $\alpha$, we can interpret the \LPCQ program in \Figure{fig:lpex2} as a linear program. A solution to this linear program will associate a real weight to each $\alpha$, that is, to each tuple $(f,w,b,o)$ that is a solution of $Q$ over $\db$. Intuitively, this weight is the quantity of object $o$ that factory $f$ has to produce to store in warehouse $w$ before being sent to buyer $b$. Moreover, these quantities are compatible with the constraints imposed on the capacity of each factory, warehouse and on the orders of each buyer.  Hence an optimal solution of this linear program will yield an optimal way of producing what is necessary while minimizing the transportation costs.

\subsection{Related Work}

Our result builds on well-known techniques using dynamic programming on tree decompositions of the hypergraph of conjunctive queries. These techniques were first introduced by Yannkakis~\cite{Yannakakis} who observed that so-called acyclic conjunctive queries could be answered in linear time using dynamic programming on a tree whose nodes are in correspondence with the atoms of the query. Generalizations have followed in two directions: on the one hand, generalizations of acyclicity such as notions of hypertree width~\cite{gottlob2002,gottlob1999tractable,grohe2006structure} have been introduced and on the other hand enumeration and aggregation problems have been shown to be tractable on these families of queries such as finding the size of the answer set~\cite{pichler2013} or enumerating it with small delay~\cite{bagan2007acyclic}. These tractability results can be obtained in a unified and generalized way by using factorized databases introduced by Olteanu and Závodný~\cite{olteanu2012factorised,olteanu2015size}, from which our work is inspired. Factorized databases provide succinct representations for answer sets of queries on databases. The representation enjoys interesting syntactic properties allowing to efficiently solve numerous aggregation problems on answer sets in polynomial time in the size of the representation. Olteanu and Závodný~\cite{olteanu2015size} have shown that when the fractional hypertree width of a query $Q$ is bounded, then one can construct, given a hypertree decomposition of $Q$ and a database $\db$, a factorized databases representing the answers of $Q$ on $\db$ of polynomial size. They also give a $O(1)$ delay enumeration algorithm on factorized databases. Combining both results gives a generalization of the result of Bagan, Durand and Grandjean~\cite{bagan2007acyclic} on the complexity of enumerating the answers of conjunctive queries.

Our result heavily draws inspiration from this approach as we use bottom up dynamic programming on hypertree decomposition of the input query $Q$ to construct  a partial representation of the answers set of $Q$ on database $\db$ that we later use to construct a factorized interpretation of the linear program to solve. While our approach could be made to work directly on factorized representations of queries answer sets as defined by Olteanu and Závodný~\cite{olteanu2015size}, we choose to directly work on tree decompositions. One reason for this is because our factorized interpretation uses hypertree decompositions that are slightly more constraint than the one usually used to efficiently handle complex linear programs. Namely, our tree decomposition needs extra bags for dependencies between variables that are not present in the query but only in the linear program. This constraints are not straightforward to translate into factorized databases while they are natural on tree decompositions.

\paragraph{Comparison with conference version.} This paper is a longer version of~\cite{confversion}. We have improved the presentation of some results from this old version, added new ones and added the full proofs left in the appendix in the earlier version. Some clarifications were made through slight changes in the theoretical framework that are described in the next paragraph. Our new contributions includes:
\begin{itemize}
\item A precise complexity analysis on how one can solve linear programs in \LPCQ depending on their structure, stated by explicitly using the AGM bound and fractional hypertree width of the queries involved in the linear program,  
\item New hardness results for the general case,
\item A cleaner logical framework to describe linear programs over database queries. 
\end{itemize}
The main change in the presentation comes from the introduction of the core language \LPclosed on which the factorized interpretation is described. It allows for a cleaner analysis of the complexity of our approach where we can separate the explanation of interpreting \LPclosed languages, now called the \emph{natural interpretation} (\emph{naive interpretation} in the conference paper), and the unfolding of quantifiers in \LPCQ. Indeed, in the conference version, we started by unfolding quantifiers before performing the analysis. This unfolding is now made explicit by the closure operation of \LPCQ programs which produces an \LPclosed program that can then be solved using our techniques. It allows us to properly separate the unfolding phase from the interpretation phase and to describe their complexity independently. In particular, in the conference version of the paper, the tractability result holds only for a fragment of \LPCQ where we are able to bound the size of the unfolding. Thanks to the introduction of \LPclosed and to a notion of normal form for \LPCQ programs, we are now able to state precise complexity bounds for every program in \LPCQ depending on their structure without assuming anything more. We also removed one constructor from the definition of \LPCQ. Indeed, in the conference version of the paper, \LPCQ programs could use an expression of the form $\weight{Q}{\Xtup:Q'}$ to generate a linear sum depending on both $Q$ and $Q'$. However, our tractability results worked only when $Q'$ has a very particular form, namely, $Q'$ needed to be of the form $\Xtup\LEq\Ytup$. We hence removed this constructor from the definition of \LPclosed language which has now only constructors of the form $\weight{Q}{\Xtup \LEq \Xtup[c]}$ for a vector of database constants $\Xtup[c]$. Consequently we do not need to introduce a fragment of the general language anymore to recover tractability since the tractable case is now the only one possible in \LPclosed. It may appear that we lost some expressivity along the way but it turns out that we can recover the same behavior using constructors in \LPCQ. Namely $\weight{Q}{\Xtup:Q'}$ can now be expressed as $\sum_{\Ytup:Q'} \weight{Q}{\Xtup \LEq \Ytup}$.

\subsection{Organization of the paper}

Section~\ref{sec:prelim} contains the necessary definitions to understand the paper. Section~\ref{lp-closed} presents the language \LPclosed of linear programs parameterized by conjunctive queries and gives its semantics by interpreting programs in \LPclosed as linear programs, which we call the natural interpretation. This language is very simple and does not allow universal quantification as used in Section~\ref{sec:introexample}. We show in Section~\ref{sec:eff_fragment} that one can exploit hypertree decomposition to compute the optimal value of \LPclosed programs efficiently by interpreting them as more succinct linear programs, via an interpretation that we call factorized interpretation. The soundness of this approach is delayed to Section~\ref{sec:weightings} as it contains results on weightings of conjunctive queries that are of independent interest. We then proceed to defining the language \LPCQ in Section~\ref{sec:closure}. The language is more expressive than \LPclosed as it allows for universal quantification over the database, as it is hinted in the previous example. We give its semantics via a closure operation that transforms an \LPCQ program to an \LPclosed program. We analyze the complexity of solving \LPCQ programs and show how one can leverage the results on \LPclosed to this program in Section~\ref{sec:lpcqcomplexity}. We present some preliminary experimental results in Section~\ref{sec:results}. Finally, Section~\ref{sec:applications} presents some applications of \LPCQ.

\section{Preliminaries}
\label{sec:prelim}

\textbf{Sets, Functions and Relations.}
Let $\Bool=\{0,1\}$ be the set of Booleans, $\N$ the set of natural numbers including $0$, 
$\Rp$ be the set of positive reals including $0$ and subsuming $\N$, and $\Reals$ the set of all reals.

Given any set $S$ and $n\in\N$ we denote by $S^n$ 
the set of all $n$-tuples over $S$ and by $S^*=\cup_{n\in\N} S^n$ the set of
all words over $S$. A \emph{weighting} on $S$ is a (total) function $f:S\to \Rp$.

Given a set of (total) functions $A\subseteq \domain^S = \{f \mid f:S\to \domain\}$ 
and a subset $S'\subseteq S$, we define the set of 
restrictions $ A_{|S'} = \{f_{|S'}\mid f\in A\} $.
For any binary
relation $R\subseteq S\times S$, we denote its
 transitive closure by $R^+\subseteq S\times S$
and the reflexive transitive closure by $R^*=R^+\cup\{(s,s)\mid s\in S\}$.

\medskip \noindent\textbf{Variable assignments.}
We fix a countably infinite set of (query) variables $\Vars$.
For any set $\domain$ of database elements, an 
assignment of (query) variables to database elements is a 
function $\alpha : X \to \domain$ that maps elements of a finite 
subset of variables $X \subseteq \Vars$ to values of $\domain$.
For any two sets of variable assignments $A_1 \subseteq \domain^{X_1}$ and $A_2 \subseteq \domain^{X_2}$ 
we define their join $A_1 \bowtie A_2 = \{\alpha_1 \cup \alpha_2 \mid \alpha_1 \in A_1, \alpha_2 \in A_2, 
    \projt{\alpha_1}{I} = \projt{\alpha_2}{I} \}$ where $I = X_1 \cap X_2$.

We also use a few vector notations.
Given a vector of variables $\Xtup = (x_1, \dots, x_n)\in\Vars^n$ 
we denote by $\Xset = \{x_1, \dots, x_n\}$ the set of the elements of $\Xtup$.
For any variable assignment $\alpha:X\to \domain$ with $\Xset\subseteq X$ we denote the application of the assignment $\alpha$ on $\Xtup$ 
by $\alpha(\Xtup) = (\alpha(x_1), \dots, \alpha(x_n))$. 

\subsection{Linear Programs}
\begin{figure}[h]
\[
\begin{array}{llcl}
\text{Linear expressions } & S,S' \in \SSUM 
    & ::= & r \mid \vlp \mid rS \mid S+S' \\
\text{Linear constraints } & C , C' \in \SLC
    & ::= & S \le S' \mid C \wedge C' \mid  \true\\
\text{Linear programs } & L \in \SLP 
    & ::= & \MAXIMIZE S ~ \SUBJECTTO C \\
\end{array}
\]

\caption{The set of linear programs $\SLP$  with variables $\vlp \in \lpvarset$ and constants $r\in \R$.}
\label{fig:lpsyntax}
\end{figure}
\newcommand{\minus}{\scalebox{0.75}[1.0]{$-$}}

Let $\lpvarset$ be a set of linear program variables. In
\Figure{fig:lpsyntax}, we recall the definition of the sets 
of linear expressions $\SSUM$, linear constraints
$\SLC$, and linear programs $\SLP$ with variables 
in $\lpvarset$.  We consider the usual linear equations $S \EQU S'$ as
syntactic sugar for the constraints $S \leq S' \wedge S' \leq S$.
For any linear program $$L= \MAXIMIZE S \SSUBJECTTO C$$ we call $S$ the
objective function of $L$ and $C$ the constraint of $L$. Note that
the linear program $\MINIMIZE S \SSUBJECTTO C$ can be expressed 
by $$\MAXIMIZE \minus 1 \cdot S \SSUBJECTTO C.$$

\noindent We recall the formal semantics of linear programs in \Figure{fig:lpeval}.

\begin{figure}[h]
\[
\begin{array}{ll}
    \eval{\w}{r}        
        & = r \phantom{\w(\vlp)}\text{\ \ \ with $r\in\R$, $\w:\lpvarset\to\R$}\\
    \eval{\w}{\vlp} 
        & = \w(\vlp) \phantom{r}\text{\ \ \ with $\vlp\in\lpvarset$}\\
    \eval{\w}{rS} 
        & = r \cdot \eval{\w}{S} \\
    \eval{\w}{S + S'} 
        & = \eval{\w}{S} + \eval{\w}{S'} \\

    \\ 

    \sem{\true}
        & = \{ \w \mid \w : \lpvarset \to \Rp\}\\

    \sem{S \leq S'} 
        & = \{ \w \mid
              \eval{\w}{S} \leq \eval{\w}{S'}\} \\

    \sem{C \wedge C'} & = \sem{C} \cap \sem{C'} \\

\end{array}
\]
\[
\begin{array}{l}
\opsem{\MAXIMIZE S ~ \SUBJECTTO C} \\
       \qquad  = max(\{\eval{\w}{S} \mid \w : \lpvarset \to \Rp,
         \w \in \sem{C}\})
        \end{array}
\]
\caption{Semantics of linear expressions, constraints and programs.}
\label{fig:lpeval}
\end{figure}

For any weightings $\w:\lpvarset\to
\Rp$, the value of a sum $S \in \SSUM$ is the real number $\eval{\w}{S} \in
\Reals$. We denote the solution set of a
constraint $C \in \SLC$ by $\sem{C}\subseteq \{\w\mid
\w:{\lpvarset}\to \Rp\}$. The optimal value
$\opsem{L}\in\R$ of a linear program $L$
with objective function $S$ and constraint $C$ is:
$$
\opsem{L}=\max\{ \eval{\w}{S} \mid \w:\lpvarset\to\Rp,\
\w\in\sem{C}\}
$$
The \emph{size} $|L|$ of a linear program $L$ is defined to be the number of symbols needed to write it down. It is well-known that the optimal solution of a linear program $L$ can be computed in polynomial time in $|L|$~\cite{DBLP:journals/combinatorica/Karmarkar84}.

Observe that we are only interested in non-negative weightings, without explicitly imposing positivity constraints. It is a usual assumption in linear programming since it is well known that one can transform any linear program $L$ into $L'$ of size at most $2|L|$ so that the feasible points of $L'$ over $\Rp$ are exactly the feasible points of $L$ over $\R$, by simply replacing every occurrence of a variable $x$ in $L$ by $x^+-x^-$.

\subsection{Conjunctive Queries}
\label{sec:conjQueries}

A \emph{relational signature} is a pair $\Sigma=(\Rels,\Consts)$
where $\Consts$ a finite set of constants ranged over by $c$
and $\Rels=\cup_{n\in\Nat}\RelsN$ is a finite set of relation symbols.
The elements $R\in\RelsN$ are called relation symbols of arity $n\in\Nat$.

\begin{figure}[h]
\arraycolsep=2pt
\[
\begin{array}{llcl}
\text{Expressions} & E_1,\ldots,E_n \inExpr 
    & ::= &x \mid c  \\
\text{Conjunctive queries} & Q ,Q' \in \CQ 
    & ::= & E_1\EQU E_2 \mid  R(E_1,\ldots,E_n)  
  \\

& &\mid &Q \wedge  Q' \mid \exists x. Q  \mid  \true 
\end{array}
\]
\caption{\label{fig:CQ}The set of conjunctive queries $\CQ$ with signature
  $\Sigma=((\RelsN)_{n\in\N},\Consts)$ where $x\in\Vars$, $c\in \Consts$, and
  $R\in\RelsN$. } 
\end{figure}
In \Figure{fig:CQ} we recall the notion of conjunctive queries.
An expression $E\inExpr$ is either a (query) variable $x\in \Vars$ or 
a constant $a\in \Consts$. The set of conjunctive queries $Q\in\CQ$
is built from equations $E_1\EQU E_2$, atoms
$R(E_1,\ldots, E_n)$, 
the logical operators of conjunction $Q\wedge Q'$ and
existential quantification $\exists x. Q$. Given a vector
 $\Xtup=(x_1,\ldots, x_n)\in \Vars^n$ and a query $Q$, we write $\exists \Xtup. Q$ instead of $\exists x_1.\ldots.\exists
 x_n.  Q$. For any sequence of constants
 $\Xtup[c]=(c_1,\ldots,c_n)\in\Consts^n$
 we write $\Xtup\EQU\Xtup[c]$ instead of
 $x_1\EQU c_1\wedge\ldots\wedge x_n\EQU c_n$.
 If $n=0$ then $\Xtup\EQU\Xtup[c]$ is equal to $\true$.

 The set of free variables $\fv(Q)\subseteq \Vars$ are those variables
 that occur in $Q$ outside the scope of an existential quantifier:
 $$
\begin{array}{l@{\qquad}l}
    \fv(R(E_1, \ldots, E_n)) = \bigcup_{i=1}^n \fv(E_i) &
    \fv(E_1 \LEq E_2) = \fv(E_1) \cup \fv(E_2) \\
    \fv(\Query \wedge \Query') = \fv(\Query) \cup \fv(\Query') &
    \fv(\exists x . \Query) = \fv(\Query) \setminus \{x\} \\
    \fv(x) = \{x\} &
    \fv(c) = \emptyset \\
\end{array}
$$

A  conjunctive query $Q$ is said to be \emph{quantifier free} if it does
 not contain any existential quantifier.  In the literature, such queries are sometimes also called full queries.

\ignore{The set of bound variables $bv(Q)$ is the set of variables $x$
introduce by some existential quantifier $\exists x. Q'$ in
some subexpression of $Q$. 
A query $Q$ is called \emph{renamed apart} if $\fv(Q)$ is
disjoint from $\bv(Q)$ and no variables $x$ is introduced twice by existential quantifiers. A \emph{renaming
of a query $Q$} is a query obtained from $Q$ by 
capture free renaming of bound variables.}

We can define operations to extend queries with additional variables $\Xtup$
such that for all $\Query\in\CQ$:
$$
\begin{array}{l}
ext_{\Xtup}(\Query) = \bigwedge_{x\in \Xset \setminus \fv(\Query)} x \LEq x \wedge \Query
\end{array}
$$
For any $n\ge 0$ and vector of constants $\Xtup[c]\in\Consts^n$  and vector of variables $\Xtup\in \Vars^n$
we define an operator $\subs{[\Xtup/\Xtup[c]]}$ on conjunctive queries, that
substitutes any variable in a vector $\Xtup$ by the constant at the
same position in vector $\Xtup[c]$, so that for all queries $\Query\in\CQ$:
$$
\begin{array}{l}
\subs{[\Xtup/\Xtup[c]]}(\Query) =  \Query[\Xtup/\Xtup[c]],
\end{array}
$$
i.e., where all occurrences in $\Query$ of variables in $\Xtup$ are replaced by the corresponding elements of $\Xtup[c]$.

\subsection{Relational Databases}

A relational $\Sigma$-structure is a tuple $\db=(\Sigma,D,\cdot^\db)$, where
$\Sigma=((\RelsN)_{n\ge 0},\Consts)$ is a relational signature, $D$ a finite set, $c^\db\in D$ an
element for each constant $c\in\Consts$ and $R^\db\subseteq D^n$
a relation for any relation symbol $R\in \RelsN$ and $n\ge 0$.
We also define the structures' domain $dom(\db)=D$.
  A \emph{(relational) database} $\db$ is a finite relational
  $\Sigma$-structure, i.e., all its components are finite.  We denote the set of all databases by $\DB$.

For any conjunctive query $Q\in \CQ$, set $X\supseteq \fv(Q)$
and relational database $\db\in\DB$ we define the answer set
$\SoldbX{Q}$
in \Figure{sem:cq}. It contains all those assignments $\alpha:X\to \dom$ for which $Q$ becomes
true on $\db$.

\begin{figure}[h]
\renewcommand\arraystretch{1.2}
\[
\begin{array}{rcl}
\assignVal\alpha\db(x)&=&\alpha(x) \phantom{c^\db}\text{\ \ \ (with $x\in\Vars$ a variable)}\\
\assignVal\alpha\db(c)&=&c^\db  \phantom{\alpha(x)}\text{\ \ \ (with $c\in\Consts$ a constant)}\\
\SoldbX{E_1\EQU E_2} &=& 
    \{ \alpha:X\to D \mid 
    \assignVal\alpha\db(E_1)=\assignVal\alpha\db(E_2) \} \\
\SoldbX{R(E_1, \ldots, E_n)} &= &
    \{ \alpha:X\to D \mid 
    (\assignVal\alpha\db(E_1), \ldots, \assignVal\alpha\db(E_n)) \in R^\db \} \\
\SoldbX{Q_1\wedge Q_2} &=& 
     \SoldbX{Q_1}\cap \SoldbX{Q_2}
\\
\SoldbX{\exists x. Q}& =& 
    \{\projt{\alpha}{X} \mid \alpha \in
                          \Soldbv{X\cup\{x\}}{Q} \} \qquad \text{if }
                          x\not\in X\\\
\SoldbX{\true} &=& X^D
\end{array}
\]
\caption{\label{sem:cq}The answer set of a conjunctive query $Q\in\CQ$ on a
  database $\db\in\DB$ for a  set of variables $X\supseteq \fv(Q)$.}
\end{figure}

\noindent We define the semantics of a query by:
$$
\sem{Q}^\db= \Soldbv{\fv(Q)}{Q}
$$
In particular, observe that the semantics of existential quantifiers is the projection
of the answer set, that is:
$\sem{\exists
\Xtup. Q}^\db = \projs{\sem{Q}^\db}{\fv(Q)\setminus\Xset}$.

\subsection{Hypertree Decompositions}
\label{sec:htd}

Hypertree decompositions of conjunctive queries are a way of laying out
the structure of a conjunctive query in a tree. It allows to solve
many aggregation problems (such as checking the existence of a
solution, counting or enumerating the solutions etc.) on
quantifier free conjunctive
queries in polynomial time where the degree of the polynomial is
given by the width of the decomposition.

A digraph is a pair $(\Vert,\Edg)$ with node set
$\Vert$ and edge set $\Edg\subseteq\Vert\times\Vert$.
A digraph is acyclic if
there is no $v\in\Vert$ for which $(v,v)\in\Edg^+$.
For any node $u \in \Vert$, we denote by
$\descorequal{u}=\{v\in\Vert\mid (u,v)\in\Edg^*\}$ the set of nodes in
$\Vert$ reachable over some downwards path from $u$, 
and we define the context of $u$, denoted $\context{u}$, by $\context{u}= (\Vert \setminus \descorequal{u}) \cup \{u\}$.
The digraph $(\Vert,\Edg)$ is a forest if it is acyclic and for all $u,u',v\in \Vert$ there holds that $(u,v),(u',v)\in \Edg$ implies $u=u'$.  Moreover, $(\Vert,\Edg)$ is a tree if there exists a node $r\in\Vert$ such that $\Vert=\descorequal{r}$.
In this case, $r$ is unique and called the root
of the tree.  If for $v\in\Vert$ it holds that $\descorequal{v}=\{v\}$, then $v$ is called a leaf.
Observe that in this tree, the paths are oriented from the root to the leaves of the tree.

\begin{defi}
    Let $X \subseteq \Vars$ be a finite set of variables.
    A \decomptree{} $T$ of $X$ is a tuple $(\Vert, \Edg, \Bag)$ such that:
    \begin{itemize}
        \item[-] $(\Vert,\Edg)$ is a finite directed rooted tree with edges from the root to the leaves,
        \item[-] the bag function $\Bag:\Vert\to 2^X$ maps nodes to subsets of variables
          in $X$,
        \item[-] for all $x \in X$ the subset of nodes $\{u \in \Vert \mid x \in \Bag(u)\}$ 
            is connected in the tree $(\Vert, \Edg)$,
        \item[-] each variable of $ X$ appears in some bag, that is $\bigcup_{u \in \Vert} \Bag(u) = X$.
    \end{itemize}
\end{defi}

Now a hypertree decomposition of a quantifier free conjunctive query
is a decomposition tree where for each atom of the query there is at least one bag 
that covers its variables.

\begin{defi}[Hypertree width of quantifier free conjunctive queries] 
Let $Q \in \CQ$ be a quantifier free conjunctive query. A \emph{generalized hypertree  decomposition of $Q$} 
is a decomposition tree $T=(\Vert,\Edg,\Bag)$ of $\fv(Q)$ such that 
for each atom $R(\Xs)$ of $Q$ there is a vertex $u \in \Vert$ such
that $\Set\Xs \subseteq \Bag(u)$.  The \emph{width} of $T$ with respect to $Q$ is 
the minimal number $k$ such that every bag of $T$ can be covered by
the variables of $k$ atoms of $Q$. 
The \emph{generalized hypertree width of a query $Q$} is the minimal
width of a tree decomposition of $Q$.
\end{defi}

For example, the query $R(x,y) \wedge R(y,z)$ has a
generalized hypertree decomposition of hypertree width $1$: $(\Vert,\Edg,\Bag)$ with
vertices $\Vert=\{1,2,3\}$, edges $\Edg=\{(1,2),(1,3)\}$, and bags $\Bag=[1/\{y\}, 2/\{x,y\},3/\{y,z\}]$.

While hypertree width allows to obtain efficient algorithms on conjunctive queries, our results will also work for the more general notion of \emph{fractional hypertree width}, which consists in a fractional relaxation of the hypertree width. We let $Q \in \CQ$ be a quantifier free conjunctive query, $A$ be the atoms of $Q$ and let $X \subseteq \fv(Q)$. A \emph{fractional cover of $X$} is a function $c \colon A \rightarrow \Rp$ assigning positive weights to the atoms of $Q$ such that for every $x \in X$, $\sum_{R \in A, x \in \fv(R)} c(R) \geq 1$. The \emph{value of a fractional cover $c$} is defined as $\sum_{R \in A} c(R)$.

For example, consider the query $Triangle = R(x,y) \wedge S(y,z) \wedge T(z,x)$ and $X = \{x,y,z\}$. The function $c$ such that $c(R) = c(S) = c(T) = 1/2$ is a fractional cover of $X$ of value $3/2$.

\begin{defi}
\label{def:fhtw} Let $Q$ be a conjunctive query and $T = (\Vert,\Edg,\Bag)$ be a generalized hypertree decomposition of $Q$. The \emph{fractional hypertree width} of $T$ is the smallest $k$ such that for every $u \in \Vert$, there exists a fractional cover of $\Bag(u)$ of value smaller than $k$. The \emph{fractional hypertree width} of $Q$, denoted by $\fhtw{Q}$, is the smallest $k$ such that $Q$ has a generalized hypertree decomposition of fractional hypertree width $k$. 
\end{defi}

\newcommand\bproj[1]{(\projs{#1}{\Bag(u)})_{u \in \Vert}}

From now on, we will only write the width of $T$ in place of the fractional hypertree width. The key observation making fractional hypertree width suitable for algorithmic purposes is due to Grohe and Marx~\cite{grohe2014constraint} who proved that if a quantifier free conjunctive query is such that $\fv(Q)$ has a fractional cover of value $k$, then $|\Soldb{Q}| \leq |\db|^k$. Hence, if $T = (\Vert, \Edg, \Bag)$ is a tree decomposition of $Q$ of width $k$, then $\Soldb{Q}_{|\Bag(u)}$ is of size at most $|\db|^k$. Moreover, it can be computed efficiently:

\begin{lem}
  \label{lem:computesol}
  Given a tree decomposition $T=(\Vert,\Edg,\Bag)$ of a quantifier
  free  conjunctive query $Q\in\CQ$ 
of width $k$ and a database $\db\in\DB$, one can compute the collection
of bag projections $\bproj{ \Soldb{Q}}$ 
in time $O((\size{\db}^k\log(\size{\db})) \cdot \size{\Vert})$. Moreover, for every $u \in \Vert$, $\Soldb{Q}_{|\Bag(u)}$ is of size at most $|\db|^k$.
\end{lem}

Lemma~\ref{lem:computesol} is folklore: it can be proven by computing
the semi-join of every bag in a subtree in a bottom-up fashion, as it
is done in \cite[Theorem 6.25]{libkin2013elements} and using a worst-case optimal join algorithm such as Triejoin~\cite{triejoin} for computing the relation at each bag. This yields a
superset $S_u$ of $\Soldb{Q}|_{\Bag(u)}$ for every $u$. Then, with a
second top-down phase, one can remove tuples from $S_u$ that cannot be
extended to a solution of $\Soldb{Q}$.

Following the previously mentioned upper bound of~\cite{grohe2014constraint}, Atserias, Grohe and Marx proved in~\cite{agm} that the bound given by an optimal fractional cover of $Q$ is tight (up to polynomial factors). This bound is now usually referred to as the AGM bound. More precisely, it says that if $\agm(Q)$ denotes the smallest value over every fractional cover of $Q$, then for every $\db$, $\Soldb{Q}$ is of size at most $|\db|^{\agm(Q)}$ and there exists a database $\db^*$  such that $\sem{Q}^{\db^*}$ is of size greater than $|\db^*|^{\agm(Q)} \over poly(|Q|)$. Hence, even if $Q$ is of width $k$, the size of $\Soldb{Q}$ could be order of magnitudes bigger than $|\db|^k$ when $k < \agm(Q)$. Hence, Lemma~\ref{lem:computesol} gives a succinct way of describing the set of solutions of $Q$ that we exploit in this paper. 

Parts of our result will be easier to describe on so-called
normalized decomposition trees:

\begin{defi}
Let $T=(\Vert,\Edg,\Bag)$ be a \decomptree.
We call a node $u \in \Vert$ of $T$:
\begin{description}
\item[- an \addnode node] if it has a single child $u'$ 
and $\Bag(u) = \Bag(u') \cup \{x\}$ for some $x \in \Vars \setminus \Bag(u')$,
\item[- a \forgnode node] if it has a single child $u'$ 
and $\Bag(u) = \Bag(u') \setminus \{x\}$ for some $x \in \Vars\setminus \Bag(u)$, 
\item[- a \joinnode node]  if it has $k\ge 1$ children
$u_1, ..., u_k$ with $\Bag(u) = \Bag(u_1) = ... = \Bag(u_k)$.
\end{description}
We call $T$ \emph{normalized}\footnote{In the literature this property is referred to as ``nice'' tree decompositions.} 
    if all its nodes in $\Vert$ are either \addnode{} nodes, \forgnode{} nodes, \joinnode{} nodes, or leaves. 
\end{defi}

It is well-known that tree decompositions can always be normalized
without changing the width. Thus 
normalization
does not change the asymptotic complexity of the algorithms.

\begin{lem}[Lemma of 13.1.2 of \cite{kloks1994treewidth}]
  For every tree decomposition of $T = (\Vert, \Edg, \Bag)$ of $Q$ of
  width $k$, there exists a normalized tree decomposition
  $T' = (\Vert', \Edg', \Bag')$ having width $k$. Moreover, one can
  compute $T'$ from $T$ in polynomial time.
\end{lem}

\section{Linear Programs with Closed Weight Expressions}
\label{lp-closed}

In this section,  we introduce the language $\LPclosed$ to express linear programs parameterized by conjunctive queries. This language is deliberately kept simple, which allow us to design efficient algorithms for it. An element of $\LPclosed$ is called a \emph{closed $\LP(\CS)$ program}. We refer to such programs as ``closed'' because they do not contain quantification in the linear program part, which contrasts with the more general definition of  $\LP(\CS)$ given  in Section~\ref{sec:lpgeneral}, which allows to express more interesting linear program. The case of closed $\LP(\CS)$ programs is however central in this work as this is the class of optimization problems for which we propose an efficient algorithm. The more general case of $\LP(\CS)$ programs is dealt with using a ``closure'' procedure which transforms any programs from $\LP(\CS)$ into a closed program $\LPclosed$. 

Let $\Sigma$ be a relational signature. A \emph{closed weight expression} is an expression of the form $\weight{\Query}{\Xs\LEq \Xtup[c]}$ where $Q$ is a conjunctive query, $\Xset \subseteq \fv(\Query)$, variables in $\Xs$ are pairwise distinct and $\Xtup[c]$ are database constants. An $\LPclosed$ program is intuitively a linear program whose variables are closed weight expressions. A formal definition is given in \Figure{fig:lp-closed}. 

\begin{figure}[H]
$$
\begin{array}{llcl}
 
\text{Linear sums } & S,S' \in \LSclosed 
  & ::= &  \weight{\Query}{\Xs\LEq \Xtup[c]}\mid  \real S \mid S+S' \mid \real \\
                    &&&\qquad \text{where } \Xset \subseteq \fv(\Query),\\
                    &&&\qquad \text{variables in } \Xs \text{ are pairwise distinct}\\
                    &&& \qquad \Xtup[c] \text{ are database constants} \\
\text{Linear constraints } & C , C' \in\LCclosed 
    & ::= & S \le S' \mid S\LEq S' \mid C \wedge C' \mid  \true\\
   \text{Linear programs } & L \in \LPclosed
    & ::= & \MAXIMIZE S \  \SUBJECTTO C \\
   \end{array}
$$
\caption{Linear sums, constraints, and programs with
  closed weight expressions containing conjunctive
  queries $\Query\in \CQ$ where $\real\in\R$.}
\label{fig:lp-closed}
\end{figure}

\noindent Such linear programs can be interpreted as standard linear
programs for any database $\db$ with numerical values.
In order to do so, we fix for any query $\Query\in\CQ$
a set $\tups[Q]  $ of fresh linear program variables $\theta^\alpha_\Query
$ arbitrarily:
$$
 \tups[Q] =  \{\theta^\alpha_\Query\mid \alpha\in\sem{\Query}^\db\}
$$
We can then
map each closed weight expression to  a linear sum
with variables in $\tups[Q]$ as follows:
\[
    \interpret{\weight{\Query}{\Xs\LEq \Xtup[c]}}^{\db}
    = \sum\limits_{\substack{\alpha \in \sem\Query^\db\\\alpha(\Xs)=\Xtup[c]}}
    {\theta_{\Query}^\alpha }
  \]
  Note that our assumption $\Xset \subseteq \fv(Q)$ ensures that
  $\alpha(\Xs)$ is well-defined.
  
  \begin{defi}
    For any  linear program  $L \in \LPclosed$ we define the \emph{natural 
    interpretation} $\interpret{L}^\db\in\LP$
  by replacing any weight expression
  $S$ in $L$ by $\interpret{S}^\db$. By applying the
  same substitution we define
  the interpretation $\interpret{S}^\db\in\LS$ of any linear
  sum $S\in\LSclosed$ and the interpretation
  $\interpret{C}^\db\in\LC$ of any linear
  constraint $C\in\LCclosed$  in analogy.
\end{defi}

The \emph{size} $|L|$ of a program $L \in \LPclosed$ is defined to be the number of symbols needed to write it down. 

\subsection{Example}
\label{sec:easyexample}
As an example we consider the conjunctive query $Q = R_1(x) \wedge
R_2(y)$ and the following program $L\in\LPclosed$:
\[
\begin{array}{rl}
\MAXIMIZE & \weight{Q}{\emptyset} \\
\SUBJECTTO 
    &  {\weight{Q}{x \EQU 0} \leq 1} \\
  \wedge &
           \weight{Q}{x \EQU 1} \leq 1
\end{array}
\]
Let $\db$ be the database $\db$ with tables $R_1^\db = \{ (0), (1) \}$ and  $R_2^\db = \{ (0), (1) \}$.
The answer set of $Q$ is $\sem{Q}^\db=  \{ \alpha \mid \alpha:\{x,y\}\to \{0,1\} \}$.
The interpretation $\interpret{L}^{\db}$ is the following linear program, where
we denote any query answer $\alpha\in \sem{Q}^\db$ by
a pair $(\alpha(x), \alpha(y))$ in the Cartesian product $\{0,1\}^2$
for simplicity:
\[
\begin{array}{rl}
\MAXIMIZE & \thetaQpar{0, 0} + \thetaQpar{0, 1} + \thetaQpar{1, 0} + \thetaQpar{1, 1} \\
\SUBJECTTO 
    & \thetaQpar{0, 0} + \thetaQpar{0, 1} \leq 1 \\
    \wedge & \thetaQpar{1, 0} + \thetaQpar{1, 1} \leq 1
\end{array}
\]
The objective function $\thetaQpar{0, 0} + \thetaQpar{0, 1} + \thetaQpar{1, 0} + \thetaQpar{1, 1}$ 
is the interpretation of the expression $\weight{Q}{\emptyset}$.
The first constraint $\thetaQpar{0, 0} + \thetaQpar{0, 1}$ 
is obtained by interpreting $\weight{Q}{x\EQU
  0}\le 1$
and the second constraint by interpreting $\weight{Q}{x \EQU
  1}\le 1$.
Note that the objective function is the sum of the lefthandsides of the two constraints,
so the three weight expressions of $L$ are semantically related.

\subsection{Complexity of solving \texorpdfstring{$\LPclosed$}{LP_clos(CQ_Σ)}}
\label{sec:complexity}
In this section, we are interested in the complexity of computing $\opsem{\interpret{L}^\db}$ given $L \in \LPclosed$ and a database $\db$. From a combined complexity point of view, that is, when both $L$ and $\db$ are assumed to be part of the input, it is not hard to see that the problem is $\NP$-hard since it requires to implicitly find the answer set of every conjunctive query appearing in $L$. We formalize this intuition in the following theorem: 
\begin{thm}
  \label{thm:complexity} The problem of deciding whether $\opsem{\interpret{L}^\db} \neq 0$ given a relational signature $\Sigma$, $L \in \LPclosed$ and a database $\db$ is both $\NP$-hard and $\coNP$-hard.
\end{thm}
\begin{proof}

It is well-known that the problem of deciding whether $\sem{Q}^\db \neq \emptyset$ given a conjunctive query $Q$ and a database $\db$ in the input is $\NP$-complete~\cite{chandra77}. We show that this problem can be reduced to the problem of deciding whether the optimal value $\opsem{\interpret{L}^\db}$ is non-zero, given a relational signature $\Sigma$, a linear program $L\in \LPclosed$ and a database $\db$ with schema $\Sigma$. The $\NP$-hardness of computing $\opsem{\interpret{L}^\db}$ is thus a direct corollary.
  
For any conjunctive query $Q$, we consider the following $\LPclosed$ program:
  $$L_Q = \textbf{maximize } \weight{Q}{true} \textbf{ subject
    to } \weight{Q}{true} \leq 1$$

We claim that 
  $$\opsem{\interpret{L_Q}^\db} \neq 0 \text{ if and only if
  }\sem{Q}^\db \neq \emptyset$$
  We first note that:
$$\interpret{L_Q}^\db = \textbf{maximize } S \textbf{ subject to }
  S \leq 1\qquad\text{ where } S = \sum_{\alpha \in \Soldb{Q}}
  \theta^\alpha_Q
  $$
  So, if $\sem{Q}^\db = \emptyset$, then $\eval\w{S}=0$
  for all $\w$, so that $\opsem{\interpret{L_Q}^\db}=0$.
  So consider the other case where $\sem{Q}^\db \neq \emptyset$.  Let
  $\alpha \in \Soldb{Q}$ and consider the weighting $\w$
  such that $\w(\theta^\alpha_Q) = 1$ and $\w(\theta^{\alpha'}_Q) = 0$ for every
  $\alpha' \in \Soldb{Q}$ such that $\alpha'\neq \alpha$. This weighting clearly
  respects the constraints of $S\le 1$, so $\w\in\sem{S\le 1}$ showing that
  $\opsem{\interpret{L_Q}^{\db}} \ge 1\not=0$.

  To show the $\coNP$-hardness, it is sufficient to observe that the same trick can be applied to reduce the problem of deciding whether $\Soldb{Q} = \emptyset$ given $Q$ and $\db$, which is $\coNP$-complete from \cite{chandra77}.
\end{proof}

\paragraph{Data complexity.} The hardness from Theorem~\ref{thm:complexity} mainly stems from the hardness of answering conjunctive queries, that is only relevant in the context of combined complexity. It is often assumed however that the size of the query is small with respect to the size of the data, hence one can study the data complexity of the problem, that is, the complexity of the problem when we fix the linear program $L$. In this case, computing $\opsem{\interpret{L}^\db}$ can be done in polynomial time in $|\db|$ using the following procedure:

\begin{itemize}
\item Compute explicitly $\Soldb{Q}$ for every $Q$ appearing in $L$,
\item Compute $\interpret{L}^\db$,
\item Solve $\interpret{L}^\db$ in time polynomial in $|\interpret{L}^\db|$ using an LP-solver.
\end{itemize}

The exact complexity of the previous procedure is however dependent on the size of $\interpret{L}^\db$ whose number of variables is the sum of $|\Soldb{Q}|$ for every $Q$ appearing in $L$. We lift the AGM bound presented in Section~\ref{sec:htd} to linear programs in $\LPclosed$ by defining $\agm(L)$ to be the maximum of $\agm(Q)$ for every query $Q$ appearing in $L$. The size of $\interpret{L}^\db$ can now be upper bounded by $|L| \times |\db|^{\agm(L)}$. Using a worst-case optimal join algorithm such as Triejoin~\cite{triejoin} to compute $\Soldb{Q}$ in time $O(|\db|^{\agm(Q)})$, we conclude that one can compute $\opsem{\interpret{L}^\db}$ in time $O((|L||\db|^{\agm(L)})^\ell)$, where $\ell$ is the best known constant to compute the optimal value of a linear program. Currently, the best known value for $\ell$ is smaller than $2.37286$ by combining a result relating the complexity of solving linear programs with the complexity of multiplying matrices~\cite{cohen2021solving} with the best known algorithm for multiplying matrices~\cite{alman2021refined}.

\begin{thm}
  \label{thm:agmcomplexity} Given a relational signature $\Sigma$, $L \in \LPclosed$ and a database $\db$, one can compute $\opsem{\interpret{L}^\db}$ in time $O(|L|^{\ell}|\db|^{\ell \cdot \agm(L)})$ with $\ell < 2.37286$.
\end{thm}

\subsection{Replacement Rewriting}
\label{sec:replrew}

In this paper, we will often introduce alternate ways of interpreting linear program of $\LPclosed$ over a database. We will say that such alternate interpretation is \emph{sound} if for any database $\db$, when interpreting the linear program using this alternate interpretation over $\db$, we obtain a linear program having the same optimal value as  $\interpret{L}^\db$ , the natural interpretation of $L$ over $\db$. It will allow us for example to construct smaller linear programs and hence speed up the computation of $\opsem{\interpret{L}^\db}$.  Formally proving the soundness of an alternate interpretation  is usually tedious as it involves transforming solutions from one interpretation to the other while proving by induction that the constraints in $L$ are all satisfied. However, every interpretation that we will consider in this paper is based on interpreting weight expressions differently with some extra equality constraints. Hence, most of the time, the reasoning may be reduced to very simple linear programs involving only equality constraints. Our goal in this section is to provide formal tools to simplify soundness proofs.

One can always construct a function $\nu$ mapping every possible weight expression $W$ into a fresh linear program variable $\nu(W)$ such that if $W \neq W'$, then $\nu(W) \neq \nu(W')$. We will often denote $\nu(\weight{Q}{\Xtup\LEq\Xtup[c]})$ by $\nu_Q^{\Xtup\LEq\Xtup[c]}$.
From now on, we assume such function $\nu$ has been fixed.
For a set of weight expressions $\wset$, we define the \emph{weight constraints of $\wset$}, denoted by $\wC[\wset]$ as the following set of linear constraints:

\[
  \bigwedge_{W \in \wset} \nu(W) = W.
\]

For for any linear sum $S\in\LSclosed$, let $\interpret{S}^\nu\in\LS$  be defined by replacing any weight expression $W$ in $S$ by $\nu(W)$ and for $C\in\LCclosed$, let $\interpret{C}^\nu\in\LP$ be the linear constraint obtained by applying the substitution to every linear sum appearing in $C$.

Consider a linear program $L=   \MAXIMIZE S \ \SUBJECTTO C$ with some linear sum $S\in\LSclosed$ and some linear constraint $C\in\LCclosed$. We denote  by $\wset(L)$ the set of weight expressions that appear in $L$. The replacement rewriting of $L$ is the following linear program $\Rho(L) \in\LP(\CQ)$:
$$
\Rho(L) = \MAXIMIZE \interpret{S}^\nu \  \SUBJECTTO \interpret{C}^\nu \wedge \wC[\wset(L)].
$$

Observe that in $\Rho(L)$, the only place where $\weight{Q}{\Xs \LEq \Xtup[c]}$ constructors appear is in the $\wC[\wset(L)]$ part. Hence, we can naturally lift the interpretation of $L$ over a database $\db$ to $\Rho(L)$ as follows:
$$
\interpret{\Rho(L)}^\db = \MAXIMIZE \interpret{S}^\nu \  \SUBJECTTO \interpret{C}^\nu 
\wedge 
\interpret{\wC[\wset(L)]}^\db.
$$

The main feature of $\Rho(L)$ is that it allows to formally separate the linear programming part from the part that is interpreted over a database, which will be helpful whenever we need to reason only on weight expressions.

\paragraph{Example.} In the example of Section~\ref{sec:easyexample}, the rewriting $\Rho(L)$ of $L$ is:
$$
\begin{array}{rl}
  \MAXIMIZE  & \nu_2 \\
  \SUBJECTTO & \nu_0\ \leq\ 1\ \wedge \\
             & \nu_1\ \leq\ 1\ \wedge \\
              \\
             & \nu_0\ \LEq\ \weight{Q}{x \EQU 0}\ \wedge \\
             & \nu_1\ \LEq\ \weight{Q}{x \EQU 1}\ \wedge \\
             & \nu_2\ \LEq\ \weight{Q}{\emptyset}\\  
\end{array}.
$$

which is then interpreted as $\interpret{\Rho(L)}^\db$ as:
$$
\begin{array}{rl}
  \MAXIMIZE  & \nu_2 \\
  \SUBJECTTO & \nu_0\ \leq\ 1\ \wedge \\
             & \nu_1\ \leq\ 1\ \wedge \\
             & \\
             & \nu_0\ \LEq\ \thetaQpar{0, 0} + \thetaQpar{0, 1}\ \wedge \\[3pt]
             & \nu_1 \ \LEq\ \thetaQpar{1, 0} + \thetaQpar{1, 1}\ \wedge \\[3pt]
             & \nu_2 \ \LEq\ \thetaQpar{0, 0} + \thetaQpar{0,1} + \thetaQpar{1,0} + \thetaQpar{1,1}
\end{array}.
$$
Note that the linear program's variables $\thetaQpar{i,j}$ do not occur in the objective
function, so they are implicitly existentially quantified. Up to
existential quantification, the above constraint
is equivalent to:
$$
   \nu_2\ \LEq\ \nu_0+\nu_1
   $$
   So we obtain a linear program with much fewer variables:
\[
\begin{array}{c}
  \MAXIMIZE  \nu_2\ \SUBJECTTO 
            \ \nu_0 \le 1 \wedge  \nu_1 \leq 1 \wedge \nu_2\ \LEq\ \nu_0+\nu_1 
\end{array}
\]
The optimal value of this new linear program is achieved with the solution $\nu_0=\nu_1=1$ and $\nu_2=2$. We can see that it directly corresponds to the optimal solution of the original linear program where $\thetaQpar{0, 0}=\thetaQpar{0, 1}=\thetaQpar{1, 0}=\thetaQpar{1, 1}={1 \over 2}$.
How to derive such factorized rewritings of constraints and how to reconstruct an optimal solution from the optimal solution of the rewritten program from a given $LP(\CQ)$ program in a systematic manner is studied in the remainder of this article.

It is easy to see that for every database $\db$ over signature $\Sigma$, the optimal value of $\interpret{\Rho(L)}^\db$ is the same as the optimal value of $\interpret{L}^\db$ which is formalized as follows:

\begin{prop}[Soundness of replacement rewriting]\label{prop:repl}
  Given a database $\db$ with signature $\Sigma$ and a linear program $L\in\LPclosed$, we have:
  $$
  \opsem{\interpret{L}^\db}=\opsem{\interpret{\Rho(L)}^\db}.
  $$
\end{prop}
\begin{proof}

  For every weight expression $\weight{Q}{\Xtup \LEq \Xtup[c]}$ of $L$ and $w \colon \tups[Q] \rightarrow \Rp$, we extend $w$ to $\nu_Q^{\Xtup\LEq\Xtup[c]}$ by defining \[w(\nu_Q^{\Xtup\LEq\Xtup[c]}) := \sum\limits_{\substack{\alpha \in \sem\Query^\db\\\alpha(\Xs)=\Xtup[c]}} w(\theta_{\Query}^\alpha ).\]

  Clearly, by definition, this extension of $w$ satisfies the weight constraint \[\nu_Q^{\Xtup\LEq\Xtup[c]} \EQU \sum\limits_{\substack{\alpha \in \sem\Query^\db\\\alpha(\Xs)=\Xtup[c]}} \theta_{\Query}^\alpha \] that appears in $\interpret{\Rho(L)}^\db$. Moreover, for any sum expression $S$ of $L$, $\eval{w}{\interpret{S}^\db}$  will give the same value as $\eval{w}{\interpret{\Rho(S)}^\db}$ as every weight expression $W$ of $S$ has been replaced in $\Rho(S)$ by $\nu(W)$ and that from what precedes $w(\nu(W))$ has the same value as $\eval{w}{\interpret{W}^\db}$. Hence, any solution of $\interpret{L}^\db$  can be extended to a solution of $\interpret{\Rho(L)}^\db$ that evaluates to the same objective value.

  On the other hand, with the same reasoning, any solution $w$ of $\interpret{\Rho(L)}^\db$ directly gives a solution of  $\interpret{L}^\db$ that evaluates to the same objective value since the weight constraints ensure that $\eval{w}{\interpret{W}^\db} = w(\nu(W))$.
\end{proof}

\subsection{Interpretations of linear programs}
\label{sec:interpretation}

In this section, we formalize the notion of interpretation of linear programs in $\LPclosed$ and give necessary conditions for an alternate interpretation to be sound. In the following, we denote by $\lcqs{L}$ the set of conjunctive queries that appear in $L$, that is, the set of $Q$ such that there is an expression of the form $\weight{Q}{\Xs \LEq \Xtup[c]}$ in $L$.

An \emph{interpretation} $I = (I_W, I_C)$ of $\LPclosed$ is a pair of functions such that, given a database $\db$ over signature $\Sigma$:
\begin{itemize}
\item $I_W$ maps every weight expression $W := \weight{Q}{\Xtup \LEq \Xtup[c]}$ and database $\db$ to a linear sum $I_W^\db(W)$ over linear program variables $X_{I,Q}^\db$,
\item $I_C$ maps every conjunctive query $Q$ over signature $\Sigma$ and database $\db$ to a set of constraints $I_C^\db(Q)$ on variables $X_{I,Q}^\db$,
\item and for every two conjunctive queries $Q,Q'$, if $Q \neq Q'$ then $X_{I,Q}^\db \cap X_{I,Q'}^\db = \emptyset$.

\end{itemize}

Given an interpretation $I$, a conjunctive query $Q$ and a set of weight constraints $\wset$ over $Q$, we denote by:
$$I_Q^\db(\wset) := I_C^\db(Q) \wedge \bigwedge_{W \in \wset} \nu(W) = I_W^\db(W)$$ where $\nu$ is a fixed function as constructed in Section \ref{sec:replrew}.
Observe that when $Q' \neq Q$, then $I_Q^\db(\wset)$ and $I_{Q'}^\db(\wset[]')$ have disjoint variables.

Given a linear program $L=   \MAXIMIZE S \ \SUBJECTTO C$ with some linear sum $S\in\LSclosed$ and some linear constraint $C\in\LCclosed$, we denote by $\wset[Q](L)$ the set of weight expressions of $L$ over $Q$. The \emph{$I$-interpretation of $L$} is the following linear program $I^\db(L)$:
$$
\begin{array}{rll}
  I^\db(L) = &  \MAXIMIZE  & \interpret{S}^\nu  \\
             & \SUBJECTTO & \interpret{C}^\nu\ \wedge\\
             & & \bigwedge_{Q\in \lcqs{L}} I_Q^\db(\wset[Q](L)).
\end{array}
$$

For example, the replacement rewriting of a linear program in $L$ can be defined by the interpretation $N=(N_W,N_C)$ such that $N_C^\db(Q) := true$ and
\[ N_W(\weight{Q}{\Xtup\LEq\Xtup[c]})^\db := \sum\limits_{\substack{\alpha \in \sem\Query^\db\\\alpha(\Xs)=\Xtup[c]}} \theta_{\Query}^\alpha. \]
It is readily verified that $N^\db(L)$ corresponds to $\Rho(L)$.

Since, for $Q \neq Q'$, the variables of $I_Q^\db(L)$ and of $I_{Q'}^\db(L)$ are disjoint, it allows us to prove the following sufficient condition for an interpretation to be sound, that only depends on the value of the interpretation on weight expressions over one conjunctive query $Q$:

\begin{prop}
  \label{prop:metasoundness} Let $I = (I_W,I_C)$ be an interpretation of $\LPclosed$ such that for every conjunctive query $Q$ and set $\wset$ of weight expressions over $Q$ we have that $\projs{\sem{I_Q^\db(\wset)}}{\nu(\wset)} = \projs{\sem{\interpret{\wC[\wset]}^\db}}{\nu(\wset)}$. Then $I$ is sound, that is, for every $L \in \LPclosed$ and database $\db$, $\opsem{\interpret{L}^\db} = \opsem{I^\db(L)}$.
\end{prop}

\begin{proof}
  Let $L=\MAXIMIZE S \  \SUBJECTTO C$. By Proposition \ref{prop:repl}, $$\opsem{\interpret{L}^\db} = \opsem{\interpret{\Rho(L)}^\db}.$$
  Hence it is sufficient to show that $\opsem{\interpret{\Rho(L)}^\db} = \opsem{I^\db(L)}$.

Now recall that $\interpret{\Rho(L)}^\db$ and $I^\db(L)$ have the same objective function $\interpret{S}^\nu$ and the same constraint $\interpret{C}^\nu$ on variables $\nu(\wset)$. Only the last part of the program is different.

$I^\db(L)$ contains additional constraints $\bigwedge_{Q\in \lcqs{L}} I_Q^\db(\wset[Q](L))$ and $\interpret{\Rho(L)}^\db$ contains $\interpret{\repl(\wset(L)) }^\db$.

However, the assumption from the statement applied to every $Q \in \lcqs{L}$ shows that $$\sem{\interpret{\repl(\wset(L)) }^\db}_{|\nu(\wset)} = \sem{\bigwedge_{Q\in \lcqs{L}} I_Q^\db(\wset[Q](L))}_{|\nu(\wset)}$$  since every $I_Q^\db(\wset[Q](L))$ contains disjoint variables.

It thus means that for any solution $w$ of constraints $\interpret{\repl(\wset(L)) }^\db$, one can construct a solution $w'$ of constraints $\bigwedge_{Q\in \lcqs{L}} I_Q^\db(\wset[Q](L))$ that assign variables $\nu(\wset)$ to the same values.  Hence, we have $\eval{w'}{\interpret{S}^\nu}=\eval{w}{\interpret{S}^\nu}$. Moreover, since $\eval{w}{\interpret{C}^\nu}$ is true by definition, and since  $w$ and $w'$ coincide on $\nu$ variables and that $\interpret{C}^\nu$ only contains $\nu(\wset)$ variables, $\eval{w'}{\interpret{C}^\nu}$ is also true. Hence $w'$ is a solution of $I^\db(L)$ and the values of both linear programs on $w$ and $w'$ respectively coincide. Taking $w$ so that it is optimal for $\interpret{\Rho(L)}^\db$ yields that $\opsem{\interpret{\Rho(L)}^\db} \leq \opsem{I^\db(L)}$.

On the other hand, we also have that given a solution  $w'$ of constraints $$\bigwedge_{Q\in \lcqs{L}} I_Q^\db(\wset[Q](L)),$$ one can construct a solution $w$ of constraints $\interpret{\repl(\wset(L)) }^\db$ that assign variables $\nu(\wset)$ to the same value. By the same reasoning, it implies that $\opsem{\interpret{\Rho(L)}^\db} \geq \opsem{I^\db(L)}$, and the equality follows.
\end{proof}

\paragraph{Example.} To illustrate the notion of interpretations, we will consider a toy alternative interpretation $I=(I_W, I_C)$ defined as follows: the linear program variables $X_{I,Q}^\db$ are defined as $\{x_{Q}^{\alpha} \mid \alpha \in \Soldb{Q}\}$ and $I_W$ is defined as:

\[
    I_W^\db(\weight{\Query}{\Xs\LEq \Xtup[c]})
    = \sum\limits_{\substack{\alpha \in \sem\Query^\db\\\alpha(\Xs)=\Xtup[c]}}
    {3 \times x_{\Query}^\alpha }
  \]

  We also define $I_C^\db$ as

  \[
    \sum\limits_{\substack{\alpha \in \sem\Query^\db}}
    { x_{\Query}^\alpha } \geq 0
  \]

  The interpretation $I^\db(L)$ of the example given in Section~\ref{sec:easyexample} is thus:

  $$
\begin{array}{rl}
  \MAXIMIZE  & \nu_2 \\
  \SUBJECTTO & \nu_0\ \leq\ 1\ \wedge \\
             & \nu_1\ \leq\ 1\ \wedge \\
             & \\
             & \nu_0\ \LEq\ 3x_Q^{[x/0, y/0]}+3x_Q^{[x/0,y/1]} \ \wedge \\[3pt]
             & \nu_1 \ \LEq\ 3x_Q^{[x/1, y/0]}+3x_Q^{[x/1,y/1]} \ \wedge \\[3pt]
             & \nu_2 \ \LEq\ 3x_Q^{[x/0, y/0]}+3x_Q^{[x/0,y/1]}+3x_Q^{[x/1, y/0]}+3x_Q^{[x/1,y/1]} \ \wedge\\[3pt]
             &  x_Q^{[x/0, y/0]}+x_Q^{[x/0,y/1]}+x_Q^{[x/1, y/0]}+x_Q^{[x/1,y/1]} \geq 0
\end{array}.
$$

The last constraint corresponds to $I_C^\db(Q)$. It is clear that this program has the same optimal value as the original one because it is obtained by substituting $\tup{Q}{\alpha}$ by $3x_Q^\alpha$ and by adding a constraint that is always true since $x_Q^\alpha \geq 0$. Actually, Proposition~\ref{prop:metasoundness} tells us that for every LPCQ $L$ and database $\db$, $I^\db(L)$ has the same optimal value as $\interpret{L}^\db$. Indeed, given a solution $w$ of $\interpret{\wC[\wset]}^\db$, we can transform it into a solution $w'$ of $I_Q^\db(\wset)$ defined as $w'(x_Q^\alpha) = {w(\tup{Q}{\alpha}) \over 3}$ and $w'(\nu(W)) = w(\nu(W))$. It is readily verified that $w'$ respects every constraint of $I_C^\db(\wset)$ and that $w'$ has the same values as $w$ on variables $\nu(\wset)$. Similarly, a solution $w'$ of $I_Q^\db(\wset)$ can be transformed into a solution $w$ of $\interpret{\wC[\wset]}^\db$ by defining $w$ as ${w(\tup{Q}{\alpha})} = 3w'(x_Q^\alpha)$ and $w(\nu(W)) = w'(\nu(W))$. Hence $\projs{\sem{I_Q^\db(\wset)}}{\nu(\wset)} = \projs{\sem{\interpret{\wC[\wset]}^\db}}{\nu(\wset)}$ and Proposition~\ref{prop:metasoundness} can be applied.
\section{Solving \texorpdfstring{$\LPclosed$}{LP_clos(CQ_Σ)} linear programs efficiently}
\label{sec:eff_fragment}

In this section, we propose an algorithm for solving linear programs in $\LPclosed$ that is better than the one given in Theorem~\ref{thm:agmcomplexity}. The proof of Theorem~\ref{thm:complexity} suggests that the main source of intractability stem from the complexity of answering conjunctive queries, which is reflected in the upper bound given in Theorem~\ref{thm:agmcomplexity} where the complexity here depends on the worst-case size of the answer sets of queries. However, for many problems on conjunctive queries such as computing the number of answers, one can get better upper bounds by exploiting the fact that they have small fractional hypertree width. By leveraging the notion of fractional hypertree width from conjunctive queries to linear programs of $\LPclosed$, we are able to lower the complexity from $O(|L|^{\ell}|\db|^{\ell \agm(L)})$ given by Theorem~\ref{thm:agmcomplexity} to $O(|L|^{\ell}|\db|^{\ell \fhtw{L}})$, where $\fhtw{L}$ denotes the (leveraged notion of) fractional hypertree width of $L$, when an optimal tree decomposition of $L$ is provided in the input.

To achieve this, we avoid the expensive step of computing $\interpret{L}^\db$ by exploiting tree decompositions $\trees{}$ of the queries of $L$ to generate a smaller linear program $\rewe{L}$ having only $O(|\db|^{\fhtw{\trees{}}})$ variables and show that the optimal value for $\rewe{L}$ is the same as the optimal value of $\interpret{L}^\db$.

\subsection{Tree decomposition of \texorpdfstring{$\LPclosed$}{LP_clos(CQ_Σ)}}

We start by lifting the concept of hypertree decompositions from conjunctive queries to linear programs in $\LPclosed$. Given a linear program $L \in \LPclosed$, recall that we denote by $\lcqs{L}$ the set of conjunctive queries that appear in $L$.

Intuitively, our notion of hypertree decomposition for $\LPclosed$ will consist in a collection of hypertree decomposition for every $\exists \Xtup[y]. Q \in \lcqs{L}$. However, we will need a stronger condition on the decomposition than the usual one:

\begin{defi}\label{def:compatible}
  Let $L \in \LPclosed$, $\exists \Xtup[y].Q \in \lcqs{L}$, with $Q$ the quantifier free part of $\exists \Xtup[y].Q$ and $\weight{\exists \Xtup[y].Q}{\Xs \LEq \Xtup[c]}$ a weight expression of $L$.

  A tree decomposition $T=(\Vert,\Edg,\Bag)$ of $\exists \Xtup[y].Q$ \emph{compatible with $\weight{\exists \Xtup[y].Q}{\Xs \LEq \Xtup[c]}$} is a tree decomposition of $Q$ such that there exists $u \in \Vert$ with $\Xset = \Bag(u)$.

  $T$ is said to be \emph{compatible with $L$} if it is a tree decomposition of $Q$ compatible with every $\weight{\exists \Xtup[y].Q}{\Xs \LEq \Xtup[c]}$ of $L$.
\end{defi}

Observe that the requirement that a tree decomposition has to be compatible with the linear program may increase the optimal width of the decomposition. For example, consider the conjunctive query $Q = R(x,y) \wedge S(y,z)$. If a linear program $L$ contains an expression of the form $\weight{Q}{x \LEq 0,z \LEq 0}$, then every tree decomposition of $Q$ compatible with $L$ has width at least $3/2$ whereas optimal tree decomposition for $Q$ have width $1$ (a decomposition of $Q$ is given in Section~\ref{sec:htd}).

\begin{defi}
  \label{def:htdec}
  Let $L$ be an \LPclosed program. A \emph{tree decomposition of $L$} is defined to be a collection $\trees{L} = (T_Q)_{Q \in \lcqs{L}}$ such that $T_Q$ is a tree decomposition of $Q \in \lcqs{L}$ that is compatible with $L$.
  
  The \emph{width of $\trees{L}$} is defined to be the maximal width of the decomposition trees in $\trees{L}$. The \emph{size of $\trees{L}$} is defined to be $|\trees{L}| = \sum_{(\Vert,\Edg,\Bag)\in \trees{} }|\Vert|$.
\end{defi}

\subsection{Factorized Interpretation of quantifier free \texorpdfstring{$\LPclosed$}{LP_clos(CQ_Σ)}}
\label{sec:factint}
In this section, we present a more succinct way of interpreting weight constraints in linear programs in $\LPclosed$, called the factorized interpretation, that exploits a tree decomposition of the queries. In this section, we assume that the linear program only contains quantifier free queries. We will explain in Section~\ref{sec:existsremove} how one can reduce the case with existentially quantified queries to the case of quantifier free queries.

From now on, we fix a linear language $L \in \LPclosed$ and $\trees{}$ a tree decomposition of $L$. We will describe an interpretation $\rewei=(\rewew, \rewec)$ of $L$ exploiting tree decompositions.

\paragraph{Interpreting weight constraints.} Given a conjunctive query $Q \in \lcqs{L}$ and $T = \trees{Q}$ the tree decomposition of $Q$ compatible with $L$ given by $\trees{}$, we define an interpretation of weight expressions on the following set of variables: 

$$
\weightbs[Q] := \{ \weightB{Q}{u}{\beta}\mid u \in \Vert,\ \beta \in \projsb{\Soldb{Q}}{u} \}.
$$

Computing the set $\weightbs[Q]$ can be done efficiently with respect to the width of the decomposition:

\begin{lem}
  \label{lem:computefactset} Let $k$ be the width of $T$. The size of $\weightbs[Q]$ is at most $|\Vert|\cdot|\db|^k$ and one can compute $\weightbs[Q]$ in time $O(|T|\cdot|\db|^k\log(|\db|))$.  
\end{lem}
\begin{proof}
  It follows directly by Lemma~\ref{lem:computesol} in Section~\ref{sec:htd}.
\end{proof}
  
We define the factorized interpretation of the weight expressions $\weight{Q}{\Xtup[x]\LEq\Xtup[c]}$ as follows:
$$
\begin{array}{ll}
        \rewet{\weight{Q}{\Xtup[x]\EQU\Xtup[c] }}{}
        =  \left\{\begin{array}{ll}
           \weightB{Q}{u}{\beta}  
                & \text{if } \beta = \Xtup[[x]/\Xtup[c]  ]
                  \in {\Soldb{Q}}_{|\Bag(u)}
\\
            0 &\text{else}
                 \end{array}\right. \\
\end{array}
$$
where $u\in\Vert$ is the vertex such that $\Xset[x] = \Bag(u)$ that is the closest to the root of $T$\footnote{Actually, any vertex $u$ such that $\Xset[x_i]=\Bag(u)$ would work but we choose it to be the closest to the root to have a deterministic definition of the factorized interpretation. It is well-defined by connectedness of tree decompositions. Indeed, if two bags $B,B'$ contain a set $S$, then their least common ancestor also contains $S$.}. If no such $u$ exists then $\rewet{\weight{Q}{\Xtup[x]\EQU\Xtup[c] }}$ is undefined. However, for every weight expression of $L$, the existence of $u$ is implied by the fact that $T$ is compatible with $L$, see Definition~\ref{def:compatible}.

For $Q \in \lcqs{L}$, we define $\rewts{\weight{Q}{\Xtup[x]\EQU\Xtup[c] }} := \reweq{\weight{Q}{\Xtup[x]\EQU\Xtup[c] }}$. It gives a function to interpret weight expressions in the sense given in Section~\ref{sec:interpretation} since if $Q$ and $Q'$ are distinct conjunctive queries, then $\rewts{\weight{Q}{\Xtup[x]\EQU\Xtup[c] }}$ and $\rewts{\weight{Q'}{\Xtup[x]\EQU\Xtup[c] }}$ will contain disjoint linear program variables.

\paragraph{Local soundness constraints.} If one simply defines the factorized interpretation as $(\rewew, true)$, it will not give a sound interpretation. We illustrate this phenomenon on the example from Section~\ref{sec:easyexample} using the tree decomposition having three nodes $r,u,v$ rooted at $r$ with $r$ connected to $u$ and $v$ and with $\Bag(r) = \emptyset$, $\Bag(u) = \{x\}$ and $\Bag(v) = \{y\}$. Interpreting only the weights without additional constraints would yield the following:

$$
\def\arraystretch{1.5}
\begin{array}{rl}
  \MAXIMIZE  & \nu_2 \\
  \SUBJECTTO & \nu_0\ \leq\ 1\ \wedge \\
             & \nu_1\ \leq\ 1\ \wedge \\
             & \nu_0\ \LEq\ \weightB{Q}{u}{[x/0]}\ \wedge \\
             & \nu_1\ \LEq\ \weightB{Q}{u}{[x/1]}\ \wedge \\
             & \nu_2\ \LEq\ \weightB{Q}{r}{[]}. \\
\end{array}
$$

One strange aspect of this program is that is does not depends on the variables $\weightB{Q}{v}{[y/c]}$ and hence on the value of $y$, because the program does not contain weight expression on variable $y$. It can be easily checked that the optimal value of this program is not the same as the on from Section~\ref{sec:easyexample}. Indeed, in the above linear program, the values of $\nu_0, \nu_1$ and $\nu_2$ are now completely independent. Hence, the optimal value of the above linear program is actually unbounded.

To make the factorized interpretation equivalent to the natural interpretation, one has to restore somehow the forgotten dependencies. One way of resolving it in the above program would be to add a new constraint $\weightB{Q}{r}{[]} \EQU \weightB{Q}{u}{[x/0]}+\weightB{Q}{u}{[x/1]}$. To achieve this, we add so-called \emph{local soundness constraints}. For every edge $e=(u,v) \in \Edg$ of $T$ and $\gamma \in \projs{\Soldb{Q}}{\Bag(u) \cap \Bag(v)}$, we define the equality constraint $\localsoundc{Q}{e}{\gamma}$ as follows:

$$
\sum\limits_{\substack{\beta \in \projsb{\Soldb{Q} }{u}\\ \gamma=\projsb{\beta}{u}}}  \weightB{Q}{u}{\beta}
\EQU
\sum\limits_{\substack{\beta' \in \projsb{\Soldb{Q} }{v}\\ \gamma=\projsb{\beta'}{v}}}  \weightB{Q}{v}{\beta'}.
$$

Intuitively, this constraint encodes the following: for expressions $\weight{Q}{\Xs \LEq \Xtup[c]}$ and $\weight{Q}{\Ys \LEq \Xtup[c']}$,  if the assignments $\beta = [\Xs/\Xtup[c]]$ and  $\beta' = [\Ys/\Xtup[c']]$ are compatible with one another, in the sense that they agree on the common variables they assign, then $\interpret{\weight{Q}{\Xs \LEq \Xtup[c]}}^\db$ and $\interpret{\weight{Q}{\Ys \LEq \Xtup[c']}}^\db$  will contain common variables from $\tups[Q]$ and hence, will not produce independent linear sums. On the other hand, in the factorized interpretation without additional constraint, they will be interpreted as independent variables $\weightB{Q}{u}{\beta}$ and $\weightB{Q}{v}{\beta'}$ for some $u,v$ in $\Vert$. If $e=(u,v)$ is an edge of $T$, then $\localsoundc{Q}{e}{\gamma}$ accounts for the missed dependency in the factorized interpretation. It turns out that these constraints are enough to make the factorized interpretation equivalent to the natural interpretation (in the sense that they have the same optimal value). 

Hence, we define $\localsound{Q}$, the \emph{local soundness constraints of $Q$ w.r.t $T$},  as follows: 
$$
\bigwedge_{e=(u, v) \in \Edg}\
\bigwedge_{\gamma \in \projs{\Soldb{Q} }{\Bag(u) \cap \Bag(v)}}
\localsoundc{Q}{e}{\gamma}
$$

Local soundness constraints can be efficiently computed with respect to the width of the decomposition:

\begin{lem}
  \label{lem:computelcs} Let $k$ be the width of $T$. The size of $\localsound{Q}$ is at most $|T|\cdot|\db|^k$ and one can compute $\localsound{Q}$  in time $O(|Q|\cdot|T|\cdot|\db|^k\log(|\db|))$. 
\end{lem}
\begin{proof}
  There is one constraint $\localsoundc{Q}{e}{\gamma}$ in  $\localsound{Q}$ for every edge $e$ of $T$ and $\gamma \in \projs{\Soldb{Q}}{\Bag(u) \cap \Bag(v)}$. Since $\Bag(u) \cap \Bag(v) \subseteq \Bag(u)$, $\projs{\Soldb{Q}}{\Bag(u) \cap \Bag(v)}$ is smaller than $\projs{\Soldb{Q}}{\Bag(u)}$ which is itself smaller than $|\db|^k$ by Lemma~\ref{lem:computesol}. Hence there are at most $|T|\cdot|\db|^k$ constraints in $\localsound{Q}$.

  Now, to compute $\localsound{Q}$, we start by computing $\bproj{ \Soldb{Q}}$ using Lemma~\ref{lem:computesol}. Now for each edge $e=(u,v)$ of $T$, we construct $\localsoundc{Q}{e}{\gamma}$ as follows:  we  start by enumerating the tuples $\beta \in \projs{\Soldb{Q}}{\Bag(u)}$ and let $\gamma = \projs{\beta}{\Bag(v)}$. If it is not yet constructed, we create an empty linear sum  $S_{u}^\gamma$ and add $\weightB{Q}{u}{\beta}$ in it. If $S_u^\gamma$ has already been created, we append $+\weightB{Q}{u}{\beta}$ to it. We do the same by enumerating $\beta' \in \projs{\Soldb{Q}}{\Bag(v)}$ and let $\gamma = \projs{\beta'}{\Bag(u)}$. If it is not yet constructed, we create an empty linear sum  $S_{v}^\gamma$ and add $\weightB{Q}{v}{\beta'}$ in it. If $S_v^\gamma$ already exists, we just append $+\weightB{Q}{v}{\beta'}$ to it. Finally, we let $\localsoundc{Q}{e}{\gamma}$ be $S_u^\gamma \EQU S_v^\gamma$ for each $\gamma \in \projs{\Soldb{Q}}{\Bag(u) \cap \Bag(v)}$  that have been found.

  To construct it, observe that each $\projs{\Soldb{Q}}{\Bag(u)}$ is listed at most once for each edges of $T$ and that the projection $\gamma$ can be constructed in time $O(|Q|)$. Hence, the total time required to construct $\localsound{Q}$ is $O(|Q|\cdot|T|\cdot|\db|^k\log(|\db|))$. 
\end{proof}

\paragraph{Factorized interpretation.} Taking local soundness constraints into account, we can now define the $\trees{}$-factorized interpretation $\rewei$ of a linear program $L$ as the pair $(\rewew, \rewec)$ where $\localsoundcalt{Q} = \localsoundcal{Q}$. When $\trees{}$ is clear from context, we will simply say ``the factorized interpretation''.

The soundness of factorized interpretation follows from Proposition~\ref{prop:metasoundness} and from results on weighting answer sets of conjunctive queries that can be of independent interest and that we give in Section~\ref{sec:correctness}:

\begin{thm}
\label{theo:main-closed}
Let $L$ be a \LPclosed program such that every conjunctive query in $\lcqs{L}$ is quantifier free, $\trees{}$ a decomposition of $L$ and $\db$ a database. The factorized  interpretation $\rewe{L}$ then has the same optimal value  as $\interpret{L}^{\db}$:
$$  \opsem{\rewe{L}}=\opsem{\interpret{L}^{\db}}  $$
Moreover, given an optimal solution $W$ of $\rewe{L}$, there exists a canonical optimal solution $\om$ of $\interpret{L}^\db$ such that given $W$ and a variable $\theta$ of $\interpret{L}^{\db}$, one can compute $\om(\theta)$ in polynomial time.
\end{thm}

The first part of Theorem~\ref{theo:main-closed} is proven by providing two explicit transformations: one to go from a solution of $\rewe{L}$ to a solution of $\interpret{L}^{\db}$ having the same value and another one to go from a solution of $\interpret{L}^{\db}$ to a solution of $\rewe{L}$ having the same value. The preservation of the values by these transformations is enough to establish that both linear programs have the same optimal value. More interestingly, it also allows us to prove the second part of the theorem, that is, that one can recover an optimal solution of $\interpret{L}^\db$ from an optimal solution of $\rewe{L}$. This transformation is described in the proof of Theorem~\ref{theo:corr}.

\paragraph{Example.} Going back to the example from Section~\ref{sec:easyexample}, and the tree decomposition previously mentioned, $\rewe{L}$ is the following program:

$$
\def\arraystretch{1.5}
\begin{array}{rl}
  \MAXIMIZE  & \nu_2 \\
  \SUBJECTTO & \nu_0\ \leq\ 1\ \wedge \\
             & \nu_1\ \leq\ 1\ \wedge \\
             & \nu_0\ \LEq\ \weightB{Q}{u}{[x/0]}\ \wedge \\
             & \nu_1\ \LEq\ \weightB{Q}{u}{[x/1]}\ \wedge \\
             & \nu_2\ \LEq\ \weightB{Q}{r}{[]} \wedge \\
  
             & \weightB{Q}{r}{\emptyassign} \EQU \weightB{Q}{u}{[x/0]} + \weightB{Q}{u}{[x/1]} \wedge \\
             & \weightB{Q}{r}{\emptyassign} \EQU \weightB{Q}{v}{[y/0]} + \weightB{Q}{v}{[y/1]} \\
\end{array}
$$

The two last lines contain the local soundness constraint $\localsound{Q}$. The last constraint mentions variables $\weightB{Q}{v}{[y/0]}$ and  $\weightB{Q}{v}{[y/1]}$ that are not used elsewhere in the program and can safely be ignored when looking for the optimal value. The other soundness constraint directly implies that $\nu_2 = \nu_0 + \nu_1$ and hence, the optimal value for this program is $2$.

\paragraph{Computing the factorized interpretation.} The factorized interpretation $\rewe{L}$ is interesting because it is smaller than the natural interpretation $\interpret{L}^\db$. Indeed, while $\interpret{L}^\db$ has $O(|\db|^{\agm(L)})$ variables and $O(|L|)$ constraints (see Section~\ref{sec:complexity}), one can show that if $k$ is the width of $\trees{}$, then the size of $\rewe{L}$ is $O(|\db|^k)$ in the data complexity model (where $L$ is considered constant). It follows from the following, more precise, combined complexity analysis:

\begin{thm}
  \label{thm:factorized-size} Given a relational signature $\Sigma$, $L \in \LPclosed$ such that every query in $\lcqs{L}$ is quantifier free, a tree decomposition $\trees{}$ of $L$ and a database $\db$, we let $k$ be the width of $\trees{}$, $t$ be the sum of the sizes of the tree decompositions in $\trees{}$ and $q$ be the sum of the sizes of the queries in $\lcqs{L}$. Then $\rewe{L}$ has at most $O(t  \cdot |\db|^k)$ variables,  $O(|L|+t  \cdot |\db|^k)$ constraints and can be computed in time $O(|L|+qt\cdot |\db|^k\log |\db|)$. Moreover, $\rewe{L}$ has size  $O(|L||\db|^k)$.
\end{thm}
\begin{proof}
  This is a direct consequence of applying Lemma~\ref{lem:computefactset} and Lemma~\ref{lem:computelcs} to each query in $\lcqs{L}$. Concerning the size of $\rewe{L}$, it comes from the fact that each $\mathbf{weight}$ construct has been replaced by at most one variables resulting in a program of size at most $|L|$  and then we add $O(t|\db|^k) = O(|L||\db|^k)$ soundness constraints.
\end{proof}

Theorem~\ref{thm:factorized-size} together with Theorem~\ref{theo:main-closed} implies that the data complexity of computing the optimal value of a linear program $L \in \LPclosed$ having only quantifier free queries is below $O(|\db|^{\ell \cdot \fhtw{L}})$ with $\ell < 2.37286$ which improves the complexity stated in Theorem~\ref{thm:agmcomplexity}.

We observe however that the factorized interpretation may be small in practice than the worst-case theoretical bound given by the fractional hypertree width of $L$. Indeed, the number of variables in the factorized interpretation is the sum of the sizes of $\projs{\sem{Q}^\db}{\Bag(u)}$ for each $u$. In particular, each one of them contains less elements than $\sem{Q}^\db$. Hence, even when $\sem{Q}^\db$ is small with respect to the worst case, the factorized interpretation will also be smaller than the worst case. 

We summarize this discussion in the following theorem, which mirrors Theorem~\ref{thm:agmcomplexity}:

\begin{thm}
  \label{thm:factcomplexity} Given a relational signature $\Sigma$, $L \in \LPclosed$, a database $\db$ and optimal tree decompositions $\trees{}$ of $L$ of width $\fhtw{L}$,  one can compute $\opsem{\interpret{L}^\db}$ in time $O(|L|^{\ell}|\db|^{\ell \cdot \fhtw{L}})$ with $\ell < 2.37286$.

\end{thm}

\subsection{Linear Programs with Existentially Quantified Conjunctive Queries}
\label{sec:existsremove}

One drawback of Theorem~\ref{theo:main-closed} is that it only works for a linear program $L$ in $\LPclosed$ containing only quantifier free conjunctive queries. This restriction can actually be lifted since replacing existentially quantified queries of $L$ with their quantifier free part yield a linear program that has the same optimal value under the natural interpretation for any database $\db$. We formalize this approach and prove its correctness in this section.

Given a quantified conjunctive query $Q' = \exists \Xtup[y].Q$ with $Q$ a quantifier free conjunctive query, we denote by $\qfp{Q'}$ the conjunctive query $Q \wedge \Xtup[y] \EQU \Xtup[y]$. Clearly, for any database $\db$, the answer set $\Soldb{Q}$ is the same as $\Soldb{\qfp{Q'}}$ and $\qfp{Q'}$ and $Q$ have the same hypertree decompositions and hence the same width. However, in the following, we will need to be able to syntactically distinguish two queries having the same quantifier free part but different quantifier prefix and $\qfp{Q'}$ allows us to do so. Indeed,$\qfp{Q'}$ is syntactically different from $Q'' = \exists \Xtup[y'].Q$ for different $\Xtup[y]$ and $\Xtup[y']$.

We lift the definition of $\qfp{Q}$ to a linear program in $L \in \LPclosed$: we denote by $\qfp{L}$ the $\LPclosed$ language obtained by replacing every expression $\weight{Q'}{\Xtup \LEq \Xtup[c]}$ in $L$  by $\weight{\qfp{Q'}}{\Xtup \LEq \Xtup[c]}$. It is clear that $\qfp{L}$ now only contains quantifier free conjunctive query.

\paragraph{Example.} We adapt the example from Section~\ref{sec:easyexample} by considering the conjunctive query $Q' = \exists y. R_1(x) \wedge
R_2(y)$ and the following program $L' \in\LPclosed$:
\[
\begin{array}{rl}
\MAXIMIZE & \weight{Q'}{\emptyset} \\
\SUBJECTTO 
    &  {\weight{Q'}{x \EQU 0} \leq 1} \\
  \wedge &
           \weight{Q'}{x \EQU 1} \leq 1
\end{array}
\]

Over the database given in Section~\ref{sec:easyexample}, we have $\interpret{L'}^\db$:
\[
\begin{array}{rl}
\MAXIMIZE & \tup{Q'}{0} + \tup{Q'}{1} \\[3pt]
\SUBJECTTO 
    & \tup{Q'}{0} \leq 1 \\[3pt]
    \wedge & \tup{Q'}{1} \leq 1
\end{array}
\]
It has optimal value $2$.

On the other hand, we have $\qfp{Q'}=R_1(x) \wedge R_2(y) \wedge y \EQU y$ and $\qfp{L'}$ is:
\[
\begin{array}{rl}
\MAXIMIZE & \weight{\qfp{Q'}}{\emptyset} \\
\SUBJECTTO 
    &  {\weight{\qfp{Q'}}{x \EQU 0} \leq 1} \\
  \wedge &
           \weight{\qfp{Q'}}{x \EQU 1} \leq 1
\end{array}
\]

which is clearly equivalent from the linear program $L$ from Section~\ref{sec:easyexample} which itself has optimal value $2$ when interpreted on the database given in the same section.

\paragraph{Soundness of quantifier elimination.} It turns out that it is equivalent to $L$ in the following sense:

\begin{prop}
  \label{prop:quantifierelim} For every $L \in \LPclosed$ and $\db$ on signature $\Sigma$, we have $$  \opsem{\interpret{\qfp{L}}^\db}=\opsem{\interpret{L}^{\db}}.  $$
\end{prop}
\begin{proof}
By Proposition~\ref{prop:repl}, it is sufficient to show $$\opsem{\interpret{\Rho(\qfp{L})}^\db}=\opsem{\interpret{\Rho(L)}^{\db}}.$$
  
We start by observing that $\wset[Q](L)$ is in one to one correspondence with $\wset[\qfp{Q}](\qfp{L})$ by definition since we replaced every occurrence of $Q$ in $L$ by $\qfp{Q}$ in $\qfp{L}$. Moreover, observe that if $Q$ and $Q'$ are distinct queries of $\lcqs{L}$, then $\qfp{Q}$ and $\qfp{Q'}$ are also distinct. Indeed, either $Q$ and $Q'$ have distinct quantifier free part and it is obvious, or they have distinct quantifier prefix $\Ytup$ and $\Ytup'$ respectively, in which case $\qfp{Q}$ and $\qfp{Q'}$ will be distinct since $\qfp{Q}$ contains $\Ytup \EQU \Ytup$  and $\qfp{Q'}$ contains $\Ytup' \EQU  \Ytup'$.

Hence, exploiting this one to one correspondence, for $$L=   \MAXIMIZE S \ \SUBJECTTO C$$ with some linear sum $S\in\LSclosed$ and some linear constraint $C\in\LCclosed$, we can see $\interpret{\Rho(\qfp{L})}^\db$ as:

$$
\MAXIMIZE \interpret{S}^\nu \  \SUBJECTTO \interpret{C}^\nu \wedge \bigwedge_{W \in \wset(L)} \nu(W) = M^\db(W)
$$

where $M^\db$ maps weight expressions of $L$ to a sum expression. In other words, one can see that $\interpret{\Rho(\qfp{L})}^\db$ as $I^\db(L)$  where $I = (M, true)$ is the alternate interpretation in the sense given in Section~\ref{sec:interpretation} defined as follows: for a weight expression $W = \weight{Q'}{\Xtup \LEq \Xtup[c]}$ where $Q' = \exists \Ytup.Q$ and $Q$ is quantifier free, $M_Q^\db(W)$ is defined over variables
$$X_{Q'}^\db :=  \{\chi^\alpha_{Q'} \mid \alpha\in\sem{Q}^\db\}$$
by the following sum expression:

$$ M_Q^\db(W) := \sum\limits_{\substack{\alpha \in \sem{Q}^\db\\\alpha(\Xs)=\Xtup[c]}} \chi_{Q'}^\alpha.$$

By Proposition~\ref{prop:metasoundness}, it is thus sufficient to show that for a set  $\wset$  of weight expressions over a conjunctive query $Q' = \exists \Ytup. Q$ with $Q$ a quantifier free conjunctive query, we have $\projs{\sem{I_{Q'}^\db(\wset)}}{\nu} = \projs{\sem{\interpret{\wC[\wset]}^\db}}{\nu}$. We let $V$ be the free variables of $Q'$.

So let $w \in \sem{I_{Q'}^\db(\wset)}$. It maps variables $\chi^\alpha_{Q'}$ to a positive real number, where $\alpha$ is in $\sem{Q}^\db$. For $\beta \in \sem{Q'}^\db$, we define $w'(\theta^\beta_{Q'}) := \sum\limits_{\alpha \in \sem{Q}^\db \mid \alpha_{|V}=\beta} w(\chi^\alpha_{Q'})$. For a weight constraint $W$, we set $w'(\nu(W)) = w(\nu(W))$. Clearly, $w$ and $w'$ coincide on $\nu$ variables. It remains to show that $w' \in \sem{\interpret{\wC[\wset]}^\db}$, that is, for a weight expression $W \in \wset$ of the form $\weight{Q'}{\Xtup\LEq\Xtup[c]}$, $w'(\nu(W)) = w'(\interpret{W}^\db)$, that is:

$$w'(\nu(W)) = \sum\limits_{\substack{\beta \in \sem{Q'}^\db\\\alpha(\Xs)=\Xtup[c]}} w'(\theta_{Q'}^\beta).$$

By definition of $w'$, the right-hand side of this equation rewrites to:
\begin{align*}
  \sum\limits_{\substack{\beta \in \sem{Q'}^\db\\\alpha(\Xs)=\Xtup[c]}} w'(\theta_{Q'}^\beta) & =  \sum\limits_{\substack{\beta \in \sem{Q'}^\db\\\beta(\Xs)=\Xtup[c]}} \sum\limits_{\substack{\alpha \in \sem{Q}^\db \\ \alpha_{|V}=\beta}} w(\chi^\alpha_{Q'}) \\
                                                                                              & = \sum\limits_{\substack{\alpha \in \sem{Q}^\db \\ \alpha(\Xtup)=\Xtup[c]}} w(\chi^\alpha_{Q'}) \text{ since } \Xset \subseteq V \\
                                                                                              & = w(W) \text{ since } w \in \sem{I^\db_{Q'}(\wset)}.
\end{align*}
Hence, $w' \in \sem{\interpret{\wC[\wset]}^\db}$.

For the other way around, let $w' \in \sem{\interpret{\wC[\wset]}^\db}$. For $\alpha \in \Soldb{Q}$, we let $\beta = \projs{\alpha}{V}$ and define $w(\chi_{Q'}^\alpha) := {1 \over N_\beta}w'(\theta^\beta_{Q'})$, where $N_\beta$ is the number of $\gamma \in \Soldb{Q}$ such that $\projs{\gamma}{V} = \beta$. For every weight expression $W \in \wset$, we let $w(\nu(W)) := w'(\nu(W))$. Clearly $w$ and $w'$ coincide on $\nu$ variables. Now, it remains to show that $w \in \sem{I_{Q'}^\db(\wset)}$ that is for a weight expression $W \in \wset$ of the form $\weight{Q'}{\Xtup\LEq\Xtup[c]}$, $w(\nu(W)) = w(M_Q^\db(W))$. By definition, we have:
\begin{align*}
  w(M_Q^\db(W))  &  = \sum\limits_{\substack{\alpha \in \sem{Q}^\db\\\alpha(\Xs)=\Xtup[c]}} w(\chi_{Q'}^\alpha) \\
                 & = \sum\limits_{\substack{\alpha \in \sem{Q}^\db\\\alpha(\Xs)=\Xtup[c]}} {1 \over N_\beta}w'(\theta^\beta_{Q'}) \text{ with } \beta=\projs{\alpha}{V}  \\
                 & = \sum\limits_{\substack{\beta \in \sem{Q'}^\db\\\beta(\Xs)=\Xtup[c]}} \sum\limits_{\substack{\alpha \in \sem{Q}^\db\\\projs{\alpha}{v}=\beta}}{1 \over N_\beta}w'(\theta^\beta_{Q'})   \\
                 & = \sum\limits_{\substack{\beta \in \sem{Q'}^\db\\\beta(\Xs)=\Xtup[c]}} w'(\theta^\beta_{Q'}) \text{ by definition of } N_\beta\\
                 & = w'(\nu(W)) \text{ since } w' \in \sem{\interpret{\wC[\wset]}^\db} \\
                 & = w(\nu(W)) \text{ by definition of } w. \tag*{\qedhere}
\end{align*}
\end{proof}

Recall that by definition, a hypertree decomposition $\trees{}$ of width $k$ of a linear program $L$ in $\LPclosed$ consists of a collection of tree decompositions for the quantifier free part of each query in $\lcqs{L}$ of width at most $k$. For $Q \in \lcqs{L}$, $\trees{Q}$ is then also a decomposition of $\qfp{Q}$ with width at most $k$. Hence, $\trees{}$ is a hypertree decomposition of $\qfp{L}$ of width $k$. The fractional hypertree width of $\qfp{L}$ is thus the same as the fractional hypertree width of $L$. Hence, one can compute the optimal value of $L$ in $O(|\db|^{\ell\cdot\fhtw{L}})$ with $\ell < 2.37286$ in data complexity by computing the optimal value of the factorized interpretation of $\qfp{L}$ using Theorem~\ref{theo:main-closed} and Theorem~\ref{thm:factorized-size}, even when the program contains existentially quantified conjunctive queries.

This is wrapped up in the following theorem which is an improvement over Theorem~\ref{thm:agmcomplexity}:
\begin{thm}
  \label{thm:complexity-closed} 
Given a relational signature $\Sigma$, $L \in \LPclosed$, a tree decomposition $\trees{}$ of $L$ of width $k$ and a database $\db$, there exists some $\ell < 2.37286$ such that one can compute $\opsem{\interpret{L}^\db}$ in time $O((|L|+tq|\db|^{k})^\ell)$ where $t$ is the sum of the sizes of tree decompositions in $\trees{}$ and $q$ the sum of the sizes of the conjunctive queries in $\lcqs{L}$.
\end{thm}

\section{Linear Programs with Open Weight Expressions}
\label{sec:lpgeneral}

While we have shown that $\LPclosed$ programs can be solved efficiently by exploiting tree decompositions of the input conjunctive queries, it is not yet  powerful enough to express interesting linear program such as the one presented in Section~\ref{sec:introexample}. The missing feature in $\LPclosed$ for is their inability to quantify over values in the database to create new constraints. This is especially useful in the example of Section~\ref{sec:introexample}. The last quantified constraint states that the storing limit of every warehouse in the database will not be exceeded in the solution of the linear program. This expressivity is enabled by the fact that one can universally quantify over warehouses given the table $store$, which will be interpreted as generating one constraint for each. Moreover, observe that this constraint also draws a numerical value $\num(l)$ from the database.

\newcommand\CSLong{\CS = (\constraints, X, D, \sem{ . }, \fv, ext,\subs{},num)}
\newcommand\CSprimeLong{\CS' = (\constraints, X, D, \sem{ . }, \fv, ext,\subs{},num)}

In this section, we introduce the language $\LP(\CQ)$ allowing to universally quantify and sum over answers set of database queries. The syntax of the language is presented in Figure~\ref{fig:lp-syntax}. It includes definitions for linear sums, linear constraints and linear programs. The semantics of programs in $\LP(\CQ)$ will be given in Section~\ref{sec:closure} via a closure operation, transforming an $\LP(\CQ)$ into a program in $\LPclosed$.

\begin{figure}[h]
$$
\begin{array}{llcl}
  \text{Domain expressions} & E \in\EXP   & ::= &  x \mid c     \\
  \text{Constant numbers }  & N \in\NUM   & ::= &  \real \mid \num(E) \\
  \text{Linear sums } & S,S' \in \SUM     & ::= &  {\weight{Q}{\Xtup[z]:\Xtup[x] \LEq \Xtup[y]}} \\
                            & & & \text{ where } \Xset \subseteq \fv(Q) \subseteq \Xset[z], \\
                            & & & \Yset \text{ contains variables and constants} \\
                            & & & \text{ and } \Yset \cap \Xset[z] = \emptyset \\
    &&\mid & \sum_{\Xtup: \C} S  \\
  & & \mid&  N S \mid S+S' \mid N \\
\text{Linear constraints } & C , C' \in\LC(\CS) 
    & ::= & S \le S' \mid S\LEq S' \mid C \wedge C' \mid  \true\\
     &&
\mid 
& 
\forall \XwhereQ{\Xtup}{\C}.  C\\
   \text{Linear programs } & L \in \LP(\CS) 
    & ::= & \MAXIMIZE S \  \SUBJECTTO C \\
                 &&& \qquad\text{where $\fv(S)=\fv(C)=\emptyset$.}  \\
   \end{array}
$$
\caption{Linear sum, constraints, and programs with
  open weight expressions over conjunctive queries
  $\C,\C'\in \CQ$, with variables $x\in X$,
  sequences of variables $\Xtup\in X^*$, constants
  $c\in C$, and reals $r\in\R$. }
\label{fig:lp-syntax}
\end{figure}

Apart from the addition of an operator $\num(E)$ which will intuitively allow to get numerical constants from the database, universal quantifiers and sums ranging over a conjunctive query, the main difference with $\LPclosed$ programs is that non-constant values are allowed in $\weight{Q}{\Xtup[z]:\Xtup \LEq \Ytup}$ expressions. We will call such weight expression \emph{open weight expressions}, as opposed to closed weight expressions where $\Ytup$ only contains constants. Intuitively, variables in $\Yset$ will be replaced by database constants in the closure of an $\LP(\CQ)$.

A valid $\LP(\CQ)$ program $L$ does not have free variables, that is, every variable has to be bound by one of the new operator: either a linear sum $\sum_{\Xtup: \C} S$ or a universal quantifiers $\forall \XwhereQ{\Xtup}{\C}.  C$. To formalize this notion, we give in Figure~\ref{fv:lp} the definition of the free variables of an $\LP(\CQ)$ program.

\begin{figure}[h]
\[
\begin{array}{llll}
   \fv(c)= \emptyset &
   \fv(\num(E)) =\fv(E)
\\
    \fv(\weight{Q}{\Xtup[z]: \Xtup \LEq \Ytup})= (\fv(Q) \cup \Yset) \setminus \Xset[z]
&
   \fv(\sum_{\Xwheredelta{\Xtup}{\Query}} S) = \fv(S) \cup\fv(\Query) \setminus\Xset
\\
\fv(N S) = \fv(N)\cup  \fv(S) & 
    \fv(S \le S') =\fv(S)\cup  \fv(S')  
\\
   \fv(S+S') =\fv(S)\cup  \fv(S')  
&
   \fv(S \LEq S')=\fv(S)\cup  \fv(S')
\\
   \fv(\forall \Xwheredelta{\Xtup}{\Query}.\ C) =\fv(\Query)\cup\fv(C)
  \setminus\{\Xtup\}   
&
   \fv(C \wedge C')=\fv(C)\cup  \fv(C')
\\
\fv(\MAXIMIZE S \  \SUBJECTTO C)=\emptyset  
&  \fv(\true)=\emptyset  
\end{array}
\]
\caption{\label{fv:lp} Free variables of expressions, constraints, and programs, where $var(\Ytup)$ denotes the elements of $\Yset$ that are not constants. }
\end{figure}

Observe in particular that for $\weight{Q}{\Xtup[z]:\Xtup \LEq \Ytup}$, only variables in $\Yset$ are considered free since $\fv(Q) \subseteq \Xset[z]$. The free variables of the conjunctive queries (and hence in $\Xset$ since $\Xset \subseteq \fv(Q)$) are \emph{not} considered free variables. In other words, variables bounded through universal quantifiers and sums over conjunctive queries in $\LP(\CQ)$ will not introduce constants in conjunctive queries appearing in weight expressions.\footnote{This restriction was not present in the conference version of this paper~\cite{confversion} which may lead to counter intuitive behaviors.} If one follows Barendregt's variable convention, it means that the variables appearing in the conjunctive queries of $L \in \LP(\CQ)$ may be considered disjoint from the variables used in the linear program part.

\subsection{Closure and semantics}
\label{sec:closure}

We define the semantics of linear programs with open weight expressions over a database $\db$ by mapping it to a linear program with closed weight expression. Intuitively, the queries in $\sum_{\Xtup: \C} S$ and a universal quantifiers $\forall \XwhereQ{\Xtup}{\C}.  C$ get interpreted over $\db$ and it maps $\Xtup$ to some possible values that passed into a context. The details are given in \Figure{fig:closure}.

 \begin{figure}[h]
 \renewcommand{\arraystretch}{1.2}
   $$
\begin{array}{l@{\hspace{-.2cm}}l}
\begin{array}{l@{\qquad}l}
     \closure{\real}{\mapf}{\gamma} = \real
   \\
     \closure{\num(d)}{\mapf}{\gamma} = d
\\
     \closure{\num(x)}{\mapf}{\gamma} = \num^\db(\gamma(x)) 
  \\[.5em]
  {\closure{\weight{Q}{\Xtup: \Xtup[z] \LEq \Ytup}}{\mapf}{\gamma} 
    = \weight{Q}{\Xtup[z] \LEq \gammabar(\Ytup)}} \\
   \closure{S_1+S_2}{\mapf}{\gamma} = \closure{S_1}{\mapf}{\gamma} +
  \closure{S_2}{\mapf}{\gamma}
  \\
\closure{N S}{\mapf}{\gamma} = \closure{N}{\mapf}{\gamma}
  \closure{S}{\mapf}{\gamma}
   \\
\closure{ \sum_{\Xtup : \C} S}{\mapf}{\gamma} 
    = \sum_{\gamma' \in
  \Soldb{ext_{\Xtup}(\subs\gammabar(\C))}}
  \closure{S}{\mapf}{\gammabar \cup \gamma'}
  \\
\end{array}
  \\[.5em]
\begin{array}{l@{\qquad}l}
\closure{\FORALL{\Xtup}{\C}{C}}{\mapf}{\gamma} 
    = \bigwedge_{\gamma' \in \Soldb{ext_{\Xtup}(\subs\gammabar(\C))}}
  \closure{C}{\mapf}{\gammabar \cup \gamma'} 
\\

\closure{S_1 \le S_2}{\mapf}{\gamma} = \closure{S_1}{\mapf}{\gamma} \le \closure{S_2}{\mapf}{\gamma} \\ 

\closure{C_1 \wedge  C_2}{\mapf}{\gamma} = \closure{C_1}{\mapf}{\gamma} \wedge  \closure{C_2}{\mapf}{\gamma} \\

  \closure{true}{\mapf}{\gamma} = \true
  \\[.5em]
\closureno{\textbf{maximize } S \textbf{ subject to } C}\mapf\db\\
    = \textbf{maximize } \closure{S}\mapf \emptyset\textbf{ subject to }
  \closure{C}\mapf\emptyset
\end{array}
\end{array}
$$
\caption{Closure $\closure{F}{\mapf}{\gamma}$ of linear programs $F$ with relational descriptors $L\in LP(\CS)$
to linear programs with set descriptors $\closureno{L}\mapf\db\in \LPclosed$.
Furthermore,     $\gamma:Y\to D$ a variable assignment  
 with  $\fv(F) \subseteq Y\subseteq \varx$, and
 $\gammabar=\gamma_{|Y\setminus \Xset}$. Recall that $\subs{}$ is the substitution operator and $ext_{\Xtup}$ is the extension operator as defined in Section \ref{sec:conjQueries}}
\label{fig:closure}
\end{figure}

The closure is defined in a somewhat classical way: one inductively evaluates a linear program $L$ over a database $\db$ and an environment $\gamma$ mapping free variables of $L$ to values in the domain. We illustrate this on by detailing the closure of $(\FORALL{\Xtup}{\C}{C})$ over a database $\db$ given an environment $\gamma$. The closure of $\sum_{\Xtup : \C} S$ and of $\weight{Q}{\Xtup: \Xtup[z] \LEq \Ytup}$ being very similar.

The closure of $(\FORALL{\Xtup}{\C}{C})$ generates one set of constraint for each answer of $Q$ over $\db$ (over variables $\Xtup$). It intuitively means that for every answer of $Q$, we want the constraints of the closure of $C$ to be satisfied. However, when evaluated the closure inductively, $Q$ may contain free variables. These variables must have a value set by $\gamma$. We start by mapping the variables of $Q$ that are free to their value in the environment using $\subs\gammabar(\C)$, where $\gammabar$ is the restriction of $\gamma$ to the free variables of $Q$. Indeed, it may be that $\gamma$ assigns a variable $x$ that appears in $\Xtup$. In this case, we consider that $x$ is not bounded by $\gamma$ since it is not free in $(\Xtup:Q)$. For example, $\forall x:Q_1.\forall x:Q_2.C$ has the same closure as $\forall x:Q_1.\forall y:Q_2.C$ since the value of $\gamma$ over $x$ when computing the closure of $\forall x:Q_2.C$ is not used in $\gammabar$. To summarize, the closure of $(\FORALL{\Xtup}{\C}{C})$ over a database $\db$ in an environment $\gamma$ generates a set of constraints $\closure{C}{\mapf}{\gammabar \cup \gamma'}$ for every $\gamma'$ that is a solution of $Q$ over variables $\Xtup$ where each variable of $Q$ that are not in $\Xtup$ have been replaced by their value under $\gamma$, which is formally written as $\gamma' \in  \Soldb{ext_{\Xtup}(\subs\gammabar(\C))}$.

Observe that the closure is always well-defined due to the syntactic restrictions on linear programs.  
Also observe that one needs to be able to interpret some database constants as numerical values because of the $\num$ operation. Hence, one needs to use databases that can contain real numbers. This is formalized in the following definition: a \emph{(relational) database with real numbers} is a tuple $\db=(\Sigma,dom(\db),\cdot^\db,\num^\db)$ such that $\db=(\Sigma,dom(\db),\cdot^\db)$ is a relational $\Sigma$-structure and $\num^\db$ a partial function from $dom(\db)$ to $\R$.

Since a linear programs $L \in \LP(\CQ)$ do not have free variables, the closure of $L$ indeed produces a linear program in $\LPclosed$, as stated below:
\begin{prop}
  For any linear programs $L\in \LP(\CQ)$ with open weight
  expressions and database  with
  numerical values $\db=(\Sigma,D,\sem{\cdot}^\db,num^\db)$
  such that $D\subseteq\Consts$,
  the closure $\closureno{L}\mapf\db$ is a linear program in $\LPclosed$.
\end{prop}

The \emph{natural interpretation $\interpret{L}^\db$ of an $\LP(\CQ)$ $L$ over a database $\db$}  is defined to be the natural interpretation of its closure, that is, $\interpret{L}^\db$ is defined to be $\interpret{\closureno{L}\mapf\db}^\db$.

\paragraph{Example.} As an example, we reconsider the conjunctive query $Q = R_1(x) \wedge R_2(y)$ from Section~\ref{sec:easyexample}. Assume we also have a unary relation $S$ and a binary relation $T$ in $\db$ with $S^\db=\{0,1\}$ and $T^\db=\{(0,0.4), (0,0.6), (1, 0.3)\}$.
We then consider the following \LPCQ program $L$:
\[
\begin{array}{ll}
\MAXIMIZE & \weight{Q}{(x,y):\true} \\
\theSUBJECTTO  & \FORALL{(z)}{S(z)} {\weight{Q}{(x,y):x \EQU z} \leq \sum_{y:T(z,y)} \num(y)} \\
\end{array}
\]
To compute the closure over $\db$ of this program, we start by unfolding the universal quantifier. Hence $\closure{L}{\db}{\emptyassign}$ is equal to:
\[
\begin{array}{ll}
\MAXIMIZE     & \weight{Q}{(x,y):\true} \\
\theSUBJECTTO & \closure{\weight{Q}{(x,y):x \EQU z}}{\db}{z \mapsto 0} \leq \closure{\sum_{y:T(0,y)} \num(y)}{\db}{z \mapsto 0} \\
\wedge        &  \closure{\weight{Q}{(x,y):x \EQU z}}{\db}{z \mapsto 1} \leq \closure{\sum_{y:T(0,y)} \num(y)}{\db}{z \mapsto 1} \\
\end{array}
\]

By evaluating the closure of the weights and sum we then have:
\[
\begin{array}{ll}
\MAXIMIZE     & \weight{Q}{(x,y):\true} \\
\theSUBJECTTO & \weight{Q}{x \EQU 0} \leq \closure{\num(y)}{\db}{z\mapsto 0, y \mapsto 0.4}+\closure{\num(y)}{\db}{z\mapsto 0, y \mapsto 0.6} \\
\wedge        & \weight{Q}{x \EQU 1} \leq \closure{\num(y)}{\db}{z\mapsto 1, y \mapsto 0.3} \\
\end{array}
\]

And finally:
\[
\begin{array}{ll}
\MAXIMIZE     & \weight{Q}{(x,y):\true} \\
\theSUBJECTTO & \weight{Q}{x \EQU 0} \leq 0.4+0.6 \\
\wedge        & \weight{Q}{x \EQU 1}\leq 0.3 \\
\end{array}
\]

\subsection{Complexity of solving \texorpdfstring{$\LP(\CQ)$}{LP(CQ_Σ)} programs}
\label{sec:lpcqcomplexity}
In this section, we are interested in the complexity of solving $\LP(\CQ)$ programs.

\paragraph{Hardness.} Universal quantifiers make the complexity of solving programs in $\LP(\CQ)$ much harder than for $\LPclosed$:

\begin{thm}
  \label{thm:lpcqhardness} The problem of computing $\interpret{L}^\db$ for an $\LP(\CQ)$ $L$ and a database $\db$ with real values given in the input is $\sP$-hard.
\end{thm}
\begin{proof}
  Given a conjunctive query $Q$ on variables $\Xtup$, we define $L_Q$ to be the following $\LP(\CQ)$ program:
\[
\begin{array}{lll}
& \MAXIMIZE & \weight{Q}{\true} \\
& \theSUBJECTTO 
& \FORALL{\Ytup}{Q(\Ytup)}
    \weight{Q}{\Xtup:\Xtup \LEq \Ytup} \leq 1. \\
\end{array}
\]

We claim that given a database $\db$, the optimal value of $\interpret{L}^\db$ is the size of $\Soldb{Q}$. Indeed, $\closureno{L}\mapf\db$ contains one constraint $\weight{Q}{\Xtup \LEq \alpha(\Xtup)} \leq 1$  for each $\alpha \in \Soldb{Q}$. Hence, it is translated as a constraint $\theta_Q^\alpha \leq 1$ in $\interpret{L}^\db$, while the objective function is  $\sum_{\alpha \in \Soldb{Q}} \tup{Q}{\alpha}$. Assigning every $\tup{Q}{\alpha}$ yields a solution of $\interpret{L}^\db$ whose value is the size of $\Soldb{Q}$. Moreover, since every constraint in $\interpret{L}^\db$ is saturated, it is also optimal.

The problem of computing the size of $\Soldb{Q}$ when both $Q$ and $\db$ are given in the input is $\sP$-complete~\cite{pichler2013}, hence, computing $\opsem{\interpret{L}^\db}$ is $\sP$-hard.
\end{proof}

\paragraph{Data complexity.} When considering the input linear program size to be constant, that is, in the data complexity model, it turns out that one can compute the optimal value of the natural interpretation of an $\LP(\CQ)$ program $L$ over $\db$ in time polynomial in the size of the database $\db$. To analyze the precise complexity of solving programs in $\LP(\CQ)$, we start by defining a normal form for $\LP(\CQ)$ programs that will be helpful.

An \emph{atomic linear constraint} is a linear constraint of the form $\FORALL{\Xtup}{Q}{S \leq S'}$, $\FORALL{\Xtup}{Q}{S=S'}$, $S \leq S'$ or $S=S'$ where $S$ and $S'$ are linear sums.  We insist on the fact that an atomic linear constraint has at most one conjunctive query on which the universal quantifier applies. An \emph{atomic linear sum} is a linear sum of the form  $N$, $N \sum_{\Xtup[z] : \C'} 1$, $\sum_{\Xtup[z] : \C'} \weight{Q}{\Xtup[z]: \Xtup \LEq \Ytup}$ or $N\sum_{\Xtup[z] : \C'} \weight{Q}{\Xtup[z]:\Xtup \LEq \Ytup}$ with $N$ a constant number (of the form $r\in\R$ or $\num(E)$).

A linear sum is said to be in \emph{normal form} if it is written as a sum of atomic linear sum. A linear constraint is in normal form if it is written as a conjunction of atomic linear constraints on linear sum in normal form. Finally, an $\LP(\CQ)$ is in normal form if its objective is a linear sum in normal form and if its constraint is in normal form.

It turns out that every $\LP(\CQ)$ program can be written as an equivalent linear program in normal form of polynomial size.

\begin{thm}
  \label{thm:normalizelpcq} Let $L$ be an $\LP(\CQ)$. There exists a normal form $\LP(\CQ)$ $L'$ such that $L'$ is of size at most $|L|^3$ and such that for every database $\db$, $\closureno{L}\mapf\db$ and $\closureno{L'}^\db$ have the same constraints (up to permutations).
\end{thm}
\begin{proof}
  We first assume that $L$ is written using Barendregt's variable convention~\cite{barendregt} to avoid variables capture. First we observe that a linear constraint can always be written as a conjunction of constraints of the form $$\forall{\Xtup_1:Q_1}\dots\forall{\Xtup_k:Q_k}. B$$ where $B$ is of the form $S=S'$ or $S \leq S'$ (we consider that $k=0$ corresponds to  the case without quantifier). It comes from the fact that the closure of $\forall{\Xtup:Q}. (C_1 \wedge C_2)$ will generate the same constraints as $(\forall{\Xtup:Q}. C_1) \wedge (\forall{\Xtup:Q}. C_2))$. One can hence apply this transformation until all constraints are of the desired form. The transformation applied to $L$ results in an $\LP(\CQ)$ $L_1$ of size at most $d_\forall|L|$ where $d_\forall$ is the depth of universal quantifiers of $L$, that is, the maximal number of universal quantifiers that are enclosed in one another. Indeed, one can see that $L_1$ will contain a conjunction of constraints of the form $\forall{\Xtup_1:Q_1}\dots\forall{\Xtup_k:Q_k}. B$ where $B$ is of the form  $S=S'$ or $S \leq S'$ and appears in $L$ enclosed in several quantifiers $\forall{\Xtup_1:Q_1}, \dots, \forall{\Xtup_k:Q_k}$.

  Similarly, one can rewrite each linear sum $S$ as a sum of expressions of the form  $$\sum_{\Xtup_1:Q_1}\dots\sum_{\Xtup_k:Q_k}B$$ where $B$ is of the form $N$, $\weight{Q}{\Xtup[z]:\Xtup\LEq\Ytup}$ or $N\weight{Q}{\Xtup[z]:\Xtup\LEq\Ytup}$ (again, the case $k=0$ corresponds to the case without $\sum$). Indeed, we proceed similarly by observing that the closure of $(\sum_{\Xtup:Q}(S_1+S_2))$ produces the same terms as $(\sum_{\Xtup:Q}S_1) + (\sum_{\Xtup:Q}S_2)$. Like above, the transformation applied to $L_1$ results in an $\LP(\CQ)$ $L_2$ of size at most $d_\Sigma|L_1| \leq d_\Sigma d_\forall|L|$ where $d_\Sigma$ is the maximal number of enclosed $\sum$ expressions of $L$.

  It remains to observe that a constraint of the form $\forall{\Xtup_1:Q_1}\dots\forall{\Xtup_k:Q_k}. B$ where $B$ is of the form  $S=S'$ or $S \leq S'$ can be rewritten as $$\forall{(\Xtup_1,\dots,\Xtup_k):(Q_1\wedge \dots\wedge Q_k)}. B.$$
  This is only true when $\Set{\Xtup_1},\dots,\Set{\Xtup_k}$ are pairwise disjoint, which is ensured by the fact that we adopted Barendregt's variables convention. Indeed, we can always rename variables that are bounded by a quantifier so that they have different names. We show it for the case $k=2$, the general case being a straightforward induction from there.

  Let $\db$ be a database and let  
  \begin{itemize}
  \item $E = \forall{\Xtup_1:Q_1}\forall{\Xtup_2:Q_2}. B$ and
  \item   $F = \forall{(\Xtup_1,\Xtup_2):(Q_1\wedge Q_2)}. B$,
  \end{itemize}

  By definition, for $\alpha$ mapping every free variables of $E$ and $F$ (they have the same free variables by definition), we have:

  \begin{align*}
    \closureno{E}\mapf{\db,\alpha} & = \bigwedge_{\gamma_1 \in \Soldb{Q_1'}} \bigwedge_{\gamma_2 \in \subs{\gamma_1}(\Soldb{Q_2'})} \closureno{B}\mapf{\db, \tilde{\alpha}\cup \gamma_1\cup\gamma_2} \\
    \closureno{F}\mapf{\db, \alpha} & = \bigwedge_{\gamma \in \Soldb{Q_1' \wedge Q_2'}} \closureno{B}\mapf{\db,  \tilde{\alpha}\cup \gamma}                           
  \end{align*}
  where $Q_1'=ext_{\Xtup_1}({\subs{\tilde{\alpha}}(Q_1)})$ and  $Q_2'=ext_{\Xtup_2}({\subs{\tilde{\alpha}}(Q_2)})$
  
  Hence, the proof follows from the fact that $\Soldb{Q_1' \wedge Q_2'}$ are the same as the set of $\gamma$ such that:
  \begin{itemize}
  \item $\projs{\gamma}{\Set{\Xtup_1}} = \gamma_1$ is in $\Soldb{Q_1'}$,
  \item $\projs{\gamma}{\Set{\Xtup_2}} = \gamma_2$ is such that $\gamma_1 \cup \gamma_2$ is in $\Soldb{Q_2'}$
  \end{itemize}
  which is clear from the definition of conjunctive queries and the fact that $\Set{\Xtup_1} \cap \Set{\Xtup_2}$.

  Similarly, the closure of linear sums of the form $\sum_{\Xtup_1:Q_1}\dots\sum_{\Xtup_k:Q_k} S$ will generate the same terms as $$\sum_{(\Xtup_1,\dots,\Xtup_k):(Q_1\wedge \dots\wedge Q_k)} S.$$

  The size of $L'$ is then at most $d_\Sigma d_\forall |L| \leq |L|^3$. \qedhere

\end{proof} 

Linear programs in $\LP(\CQ)$ in normal form allows us to upper bound the size of the closure more precisely. For an $\LP(\CQ)$ $L$ in normal form, we denote by $\lcqfo{L}$ the set of conjunctive query $Q$ that appear in $L$ in an expression of the form $\forall \Xtup:Q$ and by $\lcqsum{L}$ the set of conjunctive query $Q$ that appear in $L$ in an expression of the form $\sum_{\Xtup:Q}$. As for $\LPclosed$, we let $\lcqs{L}$ be the set of conjunctive query $Q$ that appear in $L$ in an expression of the form $\weight{Q}{\Xtup[z]:\Xtup\LEq\Ytup}$. We let $\agmfo{L}$ to be $\max_{Q \in \lcqfo{L}}\agm(Q)$ and similarly, $\agmsum{L}$ is $\max_{Q \in \lcqsum{L}}\agm(Q)$ and  $\agmw{L}$ is $\max_{Q \in \lcqs{L}}\agm(Q)$.

Observe that, given a database $\db$, an atomic linear constraint of the form $\FORALL{\Xtup}{Q}{B}$ will generate at most $|\db|^{\agm(Q)}$ constraints in the closure $\closureno{L}\mapf\db$. Similarly, an atomic linear sum of the form $\sum_{\Xtup:Q}{B}$ will generate at most $|\db|^{\agm(Q)}$ terms in  $\closureno{L}\mapf\db$. Hence, we have the following:

\begin{prop}
  \label{prop:lpcqsize} Let $L$ be an $\LP(\CQ)$ in normal form and $\db$ be a database. $\closureno{L}\mapf\db$ has at most $|L|\cdot|\db|^{\agmfo{L}}$ constraints, of size at most $|L|\cdot|\db|^{\agmsum{L}}$. In particular, $\interpret{L}^\db$ has at most $|L|\cdot|\db|^{\agmfo{L}}$ constraints, $|L|\cdot|\db|^{\agmw{L}}$ variables and can be computed in time $O(|L|\cdot|\db|^{\agmsum{L}+\agmfo{L}+\agmw{L}})$. Hence, there exists $\ell < 2.37286$ such that one can compute $\opsem{\interpret{L}^\db}$ in time $O(|L|^\ell\cdot|\db|^{\ell(\agmsum{L}+\agmfo{L}+\agmw{L})})$.
\end{prop}

Proposition~\ref{prop:lpcqsize} directly implies that $\LP(\CQ)$ can be solved in polynomial time in the data complexity. 

\paragraph{Factorized interpretation of $\LP(\CQ)$.}  One can get a better time complexity than the one stated in Proposition~\ref{prop:lpcqsize} by exploiting the factorized interpretation presented in Section~\ref{sec:eff_fragment}. Indeed, one can directly lift Definition~\ref{def:htdec} of hypertree decomposition and fractional hypertree width of $\LPclosed$ to $\LP(\CQ)$. Now, observe that for every $\LP(\CQ)$ programs for $L$ and every database $\db$, we have $\lcqs{L} = \lcqs{\closureno{L}\mapf\db}$. Hence we have the following:

\begin{lem}
  \label{lem:lpcq_td_close} Let $L$ be an $\LP(\CQ)$ program and let $\trees{}$ be a tree decomposition of $L$. Then for every $\db$, $\trees{}$ is a tree decomposition of $\closureno{L}\mapf\db$.
\end{lem}

Hence, let $L$ be an $\LPCQ$  in normal form and $\trees{}$ a tree decomposition of $L$ of width $k$. Given a database $\db$ and an $\LPCQ$ $L$ in normal form, one can compute $L'=\closureno{L}\mapf\db$ in time $O(|L|\cdot|\db|^{\agmfo{L}+\agmsum{L}})$. Then using Theorem~\ref{thm:complexity-closed}, one can compute $\opsem{L'}$ in time $O((|L'|+qt|\db|^k)^\ell)$ where $q$ and $t$ respectively denotes the sum of sizes of the queries in $\lcqs{L}$ and of the tree decompositions in $\trees{}$. Hence, in the data complexity model,  there exists $\ell < 2.37286$ such that one can compute $\opsem{\interpret{L}^\db}$ in time $O(|\db|^{\ell p})$ where $p=\max(k, \agmfo{L}+\agmsum{L})$\footnote{In the conference version of this paper~\cite{confversion}, we were restricting $\LPCQ$ to only use queries with one atom in $\lcqfo{L}$ and $\lcqsum{L}$ to get a tractable fragment of $\LPCQ$. Our assumption allowed us to born the size of the closure, which is now implicitly done by observing that in this case, $\agmfo{L}+\agmsum{L}\leq2$. This new formulation allows us to be more precise and provides better bounds.}.

While it is not clear to us how one could avoid unfolding every universal quantifiers when constructing the factorized interpretation of an $\LPCQ$, we explain how one could possibly get a complexity that sometimes may be smaller than $|\db|^{\agmsum{L}}$ when computing the closure of a linear expression. Observe that when computing the closure over a database $\db$ of an expression of the form $\sum_{\Xtup:Q} 1$, it will evaluate to the size of $\Soldb{Q}$. Now, if $Q$ has fractional hypertree width $k$, one can reduce the data complexity of this task from $O(|\db|^{\agm(Q)})$ to $O(|\db|^{k})$ using~\cite{pichler2013}. Similarly, when evaluating expression of the form $\sum_{\Xtup:Q} \weight{Q'}{\Xtup[x']: \Ytup \LEq \Xtup[z]}$, we can first exploit a tree decomposition of $Q'$ of width $k'$ to compute $\projs{\Soldb{Q'}}{\Yset}$ which will be of size at most $|\db|^{k'}$. Now observe that the closure of $\sum_{\Xtup:Q} \weight{Q'}{\Xtup[x']:\Ytup \LEq \Xtup[z]}$ over $\db$ will be of the form  $\sum_{\alpha \in \projs{\Soldb{Q'}}{\Yset}} K_\alpha \cdot \weight{Q'}{\Ytup \LEq \alpha(\Ytup)}$ where $K_\alpha$ is the number of $\gamma \in \Soldb{Q}$ such that $\gamma(\Xtup[z]) = \alpha(\Ytup)$. Again, if $Q$ has bounded fractional hypertree width $k$, one can compute $K_\alpha$ efficiently since the condition $\gamma(\Xtup[z]) = \alpha(\Ytup)$ corresponds into some condition of the values of that some variables in $\Xtup$ have to take. Hence, since computing the list of possible $\alpha$ being doable in time $O(|\db|^{k'})$ and since computing $K_\alpha$ for each $\alpha$ can be done in time $O(|\db|^k)$, we can compute the closure in time $O(|\db|^{k+k'})$ if we have the relevant tree decomposition. 

\subsection{Case study}
\label{sec:results}

The practical performances of our idea heavily depends on how linear solvers perform on factorized interpretation. We compared the performances of GLPK on both the natural interpretation and the factorized interpretation of the resource delivery problem from Section~\ref{sec:introexample} using some synthetic data.

For each run we fixed an input size $m$ as well as a domain $\domain$ of size $n = f(m)$. We then generated each input table of arity $k$ by uniformly sampling $m$ tuples from the $n^k$ possible tuples on $\domain$. The value of $k$ was defined so that the ratio of selected tuples $\frac{m}{n^k}$ was constant throughout the runs. We used Python and the Pulp library to build the linear programs as well as a hard-coded tree-decomposition of the $\deliver$ query (see Section \ref{sec:introexample}). The tests were run on an office laptop by progressively increasing the size of the generated input tables. A summary of our experiments is displayed on Figure~\ref{fig:implem}.

\begin{figure}[h]
\centering
\includegraphics[width=6.2cm]{nbvar.pdf}
\includegraphics[width=6.2cm]{naivevsfactorized.pdf}
\caption{Number of variables and performances of GLPK for natural (blue) and factorized (red) interpretation of the resource delivery problem with respect to table size.}
\label{fig:implem}
\end{figure}

Interestingly, in this example, the theoretical guarantees given by the factorized interpretation should not be much better than the theoretical guarantees given by the natural interpretation. Indeed, recall that the query considered in this example is the following one:
\[
  \deliver(f, w, b, o) = \exists q.\exists q_2.\exists c \exists c_2.\ \tfactory(f, o, q) \wedge \torder(b, o, q_2)
  \wedge \troute(f, w, c) \wedge \troute(w, b, c_2)
\]
Observe that the existentially quantified variables functionally depends on the free variables of $\deliver$. For example, in the table $\tfactory$, $q$ represents the quantity of objects $o$ that factory $f$ can produce, $q$ is hence functionally dependent on $f,o$. Hence, the AGM bound for $\deliver$ implies that for every database $\db$, $\Soldb{\deliver}$ is of size at most $|\db|^2$. Now, observe that the fractional hypertree width of $\deliver$ is also $2$. Hence, both the factorized interpretation and the natural interpretation may have up to $O(|\db|^2)$ variables. Observe however, that, in the light of Lemma~\ref{lem:computesol}, the number of variables in the factorized interpretation will never exceed the number of variables in the natural interpretation multiply by a factor depending only on the size of the tree decomposition. But, even if the decomposition has width $2$, the factorized interpretation may still be smaller in practice than the natural interpretation and that is what our experiments show. The decomposition we used for the experiments consist in a normalized version of the tree decomposition having two connected vertices $u_1, u_2$ with $\Bag(u_1) = \{f,o,q,w,b,c_2\}$ and $\Bag(u_2) = \{b,o,q_2,f,w,c\}$. Hence, the variables in the factorized interpretation will be the projection of $\Soldb{\deliver}$ over $\Bag(u_1)$ and $\Bag(u_2)$. It turns out that in the syntactical data we have experimented on, these projections are much smaller than the total size of $\Soldb{\deliver}$, leading to a factorized interpretation that is more succinct than the natural one.

As expected when comparing both linear programs we observed a larger number of constraints (due to the soundness constraints) in the factorized interpretation. We observe that the factorized interpretation has less variables, as explained in the previous paragraph. While building the natural interpretation quickly became slower than building the factorized interpretation, we did not analyze this aspect further since we are not using a database engine to build the natural interpretation and solve it directly from the tree decomposition, which may not be the fastest method without further optimizations. Most interestingly solving the factorized interpretation was faster than solving the natural interpretation in spite of the increased number of constraints thanks to the decrease in the number of variables. In particular for an instance with an input size of 2000 lines per table, the natural interpretation had roughly 1.5 million variables while the factorized interpretation had only roughly 150000. The solving time was also noticeably improved at 22s for the factorized case against 106s for the natural one.

\section{Weightings on Tree Decompositions}
\label{sec:weightings}

This section is dedicated to the proof of Theorem~\ref{theo:main-closed} stating the soundness of the factorized interpretation. 
The soundness of the factorized interpretation boils down to a purely algebraic result concerning conjunctive queries that we now explain. Let $Q$ be a conjunctive query, $\db$ a database, $T=(\Vert,\Edg,\Bag)$ a tree decomposition of $Q$. We are interested in weightings of $\Soldb{Q}$, that is, mappings $w \colon \Soldb{Q} \rightarrow \Rp$. Such a mapping naturally defines a mapping $\pi_T(w)$ from $\{\projs{\alpha}{\Bag(u)} \mid \alpha \in \Soldb{Q}, u \in \Vert\}$ to $\Rp$ as follows: for $\beta = \projs{\alpha}{\Bag(u)}$ for some $u \in \Vert$, we define the projection of $w$ on $T$ as follows: $\pi_T(w)(\beta) = \sum_{\gamma \in \Soldb{Q} \colon \projs{\gamma}{\Bag(u)}=\beta} w(\gamma)$. That is, the weight of $\beta$ is obtained by summing the weight of every answer of $Q$ compatible with $\beta$.

We show in Section~\ref{sec:correctness} that the soundness of the factorized interpretation boils down to inverting this projection. Namely, if we are given a weighting $W$ of $\{\projs{\alpha}{\Bag(u)} \mid \alpha \in \Soldb{Q}, u \in \Vert\}$, can we construct a weighting $\amalg(W)$ of $\Soldb{Q}$ such that $\pi_T(\amalg(W)) = W$? While this is not always possible, we show in Section~\ref{sec:constructweightings} that it is always possible as long as $W$ is sound, a property that roughly says that $W(\beta)$ and $W(\gamma)$ could not be independent if $\beta$ and $\gamma$ are compatible tuples (that is, they assign the same value to their common variables). The proof structure is as follows: Proposition~\ref{prop:mainsoundness} established that $\pi_T(w)$ is sound. Then Theorem~\ref{theo:corr} explains the constructions of $\amalg(W)$ from a sound $W$. The proof of this theorem relies on a good understanding on how projections of the form $\projs{\Soldb{Q}}{\Bag(u)}$ are related to one another, which is done via several intermediate lemmas presented in Section~\ref{sec:constructweightings}.

\subsection{Factorized interpretation and weightings}
\label{sec:correctness}

By Proposition~\ref{prop:metasoundness}, to prove Theorem~\ref{theo:main-closed}, it is sufficient to show that for every quantifier free conjunctive query $Q$ with tree decomposition $T=(\Vert,\Edg,\Bag)$ and for any set $\wset$ over weight expressions over $Q$, we have $\projs{\sem{\rewe{\wset}}}{\nu(\wset)}=\projs{\sem{\interpret{\wC[\wset]}^\db}}{\nu(\wset)}$.

In other words, we have to prove the following:
\begin{lem}
  \label{lem:soundprecise}\hfill 
\begin{itemize}
\item Given  $w_2 \in \sem{\interpret{\wC[\wset]}^\db}$, there exists $w_1 \in \sem{\rewe{\wset}}$ such that for every $W \in \wset$, $w_1(\nu(W)) = w_2(\nu(W))$.
\item Given $w_1 \in \sem{\rewe{\wset}}$, there exists $w_2 \in \sem{\interpret{\wC[\wset]}^\db}$ such that for every $W \in \wset$, $w_1(\nu(W)) = w_2(\nu(W))$.
\end{itemize}  
\end{lem}

Now recall that $\interpret{\wC[\wset]}^\db$ is a set of equality constraints of the form:
$$ \nu(\weight{Q}{\Xtup \LEq \Xtup[c]}) \EQU \sum\limits_{\substack{\alpha \in \sem\Query^\db\\\alpha(\Xs)=\Xtup[c]}} \theta_{\Query}^\alpha .$$
In other words, $w_2 \in \sem{\interpret{\wC[\wset]}^\db}$ is a function that maps every $\tup{Q}{\alpha}$ to a value in $\Rp$ and maps $\nu(\weight{Q}{\Xtup \LEq \Xtup[c]})$ to $\sum\limits_{\substack{\alpha \in \sem\Query^\db\\\alpha(\Xs)=\Xtup[c]}} w_2(\theta_{\Query}^\alpha)$.
We thus have a one-to-one correspondence between $\sem{\interpret{\wC[\wset]}^\db}$ and weightings of $\Soldb{Q}$ by associating $w_2 \in \sem{\interpret{\wC[\wset]}^\db}$ to the weighting $\omega$ of $\Soldb{Q}$ such that for every $\alpha \in \Soldb{Q}$, $\omega(\alpha) := w_2(\tup{Q}{\alpha})$.

Similarly, recall that $\rewe{\wset}$ is a set of equality constraints of the form:
$$ \nu(\weight{Q}{\Xtup \LEq \Xtup[c]}) = \weightB{Q}{u}{\beta}$$
and a conjunction of local soundness constraints, for every edge $e=(u,v) \in \Edg$ of $T$ and $\gamma \in \projs{\Soldb{Q}}{\Bag(u) \cap \Bag(v)}$, we define the equality constraint $\localsoundc{Q}{e}{\gamma}$ as follows:
$$
\sum\limits_{\substack{\beta \in \projsb{\Soldb{Q} }{u}\\ \gamma=\projsb{\beta}{u}}}  \weightB{Q}{u}{\beta}
\EQU
\sum\limits_{\substack{\beta' \in \projsb{\Soldb{Q} }{v}\\ \gamma=\projsb{\beta'}{v}}}  \weightB{Q}{v}{\beta'}.
$$

Hence, one can naturally associate $\sem{\rewe{\wset}}$ to a family of weightings $(W_u)_{u \in \Vert}$ where $W_u$ is a weighting of $\projs{\Soldb{Q}}{\Bag(u)}$. We do so by associating $w_1 \in \sem{\rewe{\wset}}$ to the family of weightings $W_u$ such that for every $\beta \in  \projs{\Soldb{Q}}{\Bag(u)}$, $W_u(\beta) = w_1(\weightB{Q}{u}{\beta})$.

Observe however that every family of $(W_u)_{u \in \Vert}$ cannot always be mapped back to $\sem{\rewe{\wset}}$ since it may not satisfy the local soundness constraints. One needs also $(W_u)_{u \in \Vert}$ to be \emph{sound}, that is, for every edge $(u,v) \in \Edg$ and $\gamma \in \projs{\Soldb{Q}}{\Bag(u) \cap \Bag(v)}$, we have:

$$
\sum\limits_{\substack{\beta \in \projsb{\Soldb{Q} }{u}\\ \gamma=\projsb{\beta}{u}}}  W_u(\beta)
=
\sum\limits_{\substack{\beta' \in \projsb{\Soldb{Q} }{v}\\ \gamma=\projsb{\beta'}{v}}}  W_{v}(\beta').
$$

Observe that, when interchanging $\weightB{Q}{u}{\beta}$  and $W_u(\beta)$ in the equality, it exactly corresponds to the local soundness constraints of $Q$ over $\db$. Hence, we have a one-to-one correspondence between $\sem{\rewe{\wset}}$ and sound family of weightings $(W_u)_{u\in\Vert}$. 

Hence, proving Lemma~\ref{lem:soundprecise} boils down to the following. For a query $Q$, a set of weight expressions $\wset$, a tree decomposition of $T=(\Vert,\Edg,\Bag)$ of $Q$ compatible with $\wset$, we have:

\begin{itemize}
\item Given a weighting $w$ of $\Soldb{Q}$, there exists a sound family of weightings $(W_u)_{u\in\Vert}$ such that for every $\weight{Q}{\Xtup\LEq \Xtup[c]} \in \wset$, we have:
  $$ W_u([\Xtup/\Xtup[c]]) = \sum\limits_{\substack{\alpha \in \sem\Query^\db\\\alpha(\Xs)=\Xtup[c]}} w(\alpha),$$
  where $u$ is the vertex of $T$ closest to the root such that $\Bag(u) = \Xset$.
\item Given a sound family of weightings $(W_u)_{u\in\Vert}$, there exists a  weighting $w$ of $\Soldb{Q}$ such that for every $\weight{Q}{\Xtup\LEq \Xtup[c]} \in \wset$, we have:
  $$ \sum\limits_{\substack{\alpha \in \sem\Query^\db\\\alpha(\Xs)=\Xtup[c]}} w(\alpha) = W_u([\Xtup/\Xtup[c]]), $$
  where $u$ is the vertex of $T$ closest to the root such that $\Bag(u) = \Xset$.
\end{itemize}

The existence of such weightings will be proven in Theorem~\ref{theo:corr}. Proving the first item is actually relatively straightforward as it is sufficient to define $W_u([\Xtup/\Xtup[c]])$ as $\sum\limits_{\substack{\alpha \in \sem\Query^\db\\\alpha(\Xs)=\Xtup[c]}} w(\alpha)$ and prove that it yields a sound weighting. The second item requires a bottom up inductive construction from $T$. Section~\ref{sec:constructweightings} is dedicated to proving this correspondence between weightings.

\subsection{Constructing Weightings}
\label{sec:constructweightings}

To make notations lighter, we fix in this section a relation $A \subseteq D^X=\{\alpha\mid\alpha:X \to D\}$ on a finite set of variables $X$. In this work, $A$ can be thought as the answer set of a conjunctive query $Q$ with $\fv(Q)=X$ on database with domain $D$ but the results presented in this section could apply to any relation that is \emph{conjunctive} with respect to a tree decomposition (see Definition~\ref{def:conjunctivity} for more details).

We also fix $T = (\Vert,\Edg, \Bag)$ a decomposition tree for $X$ and define a few useful notations. Given two nodes $u, v \in \Vert$ we denote the intersection of their bags by $\interset{u}{v} = \Bag(u) \cap \Bag(v)$. We denote by $\descorequal{u}$ the set of vertices $v$ such that  $v$ is in the subtree rooted in $u$ ($u$ included) and by $\context{u}$ the set of vertices containing $u$ and every vertex $v$  not in $\descorequal{u}$. We extend the notation: $\vart{u}$ (resp. $\varc{u}$) is the union of $\Bag(v)$ for $v$ in $\descorequal{u}$ (resp. $\context{u}$).

\subsubsection{Projections and Extensions}

We start by introducing a few notations to formally restrict relations and manipulate weightings on relations that will be necessary to write down the proofs.  Let $X'\subseteq X\subseteq \Vars$. For any $\alpha':X'\to D$ we define the set of its extensions into $A$ by:
$$ \extset{\alpha'}{A} = \{\alpha \in A \mid \projt{\alpha}{X'} = \alpha' \}.$$

For a weighting $\omega$ of $A$ and subset of variables $X' \subseteq X$, the projection of $\omega$ on $X'$ denoted as  $\projw{\omega}{X'}:A_{|X'}\to\Rp$ is defined for all $ \alpha' \in \projs{A}{X'}$ as:
\[
\projw{\omega}{X'}(\alpha') = \sum_{\alpha \in \extset{\alpha'}{A}} \omega(\alpha).
\]

We make a few useful observations on how extensions and projections interact with one another. Formal proofs of these statements may be found in the appendix.
        
\begin{lem}\label{lem:projdisjoint} 
For any two $\alpha_1,\alpha_2 \in \projs{A}{X'}$, if
$\alpha_1\not=\alpha_2$
then $\extset{\alpha_1}{A}\cap\extset{\alpha_2}{A}=\emptyset$.
\end{lem}

\begin{lem}
    \label{lem:projproj}
For $A \subseteq D^X$,
$X'' \subseteq X' \subseteq X$, $\alpha'' \in \projs{A}{X''}$:
    $
        \extset{\alpha''}{A} = \biguplus_{\alpha' \in \extseta{\alpha''}{X'}} \extset{\alpha'}{A}$.
\end{lem}

\begin{lem}
\label{prop:projproj} For $A \in D^X$, $\omega:A\to \Rp$, $X'' \subseteq X' \subseteq X$: 
$\projw{\omega}{X''} = \projw{\projw{\omega}{X'}}{X''}.$
\end{lem}
      
\subsubsection{Weighting Collections}

\begin{defi}
   A family $\Wfam = \Wfamlong$ is a \emph{\TWD for $A$} if it satisfies
    the following conditions for any two nodes $u, v \in \Vert$:
    \begin{description}
        \item[- $\W_u$ is a weighting of $\projsb{A}{u}$] i.e., $\W_u:\projsb{A}{u}\to\Rp$. 
        \item[- $\W_u$ is sound for $T$ at $\{u,v\}$] i.e.,
            \label{eq:local.soundness}
          $  \projwi{\W_u}{u}{v} = \projwi{\W_v}{u}{v}.$
    \end{description}
\end{defi}

Intuitively, the soundness of a \TWD is a minimal requirement for the existence of a weighting $\omega$ of $A$ such that  $\W_u$ is the projection of $\omega$ on the bag $\Bag(u)$ of $T$, that is  $\W_u = \projwb{\omega}{u}$ since we have the following:

\begin{prop}
    \label{prop:mainsoundness}
    For any weighting $\omega:A\to \Rp$, the family
    $(\projwb{\omega}{v})_{v \in \Vert}$ is a \TWD for $A$.
\end{prop}

\begin{proof} 
For any $u\in\Vert$ let $W_u=\projwb{\omega}{u}$.
The first condition on weighting projections holds trivially so we only have to show that the soundness constraint holds.
By definition of $W_u$, $\projwi{W_u}{u}{v} = \projwi{\projw{\omega}{\Bag(u)}}{u}{v}$.
Observe that $\interset{u}{v} \subseteq \Bag(u)$ 
so by Lemma~\ref{prop:projproj} $\projwi{W_u}{u}{v} = \projwi{\omega}{u}{v}$.
Similarly $\projwi{W_v}{u}{v} = \projwi{\omega}{u}{v}$.
\end{proof}

What is more interesting is the other way around. For any weighting
$\omega:A\to \Rp$ with $A\subseteq D^X$ and decomposition tree
$T=(\Vert,\Edg,\Bag)$ of $X$, let:
$$\Pi_T(\omega)=(\projwb{\omega}{v})_{v \in \Vert}
$$
So the question is then given  $\W=(\W_u)_{u \in \Vert}$ a \TWD  whether we can find a
weighting $\omega$ of $A$ such that $\W =\Pi_T(\omega) $.
It turns out that soundness is not
enough to ensure the existence of such a weighting.

\subsubsection{Conjunctive Decompositions}

However it becomes possible when $A$ is \conjunctive, as we define
next. For this, given a \decomptree{}  $\TLong$ and a subsets
$V\subseteq\Vert$
we define $\var{V} =\bigcup_{v \in V} \Bag(v)$.

\begin{defi}
\label{def:conjunctivity}
Let $\TLong$ be a \decomptree{} of $X\subseteq \Vars$.
We call a subset of variable assignments $A\subseteq D^X$
\emph{\conjunctive by $T$} if for all 
$u \in \Vert$ and $\beta \in \projsb{A}{u}$:
$$ 
   \{\alpha_1 \cup \alpha_2 \mid \alpha_1\in
  \extset{\beta}{\projsc{A}{u}} ,\  \alpha_2\in
  \extset{\beta}{\projs{A}{\vart{u}}}\} \subseteq \extset{\beta}{A}
$$
\end{defi}

Note that the inverse inclusion does hold in general in any case.

\begin{prop}
\label{prop:solqdecomp}
  For any  tree decomposition $T$ of a quantifier free conjunctive query $Q\in\CQ$ 
and database $\db\in\DB$, the answer set $\Soldb{Q}$ is \conjunctive by $T$.
\end{prop}

\begin{proof}
  Let $u$ be a node of $T$. Let $R(\Xtup)$ be an atom of $Q$ such that $\Xtup \not \subseteq \Bag(u)$. Then we either have $\Xtup \subseteq \vart{u}$ or $\Xtup \subseteq \varc{u}$. Indeed, by definition, there exists $v$ in $T$ such that $\Xtup \subseteq \Bag(v)$. Since it is not $u$, $v$ is either in $\vart{u}$ or $\varc{u}$ and the result follows. Hence $Q$ can be written as a conjunction $Q_1 \wedge Q_2$ where the variables of $Q_1$ are included in $\vart{u}$ and the variables of $Q_2$ are included in $\varc{u}$ by defining $Q_1$ as the set of atom $R(\Xtup)$ of $Q$ such that $\Xtup \subseteq \vart{u}$ and $Q_2$ to be the other atoms (observe that if $\Xtup \subseteq \Bag(u)$, $R(\Xtup)$ will appear in $Q_1$ by definition).
  
  Moreover, recall that $\Bag(u) = \vart{u} \cap \varc{u}$ by the connectedness of tree decompositions. Let $\beta \in \projs{\Soldb{Q}}{\Bag(u)}$and let $\alpha_1 \in \projs{\Soldb{Q}}{\vart{u}}[\beta]$ and $\alpha_2 \in \projs{\Soldb{Q}}{\varc{u}}[\beta]$. We have to show that $\alpha := \alpha_1\cup\alpha_2 \in \Soldb{Q}[\beta]$. Clearly, $\projs{\alpha}{\Bag(u)} = \beta$ by construction. It remains to show that $\alpha \in \Soldb{Q}$. To do so, it is sufficient to observe that $\alpha_1 \in \Soldb{Q_1}$ and $\alpha_2 \in \Soldb{Q_2}$ which is true since $\alpha_1$ (and symmetrically $\alpha_2$) is a projection of some $\alpha' \in \Soldb{Q}$ to $\vart{u}$ and that the variables of $Q_1$ are included in $\vart{u}$. It shows that $\Soldb{Q}$ is \conjunctive{} by $T$.

\end{proof} 

Proposition~\ref{prop:solqdecomp} does not hold when $Q$ is not
quantifier free. It is the reason why this technique only works when every query in the linear program are quantifier free.

Conjunctive decomposition is necessary to get clean relations between the projections of the form $\projsb{A}{u}$ for a vertex $u$ and projections of $\projsb{A}{v}$ for $v$ a child of $u$. We express these relations depending on the type of $u$ in Lemma~\ref{lem:A}, \ref{lem:B}, \ref{lem:B'} and \ref{lem:C}. These relations will be necessary to prove the correctness of our construction.

\newcommand\betaprime{\beta_{|\Bag(v)}}

\begin{lem}[Extend nodes] 
    \label{lem:A}
Let $T$ be a \decompositiontree of $X$, $u$ an 
\addnode{} node of $T$ with child $v$, and $A\subseteq D^X$ 
a subset of variable assignments. If $A$ is \conjunctive by $T$ 
then any assignment $\beta \in \projsb{A}{u}$ satisfies:
\[
\extset{\beta}{\projsv{A}{u}}_{\mid \vart{v}} =
\extset\betaprime{\projsv{A}{v}}
\]
\end{lem}

\begin{proof}
    For \ltri let $\alpha \in \extset{\beta}{\projsv{A}{u}}_{\mid \vart{v}}$.
    Since $\alpha \in \projsv{A}{v}$ and $\projtb{\alpha}{v} = \betaprime$ 
    it follows that $\alpha \in \extset{\projtb{\beta}{v}}{\projsv{A}{v}}$.

    For \rtli let $\alpha \in \extset{\betaprime}{\projsv{A}{v}}$.
    Let $\gamma \in \extset{\beta}{\projsc{A}{v}}$ be arbitrary
    and $\tau = \gamma \cup \alpha$. 

    Note that $\projtv{(\projtv{\tau}{u})}{v} = \alpha$, so it is
    sufficient to show $\projtv{\tau}{u} \in \extset{\beta}{\projsv{A}{u}}$.

    Since $u$ is an \addnode{} node with child $v$ it follows that $\varc{u} =
\varc{v}$, and thus $\gamma \in \extset{\beta}{\projsc{A}{v}}$.
    By \conjdecomp of $A$ by $T$ it follows that 
   $\tau \in \extset\beta{A}$. Hence, $\projtv{\tau}{u} \in \extset{\beta}{\projsv{A}{u}}$
    as required.
  \end{proof}

  \begin{lem}[Join nodes]
    \label{lem:B}
Let $T$ be a \decompositiontree of $X$, $u$ a \joinnode{} node of $T$ 
with children $v_1, \dots, v_k$ where $k\ge 1$, and $A\subseteq D^X$ 
a subset of variable assignments. If $A$ is \conjunctive by $T$ then
any $\beta \in \projsb{A}{u}$ satisfies:
    
\[ 
    \extset{\beta}{\projsv{A}{u}} = 
        \extset{\beta}{\projsv{A}{v_1}} \bowtie \dots \bowtie \extset{\beta}{\projsv{A}{v_k}}
\]
\end{lem}

\begin{proof}
The inclusion from the left to the right is obvious by projecting an element of $\extset{\beta}{\projsv{A}{u}}$ to
    $\vart{v_1} \dots \vart{v_k}$.

    For the inclusion from the right to the left let $\alpha_1 \in \extset{\beta}{\projsv{A}{v_1}}, \dots
        \alpha_k \in \extset{\beta}{\projsv{A}{v_k}}$.
    We show by induction that $\forall p \in [1, k]$, $\tau_p = \alpha_1 \bowtie \dots \bowtie \alpha_p
    \in \extset{\beta}{\projs{A}{Y_p}}$ where $Y_p = \bigcup_{i=1}^{p} \vart{v_i}$.

    \begin{description}

        \item[Base case] $p = 1$: Obvious.

        \item[Inductive case]

        Recall that by induction $\tau_p \in \extset{\beta}{\projs{A}{Y_p}}$ and observe that 
        $Y_p \subseteq \varc{v_{p+1}}$ so there exists $\gamma \in \extset{\beta}{\varc{v_{p+1}}}$ such that
        $\projt{\gamma}{Y_p} = \tau_p$.
        
        By \conjdecomp on $v_{p + 1}$, $\alpha = \gamma \bowtie \alpha_{p+1} \in A$. 
        Finally we have $\projt{\alpha}{Y_{p+1}} = \projt{(\gamma \bowtie \alpha_{p+1})}{Y_p \cup \vart{v_{p+1}}} 
        = \projt{\gamma}{Y_p} \bowtie \projt{\alpha_{p+1}}{\vart{v_{p+1}}}
        = \tau_p \bowtie \alpha_{p+1} = \tau_{p+1}$ so $\tau_{p+1} \in \projs{A}{Y_{p+1}}$.
        Thus $\tau_{p+1} \in \extset{\beta}{Y_p}$ because $\projtb{\tau_{p+1}}{u} = \beta$. \qedhere

      \end{description}
\end{proof}

\begin{lem}
    \label{lem:B'}
Let $T$ be a \decompositiontree of $X$ and $u$ a \joinnode{} of $T$
with children $v_1, \dots, v_k$.  If $A\subseteq D^X$ is \conjunctive by $T$ then
for any $\alpha \in \projsv{A}{v_1}$ 
    and $\beta = \projtb{\alpha}{u}$:
    
    \[ 
        \extset{\alpha}{\projsv{A}{u}} = 
        \{\alpha\} \bowtie \extset{\beta}{\projsv{A}{v_2}} \bowtie \dots \bowtie \extset{\beta}{\projsv{A}{v_k}}
    \]
\end{lem}

\begin{proof}

    Clearly $\extset{\alpha}{\projsv{A}{u}} = 
    \{ \tau \in \extset{\beta}{\projsv{A}{u}} \mid \projtv{\tau}{v_1} = \alpha \}$
    since $\beta = \projtb{\alpha}{u}$.
    Thus by Lemma~\ref{lem:B}, $\extset{\alpha}{\projsv{A}{u}} 
    = \{ \tau \in \extset{\beta}{\projsv{A}{v_1}} \bowtie \dots 
        \bowtie \extset{\beta}{\projsv{A}{v_k}} \mid \projtv{\tau}{u} = \alpha \}
    = \{\alpha\} 
        \bowtie \extset{\beta}{\projsv{A}{v_2}} \bowtie \dots 
        \bowtie \extset{\beta}{\projsv{A}{v_k}}$.
      \end{proof}

\begin{lem}[Extend node]
    \label{lem:C}
    Let $v$ be the child of an extend node $u$. It the holds for all
    $\alpha \in \projsv{A}{v}$ with $\beta = \projtb{\alpha}{v}$ that:
    \[ \extset{\alpha}{A} = \biguplus_{\beta' \in \extset{\beta}{\projsb{A}{u}}} \extset{\alpha \cup \beta'}{A} \]
\end{lem}

\begin{proof}
    For \ltri{}, let $\tau \in \extset{\alpha}{A}$ and $\beta' = \projtb{\tau}{u}$. Observe that 
    $\beta' \in \extset{\beta}{\projsb{A}{u}}$. Moreover $\vart{u} = \vart{v} \cup \Bag(u)$ so 
    $\projtv{\tau}{u} = \alpha \cup \beta'$ so $\tau \in \extset{\alpha \cup \beta'}{A}$.

    For \rtli{}, let $\tau \in \biguplus_{\beta' \in \extset{\beta}{\projsb{A}{u}}} \extset{\alpha \cup \beta'}{A}$.
    By definition $\tau \in A$ and $\projtv{\tau}{v} = \alpha$ so $\tau \in \extset{\alpha}{A}$.
\end{proof}

\subsubsection{Correspondence Theorem}

We can now establish the correspondence.

\begin{thm}[Correspondence]
\label{theo:corr}
Let $\TLong$ be a normalized \decomptree{} of $X \subseteq \Vars$ 
and $A\subseteq D^X$ be a set of variable assignments that is \conjunctive by $T$.
\begin{itemize}
    \item For every weighting $\omega$ of $A$, $\Pi_T(\omega)$    is  a \TWD{} for $A$.
    \item For any \WC $\Wfam$ on $T$ for $A$ there exists a weighting $\omega$
    of $A$ such that $\Wfam=\Pi_T(\omega)$.
  \end{itemize}
\end{thm}

\begin{proof}
  The first property was shown in Proposition \ref{prop:mainsoundness}.
  So it remains to prove the second property.
  \ignore{
\begin{prop}
\label{theo:corr-back}
Let $\TLong$ be a normalized \decomptree{} of $X \subseteq \Vars$ 
and $A\subseteq D^X$ be a set of variable assignments that is \conjunctive by $T$.
Then for any \WC $\Wfam$ on $T$ for $A$ there exists a weighting $\omega$
    of $A$ such that $\Wfam=\Pi_T(\omega)$.
\end{prop}
\begin{proof}}
Let $\W=(\W_u)_{u\in\Vert}$ be a \TWD for $A \subseteq D^X $.
We start with the construction of $\omega$ from $(\W_u)_{u \in \Vert}$. 
For this we inductively construct for any node $u \in \Vert$, a
weighting $\omusym{u} : \projsv{A}{u} \to \Rp$, always assuming
that $\omusym{u'}$ is defined for all children $u'$ of $u$. 
For any $\alpha \in \projsv{A}{u}$ and $\beta = \projtb{\alpha}{u}$
we define $\omu{u}{\alpha}$ as follows:

\begin{description}
  \item[Case $u$ is a leaf of $T$.] We define $\omu{u}{\alpha} = \W_u(\alpha)$.
\item[Case $u$ is an \addnode{} node of $T$ with a single child $v$.] We
  define:
  $$\omu{u}{\alpha} =\left\{\begin{array}{ll}
                              \frac{{\W_u(\beta)}}{\W_v(\projtb{\alpha}{v})}
                              \omu{v}{\projtv{\alpha}{v}}
                              &\text{ if $\W_v(\projtb{\alpha}{v}) >
                                0$}\\
                              0 &\text{ otherwise}
                            \end{array}\right.
                          $$

\item[Case $u$ is a \forgnode{} node of $T$ with a single child $v$.] We define
    $\omu{u}{\alpha} = \omu{v}{\projtv{\alpha}{v}}$.

\item[Case $u$ is a \joinnode{} node of $T$ with children $v_1, \dots,
  v_k$.] Then we define:

  $$\omu{u}{\alpha}= \left\{\begin{array}{ll}
                              \frac{\prod_{i=1}^k
                              \omu{v_i}{\projsv{\alpha}{v_i}}}{\W_u(\beta)^{k-1}}
                              & \text{ if $\W_u(\beta) > 0$}\\
                              0&\text{ otherwise}
                            \end{array}\right.
                          $$ 
    \end{description}
Finally, we define $\omega=\omega_r$ where $r$ is the root of
$T$. The proof that $\forall u: \W_u = \projw{\omega}{\Bag(u)}$ is
done via two inductions. The first one is a bottom-up induction to
prove that $\W_u = \projw{\omega_u}{\Bag(u)}$ for every node $u$ in the
tree decomposition, see Lemma \ref{prop:A} below.
Then, by top-down induction, one can prove that
$\omusym{u} = \projw{\omusym{r}}{\vart{u}}$. See Lemma \ref{prop:B}
below. Thus:
$$
\W_u = \projw{\omega_u}{\Bag(u)}=
\projw{\projw{\omusym{r}}{\vart{u}}}{\Bag(u)} = 
\projw{\omega_r}{\Bag(u)} = \projw{\omega}{\Bag(u)}
$$
In other words, $\W=\Pi_T(\omega)$ as stated by the proposition.
\end{proof}

\begin{lem}
    \label{prop:A} 
    For all $u \in \Vert$: $\W_u = \projw{\omusym{u}}{\Bag(u)}$.
\end{lem}
\begin{proof}
  We show by bottom-up induction on the nodes of $T$ that for
  all $u \in \Vert$ and $\beta \in \projsb{A}{u}$,
    $\sum_{\alpha \in \extsetav{\beta}{u}} \omu{u}{\alpha} = \W_u(\beta)$.

    The base case is clearly true by the definition
    of $\omusym{u}$ when $u$ is a leaf.

    \begin{description}

    \item[Case 1] $u$ is an \addnode{} node with $v$ its only child.
    
    Let $\beta \in \projsb{A}{u}$ and $\beta' = \projtb{\beta}{v}$. 

        \begin{description}

            \item[Case 1.1] $\W_v(\beta') = 0$.

            By definition $\forall \alpha \in \extsetav{\beta}{u}, \omu{u}{\alpha} = 0$.

            Recall that by soundness $\sum_{\beta'' \in \extseta{\beta'}{\Bag(u)}} \W_u(\beta'') = \W_v(\beta') = 0$.
            Observe that $\beta \in \extseta{\beta'}{\Bag(u)}$ so $\W_u(\beta) = 0 = 
            \sum_{\alpha \in \extsetav{\beta}{u}} \omu{u}{\alpha}$.

            \item[Case 1.2] $\W_v(\beta') > 0$.

            \[
            \begin{array}{rcll}
               &&\hspace{-3em} \sum_{\alpha \in \extsetav{\beta}{u}} \omu{u}{\alpha} \\
                & = & \sum_{\alpha \in \extsetav{\beta}{u}} \frac{{\W_u(\beta)}}{\W_v(\beta')} \omu{v}{\projtv{\alpha}{v}}
                    & \text{ by definition} \\
                & = & \frac{{\W_u(\beta)}}{\W_v(\beta')} \sum_{\alpha \in \extsetav{\beta}{u}} 
                    \omu{v}{\projtv{\alpha}{v}} \\
                & = & \frac{{\W_u(\beta)}}{\W_v(\beta')} \sum_{\alpha' \in \extsetav{\beta'}{v}} \omu{v}{\alpha'}
                    & \begin{array}{l}\text{conj. decomp. Lemma~\ref{lem:A}}\\\text{for extend nodes}\end{array} \\
                & = & \frac{{\W_u(\beta)}}{\W_v(\beta')} \W_v(\beta')
                    & \text{ by induction} \\
                & = & {\W_u(\beta)}
            \end{array}
            \]

        \end{description}

    \item[Case 2] $u$ is a \forgnode{} node with only child $v$.

    \[
    \begin{array}{rcll}
        &&\hspace{-3em} \sum_{\alpha \in \extsetav{\beta}{u}} \omu{u}{\alpha} \\
        & = & \sum_{\alpha \in \extsetav{\beta}{u}} \omu{u}{\alpha}
            & \text{ by definition} \\
        & = & \sum_{\beta' \in \extsetab{\beta}{v}} \sum_{\alpha' \in \extsetav{\beta'}{u}} \omu{v}{\alpha}
            & \text{ by Lemma~\ref{prop:projproj} and $\Bag(v) \subseteq \vart{u}$} \\
        & = & \sum_{\beta' \in \extsetab{\beta}{v}}  \W_v(\beta')
            & \text{ by induction and $\vart{u} = \vart{v}$} \\
        & = & \W_u(\beta)
            & \text{ by soundness at $(u, v)$}
    \end{array}
    \]

    \item[Case 3] $u$ is a \joinnode{} node with children $v_1, \dots, v_k$.

    Let $\beta \in \projsb{A}{u}$.
    \begin{description} 
    
        \item[Case 3.1] $\\W_u(\beta) = 0$.
    
        By definition $\forall \alpha \in \extsetav{\beta}{u}, \omu{u}{\alpha} = 0$.
    
        Thus $\sum_{\alpha \in \extsetav{\beta}{u}} \omu{u}{\alpha} = 0 = \W_u(\beta)$.
        
        \item[Case 3.2] $\W_u(\beta) > 0$.
\newenvironment{talign*}  
{\let\displaystyle\textstyle\csname align*\endcsname}
{\endalign}
        \begin{talign*}
             & \hspace{-2em}  \sum_{\alpha \in \extsetav{\beta}{u}} \omu{u}{\alpha} \\
            & = \sum_{\alpha \in \extsetav{\beta}{u}} \frac{\prod_{i=1}^k \omu{v_i}{\projsv{\alpha}{v_i}}}
                                                             {\W_u(\beta)^{k - 1}}
                && \text{ by definition} \\
            & = \sum_{\alpha_1 \in \extsetav{\beta}{v_1}} \dots \sum_{\alpha_k \in \extsetav{\beta}{v_k}}
                \frac{\prod_{i=1}^k \omu{v_i}{\alpha_i}}{\W_u(\beta)^{k - 1}}
                && \begin{array}{l}\text{conj. decomp. Lemma~\ref{lem:B}}\\\text{for join nodes}\end{array} \\
            & = \frac{\prod_{i=1}^k \sum_{\alpha_i \in \extsetav{\beta}{v_i}} \omu{v_i}{\alpha_i}}{\W_u(\beta)^{k - 1}} \\
            & = \frac{\prod_{i=1}^k \W_{v_i}(\beta)}{\W_u(\beta)^{k - 1}}
                && \text{ by induction} \\
            & = \frac{\W_u(\beta)^k}{\W_u(\beta)^{k - 1}} 
                && \text{ by soundness at $(u, v_i)$} \\
            & = \W_u(\beta) \tag*{\qedhere}
        \end{talign*}
    \end{description}
\end{description}
\end{proof}

\begin{lem}
    \label{prop:B}
    For all $u \in \Vert$: $\omusym{u} = \projw{\omusym{r}}{\vart{u}}$.
\end{lem}
\begin{proof}

    We show by top-down induction on the nodes of $T$ that for all $v \in \Vert$ and $\alpha \in \projsv{A}{v}$,
    $\sum_{\tau \in \extset{\alpha}{A}} \omu{r}{\tau} = \omu{v}{\alpha}$.

    The base case is clearly true when $v$ is the root $r$ of $T$.

    In the following we consider a given $\alpha \in \projsv{A}{v}$.
    and we let $\beta = \projtb{\alpha}{v}$ 

    \begin{description}

    \item[Case 1] $v$ is the only child of an \addnode{} node $u$.

    By Lemma~\ref{lem:C}, $\sum_{\tau \in \extset{\alpha}{A}} \omu{r}{\tau} =
    \sum_{\beta' \in \extsetab{\beta}{u}} \sum_{\tau \in \extset{\alpha \cup \beta'}{A}} \omu{r}{\tau}$.
    By induction this is equal to $\sum_{\beta' \in \extsetab{\beta}{u}} \omu{u}{\alpha \bowtie \beta'}$.

        \begin{description}
    
        \item[Case 1.1] $\W_v(\beta) = 0$.
       
        By definition of $\omusym{u}$, $\sum_{\beta' \in \extsetab{\beta}{u}} \omu{u}{\alpha \bowtie \beta'} = 0$.
    
        Observe that by Proposition~\ref{prop:A}, $\sum_{\alpha' \in \extsetav{\beta}{u}} \omu{v}{\alpha'} 
            = \W_v(\beta) = 0$.
        However $\omusym{v}$ is non-negative so $\omu{v}{\alpha} = 0 = \sum_{\tau \in \extset{\alpha}{A}} \omu{r}{\tau}$.
    
        \item[Case 1.2] $\W_v(\beta) > 0$.
        
        \[
        \begin{array}{rcll}
        &&\hspace{-3em} \sum_{\beta' \in \extsetab{\beta}{u}} \omu{u}{\alpha \bowtie \beta'} \\
        & = & \sum_{\beta' \in \extsetab{\beta}{u}} 
            \frac{\W_u(\beta')}{\W_v(\beta)} \omu{v}{\projtv{(\alpha \bowtie \beta')}{v}}
            & \text{by definition} \\
        & = & \frac{\sum_{\beta' \in \extsetab{\beta}{u}} \W_u(\beta')}{\W_v(\beta)} \omu{v}{\alpha} \\
        & = & \omu{v}{\alpha} 
            & \text{by soundness at $(u,v)$} \\
        \end{array}
        \] 
      
        \end{description}
 
    \item[Case 2] $v$ is the only child of a \forgnode{} node $u$.
   
    Observe that $\vart{u} = \vart{v}$ because $u$ is a \forgnode{} node so by induction:
    $$\sum_{\tau \in \extset{\alpha}{A}} \omu{r}{\tau} = \omu{u}{\alpha} = \omu{v}{\alpha}.$$
 
    \item[Case 3] $v$ is the child of a \joinnode{} node $u$.

    Let $v_1, \dots, v_n$ be the children of $u$, we assume wlog that $v$ is $v_1$. 

    By Lemma~\ref{prop:projproj}, $\sum_{\tau \in \extset{\alpha}{A}} \omu{r}{\tau}
        = \sum_{\alpha' \in \extsetav{\alpha}{u}} \sum_{\tau \in \extset{\alpha'}{A}} \omu{r}{\tau}$.

    By induction we obtain $\sum_{\alpha' \in \extsetav{\alpha}{u}} \omu{u}{\alpha'}$.

        \begin{description}
    
        \item[Case 3.1] $\W_u(\beta) = 0$.
       
        By definition of $\omusym{u}$, $\sum_{\alpha' \in \extsetav{\alpha}{u}} \omu{u}{\alpha'}= 0$.
    
        Recall that because $u$ is a \joinnode{} node, $\W_v(\beta) = \W_u(\beta) = 0$ so similarly to Case~1.2,
        $\omu{v}{\alpha} = 0 = \sum_{\tau \in \extset{\alpha}{A}} \omu{r}{\tau}$.
    
        \item[Case 3.2] $\W_v(\beta) > 0$.
        
        By definition of $\omusym{u}$, $\sum_{\alpha' \in \extsetav{\alpha}{u}} \omu{u}{\alpha'} 
        = \sum_{\alpha' \in \extsetav{\alpha}{u}} 
            \frac{\prod_{i=1}^k \omu{v_i}{\projtv{\alpha'}{v_i}}}{\W_u(\beta)^{k - 1}}$.
        Moreover by Lemma~\ref{lem:B'} we can split $\alpha'$ into $\alpha \times \alpha_2 \times \dots \times \alpha_k$
        and the sum into \\
        $\sum_{\alpha_2 \in \extsetav{\beta}{v_2}} \dots \sum_{\alpha_k \in \extsetav{\beta}{v_k}} 
            \frac{\prod_{i=1}^k \omu{v_i}{\projtv{\alpha'}{v_i}}}{\W_u(\beta)^{k - 1}}$.
        Observe that each term in the product only depends on $\alpha_i$ (or $\alpha$ for $i =1$) and that the 
        denominator only depends on the fixed $\beta$ so we can rewrite the formula into the following
        $\omu{v}{\alpha} \cdot 
            \frac{\prod_{i=2}^k \sum_{\alpha_i \in \extsetav{\beta}{v_i}} \omu{v_i}{\alpha_i}}{\W_u(\beta)^{k - 1}}$ 
        which is equal to $\omu{v}{\alpha} \cdot \frac{\prod_{i=2}^k \W_{v_i}(\beta)}{\W_u(\beta)^{k - 1}}$ 
        by Proposition~\ref{prop:A}.
        Finally observe that by soundness, $\prod_{i=2}^k \W_{v_i}(\beta) = \W_u(\beta)^{k - 1}$.
    
        Thus $\sum_{\tau \in \extset{\alpha}{A}} \omu{r}{\tau} = \omu{v}{\alpha}$. \qedhere
        
        \end{description}

    \end{description}

\end{proof}
\ignore{proof from long version:
  
  For the second item we know by Proposition~\ref{prop:A} that
$\W_u = \projw{\omusym{u}}{\Bag(u)}$
 which is equal to $\projw{\projw{\omusym{r}}{\vart{u}}}{\Bag(u)}$ by Proposition~\ref{prop:B}.
 Thus $\W_u = \projw{\omusym{r}}{\Bag(u)}$ by Lemma~\ref{prop:projproj}.
}

\section{Applications}
\label{sec:applications}

In this section, we provide two applications of our results for existing optimization problems in different areas of computer science.

\subsection{Minimizing Noise for \texorpdfstring{$\varepsilon$}{ε}-Differential Privacy.}
The strategy of differential privacy is to add noise to the relational
data before publication. Roughly speaking, the general objective of $\varepsilon$-differential privacy \cite{Dwork2014a}
is to add as little noise as possible, without disclosing more than an
$\varepsilon$ amount of information.
We illustrate this with  the
example of a set of hospitals which publish medical studies 
aggregating results of tests on patients, which are
to be kept confidential. We consider the problem of
how to compute the optimal amount
of noise to be added to each separate piece of sensitive information
(in terms of total utility of the studies)
while guaranteeing $\varepsilon$-differential privacy. We show that
this question can be solved (approximately) by computing the optimal
solution of a projecting program in \LPCQ with a single
conjunctive query that is acyclic, i.e., of hypertree with $1$. 
While the natural interpretation
yields a linear program with a quadratic number of variables
in the size of the database,
the factorized interpretation requires only a linear number.

We consider a database $\db$ with signature 
$\Sigma=\{H,\Test,St,\Priv, \Sens\}$ whose domain 
provides patients, hospitals, studies, and positive real numbers.
The relations of $\db$ are the following:
\begin{itemize}
\item $(pat, hosp)\in \reladb H$: the patient $pat$ is in the hospital $hosp$.
\item $(pat, st)\in\reladb {\Test}$: the patient $pat$ participates in the study $st$.
\item $(test,st)\in\reladb {St}$: the test $test$ is used in the study $st$.
\item $(obj, \varepsilon)\in \reladb {\Priv}$: the object $obj$ is either a
  patient or a hospital. The positive real number $\varepsilon$ indicates 
  the privacy budget for $obj$.
\item $(st,test,val)\in \reladb {\Sens}$:  the value (in terms of study results) of a patient participating in a study and contributing a unit of information on their result on test $test$.
\end{itemize}

The following query defines the sensitive information that will be 
revealed to the researchers performing the medical studies. It selects all pairs
of patients $pat$ and tests $test$, such $pat$ did the $test$ which
was then used by some study $st$.
\[
    \InSt(pat,test)= \exists st.\  \rela{\Test}(pat,test)\wedge\rela{St}(test,st) 
\]
More precisely, the sensitive information is the answer set of
 this query over the database $\db$. We want to assign a weight to 
all the pairs in the answer set. The weight of a sensitive pair states
the amount of information that may be disclosed about the pair after 
the addition of the noise. The needed amount of noise for the pair
is then inversely proportional to the amount of information that may 
be disclosed, i.e, the weight of the pair, which is also called its privacy budget. 
The weight of a patient $pat$ and a test $test$ is specified by the weight expression:
\[
    \weightq{\Para{pat',test'}}{\InSt(pat',test')}{test' \EQU test\wedge pat'\EQU pat}
\]
In an environment $\gamma$ for the global variables $pat$ and $test$ 
this weight expression is interpreted as the linear program variable:
\[
\theta^{[pat'/\gamma(pat),test'/\gamma(test)]}_{\InSt(pat',test')}
\]
The overall weight of all sensitive tests of the same patient $pat$ is
described by the weight expression:
\[
    \weightq{\Para{pat',test'}}{\InSt(pat',test')}{pat' \EQU pat}
\]
In an environment $\gamma$ for the global variable $pat$ 
this weight expression is interpreted as the following 
sum of linear program variables:
\[
    \sum_{\alpha\in\Sol\db(\InSt(pat',test') \wedge
    pat'=\gamma(pat))}\theta^{[pat'/\gamma(pat),test'/\alpha(test')]}_{\InSt(pat',test')}
\]
This sum may be represented
more compactly in factorized interpretation avoiding the enumeration
of the answer set for the database $\db$.

The \LPCQ program for this example is given in
\Figure{lpcq:privex}.

\begin{figure}[h]
Queries
\[
\begin{array}{lll}
    \InSt(pat,test)= \exists st.\  \rela{\Test}(pat,test)\wedge\rela{St}(st,test) \hspace{2cm}\\
\end{array}
\]
Constraints 
\[
\begin{array}{lll}
    & C_{\textsc{Pat}} = \FORALL{(pat, \varepsilon)}{Priv(pat,\varepsilon)} 
        \\ & \qquad 
              \weightq{\Para{pat',test'}}{\InSt(pat',test')}{pat'\EQU pat} 
        \le \num(\varepsilon)\\
    & C_{\textsc{Hosp}} = \FORALL{(hosp, \varepsilon)}{Priv(hosp, \varepsilon)} 
        {\sum_{\Para{pat}:H(pat,hosp)} \\ & \qquad 
             \weightq{\Para{pat',test'}}{\InSt(pat',test')}{pat'\EQU pat} 
        \le \num(\varepsilon)}
\end{array} 
\]
Program 
\[
\begin{array}{lll}
     & \MAXIMIZE \sum_{(st, test,val): \rela{\Sens}(st,test,val)} \\
&\qquad     \qquad     \qquad        \num(val)\ \weightq{\Para{pat',test'}}{\InSt(pat',test')}{test'\EQU test} \\
    & \theSUBJECTTO  C_{\textsc{Pat}} \wedge C_{\textsc{Hosp}} \\
\end{array}
\]
\caption{\label{lpcq:privex}An \LPCQfrag program for differential
  privacy when publishing medical studies aggregating 
  patient tests in hospitals.}
\end{figure}

The linear privacy constraints that are to be satisfied are
$C_{\textsc{Pat}}$ and $C_{\textsc{Hosp}}$. 
Constraint
$C_{\textsc{Pat}}$ states that for all patients $pat$ with
privacy requirement $\varepsilon$, i.e.,
$\forall (pat,\varepsilon): \Priv(pat,\varepsilon)$,
the sum of all weights of all sensitive pairs $(pat,test')$ in
$\InSt$ must be bounded by $\varepsilon$.
This constraint is motivated by the composition rule of differential
  privacy (DP). Suppose we have sensitive pairs $p_i=(pat_i,test_i)$.
If $p_i$ is $\varepsilon_i$-DP for $1\le i \le n$,
  then
  $\{p_1 \ldots p_n\}$ is $(\sum_{i=1}^n
  \varepsilon_i)$-DP.

Similarly, constraint $C_{\textsc{Hosp}}$ states that for all hospitals $hosp$ with
privacy requirement $\varepsilon$, i.e., $\forall (hosp,\varepsilon): \Priv(hosp,\varepsilon)$,
the sum of all weights of all sensitive pairs $(pat,test)$ in $\InSt$
where $pat$ is a patient of $hosp$ must be bounded by $\varepsilon$.
Finally, the objective function is to maximize the sum over all
triples $(st,test,val)$ in $\Sens$ of the weights of pairs $(pat',test)$
in $\InSt$ but multiplied with $\num(val)$, the utility of the information for the study.

This program is projecting, so it is a member 
of \LPCQfrag. 
Furthermore, a hypertree decomposition of width $1$
is available. While the natural interpretation over a database yields a linear program
with a quadratic number of variables (in the size of the database),
the factorized interpretation yields a linear program with a linear number
of variables.

Please note that the approach presented above is only approximate. For example, summing over noise variance in the objective function would be more accurate but would only lead to a convex program, which motivates us to extend beyond linear programs in future work.  Also, the composition rule for DP is only approximate, more advanced composition rules have been studied but they are more complex and still approximate.

\subsection{Computing the s-Measure for Graph Pattern Matching.}
A matching of a subgraph pattern in
  a graph is a graph homomorphism from the pattern to the graph.
The $s$-measure of Wang et al. ~\cite{wang_efficiently_2013} is used
in data mining to measure the frequency of matchings of subgraph 
patterns, while accounting for overlaps of different matchings. The
idea is to find a maximal weighting for the set of matchings, such that for any
node of the subgraph pattern, the set of matchings mapping it on the same 
graph node must have an overall weight less than $1$. This optimization
problem can be expressed by a projecting \LPCQ program over a database
storing the graph. The conjunctive query of this program 
expresses the matching of the subgraph pattern. The hypertree
width of this conjunctive query is bounded by the hypertree width
of the subgraph pattern. Our factorized interpretation therefore
reduces the size of the linear program for subgraph patterns
with small hypertree width.

\label{sec:smeasure}

The $s$-measure has been introduced by Wang et al.~\cite{wang_efficiently_2013}
to evaluate the frequency of matchings of a subgraph pattern in a larger
graph. Here, we consider pattern matches as graph homomorphism,
but we could also restrict them to graph isomorphisms.

A naive way of evaluating this frequency is to use the number of
pattern matches as the frequency measure. Using this value as a frequency measure is
problematic since different pattern matches may overlap, and as such
they share some kind of dependencies that is relevant from a statistical point
of view. More importantly, due to the overlaps, this measure fails to be
anti-monotone, meaning that a subpattern may be counter-intuitively matched less
frequently than the pattern itself. Therefore, the finding of better
anti-monotonic frequency measures -- also known as {\em support measures} -- has
received a lot of attention in the data mining
community~\cite{bringmann08,calders11,fiedler07}. A first idea is to count the
maximal number of non-overlapping patterns~\cite{vanetik02}. However, finding
such a maximal subset of patterns essentially boils down to finding a maximal
independent set in a graph, a notorious NP-complete problem~\cite{garey02}.

The $s$-measure is a relaxation of this idea where the frequency of
pattern  matches  is
computed as the maximum of the sum of the weights that can be
assigned to each pattern match, under the constraint that for any
node $v$ of the graph and node $v'$ in the subgraph pattern that the sum 
of the weights of the matchings
mapping $v'$ to $v$ is at most $1$. 
More formally, given two digraphs $G =(V_G,E_G)$ and $P=(V_P,E_P)$, 
we define a matching of the pattern $P$ in graph $G$ as a 
graph homomorphism $h:V_P\to V_G$. Recall that a graph homomorphism
requires for all $(v,v')\in E_p$ that $(h(v),h(v'))\in E_G$. We 
denote by $hom(P,G)$ the set of matchings of $P$ in $G$. 
The $s$-measure of $P$ in $G$ is then defines as the 
optimal value of the following linear
program with variables in $\{\theta_h \mid h \in hom(P,G)\}$
for positive real numbers:
\[
\begin{array}{ll}
    \MAXIMIZE & \sum_{h \in hom(P,G)} \theta_h \\
    \SUBJECTTO  
        & \forall v \in V_G. \forall v'\in V_P.
            \sum_{\substack{h \in hom(P,G) \\ h(v')=v} } \theta_h \le 1
\end{array}
\]
\newcommand\match[1]{\mathit{match}_{#1}}

We can consider each graph $G$ as a database $\db$ with signature
$\Sigma=\{node,edge\}$, domain $\dom(\db_G)=V_G$ and relations
$node^\db=V_G$ and $edge^\db=E_G$. Since the names of the nodes
of the pattern do not care for pattern matching, we can assume 
without loss of  generality that $V_P=\{1,\ldots,\ell\}$ for some 
$\ell\in\N$. We can then define a matching of a pattern $P$ 
by a conjunctive query $\match{P}(x_1, \dots, x_\ell)$:
\[ 
\match{P}(x_1,\dots,x_\ell) = \bigwedge_{(i,j) \in E_P} edge(x_i, x_j) 
\]
It is clear that $\alpha\in \Sol\db(\match{P}(x_1,\dots,x_\ell)$ if
and only if $\alpha\circ[1/x_1,\ldots,\ell/x_\ell]$ is a pattern
matching in $hom(P,Q)$.
One can thus rewrite the previous linear program as
\LPCQ program as follows:
\[
\begin{array}{ll}
    \MAXIMIZE & \sum_{(x):node(x)}
                \weightq{(x_1,\ldots,x_n)}{\match{P}(x_1,\ldots,x_n)}
                {x_1\EQU x} \\
    \SUBJECTTO  &
         \FORALL{(x):node(x)}{\wedge_{i=1}^\ell
                  \weightq{(x_1,\ldots,x_n)}{\match{P}(x_1,\ldots,x_n)}
               {x_i\EQU  x} \le 1}
\end{array}
\]

\ignore{\begin{align*}
    \text{maximize } & \sum\limits_{v \in V} \sxd{{x_1}}{v} \\
    \text{subject to }  &
         \forall v \in V, \forall i \leq \ell,
            \sxd{{x_i}}{v} \in [0,1].
\end{align*}}

Moreover, the hypertree width of the conjunctive query
$\match{P}(x_1,\ldots,x_\ell)$ 
is at most the (hyper)tree width of the pattern graph $P$. By our
main Theorem~\ref{theo:main-closed}, the factorized interpretation
yields a linear program with at most $(|V_G|+|E_G])^{k}$ variables,
where $k$ is the (hyper)tree width of pattern $P$. The original
linear program could have been of size $\left(\begin{array}{c} 
|V_G| \\ \ell \end{array}\right)$ which is bounded by $|V_G|^\ell$.
So the factorized interpretation will pay off if the (hyper)tree width $k$ of the
pattern is considerably smaller than the number $\ell$ of its nodes.

\section{Conclusion and future work}

In this paper we studied linear programs whose variables are the answers to conjunctive queries.  While in general it is possible to construct in this way large and hard to solve programs, we defined a tractable fragment which can benefit from factorization.
Our experiments suggest the efficiency of
factorized interpretation, in accordance with our 
complexity results.

Several directions for future work exist.
One direction is to explore how to better integrate our approach into a database
engine, in the way it is done by \texttt{SolveDB} for
example. Also, other optimization problems may
benefit from this approach such as convex optimization or integer
linear programming. It would be interesting to define languages
analogous to \LPCQ for these optimization problems and study how
conjunctive query decompositions could help to improve the
efficiency.

\bibliographystyle{alphaurl}
\bibliography{biblio}

\newcommand{\etalchar}[1]{$^{#1}$}
\begin{thebibliography}{CRVD11}

\bibitem[AGM13]{agm}
Albert Atserias, Martin Grohe, and D{\'a}niel Marx.
\newblock Size bounds and query plans for relational joins.
\newblock {\em SIAM Journal on Computing}, 42(4):1737--1767, 2013.
\newblock \href {https://doi.org/10.1137/110859440}
  {\path{doi:10.1137/110859440}}.

\bibitem[AW21]{alman2021refined}
Josh Alman and Virginia~Vassilevska Williams.
\newblock A refined laser method and faster matrix multiplication.
\newblock In {\em Proceedings of the 2021 ACM-SIAM Symposium on Discrete
  Algorithms (SODA)}, pages 522--539. SIAM, 2021.
\newblock \href {https://doi.org/10.1137/1.9781611976465.32}
  {\path{doi:10.1137/1.9781611976465.32}}.

\bibitem[Bar12]{barendregt}
H.~Barendregt.
\newblock {\em The Lambda Calculus: Its Syntax and Semantics}.
\newblock Studies in Logic and the Foundations of Mathematics. College
  Publications, 2012.
\newblock \href {https://doi.org/10.2307/2274112} {\path{doi:10.2307/2274112}}.

\bibitem[BDG07]{bagan2007acyclic}
Guillaume Bagan, Arnaud Durand, and Etienne Grandjean.
\newblock On acyclic conjunctive queries and constant delay enumeration.
\newblock In {\em International Workshop on Computer Science Logic}, pages
  208--222. Springer, 2007.
\newblock \href {https://doi.org/10.1007/978-3-540-74915-8_18}
  {\path{doi:10.1007/978-3-540-74915-8_18}}.

\bibitem[BN08]{bringmann08}
Bj{\"o}rn Bringmann and Siegfried Nijssen.
\newblock What is frequent in a single graph?
\newblock In {\em Pacific-Asia Conference on Knowledge Discovery and Data
  Mining}, pages 858--863. Springer, 2008.
\newblock \href {https://doi.org/10.1007/978-3-540-68125-0_84}
  {\path{doi:10.1007/978-3-540-68125-0_84}}.

\bibitem[CCNR22]{confversion}
Florent Capelli, Nicolas Crosetti, Joachim Niehren, and Jan Ramon.
\newblock Linear programs with conjunctive queries.
\newblock In Dan Olteanu and Nils Vortmeier, editors, {\em 25th International
  Conference on Database Theory, {ICDT} 2022, March 29 to April 1, 2022,
  Edinburgh, {UK} (Virtual Conference)}, volume 220 of {\em LIPIcs}, pages
  5:1--5:19. Schloss Dagstuhl - Leibniz-Zentrum f{\"{u}}r Informatik, 2022.
\newblock \href {https://doi.org/10.4230/LIPIcs.ICDT.2022.5}
  {\path{doi:10.4230/LIPIcs.ICDT.2022.5}}.

\bibitem[CLS21]{cohen2021solving}
Michael~B Cohen, Yin~Tat Lee, and Zhao Song.
\newblock Solving linear programs in the current matrix multiplication time.
\newblock {\em Journal of the ACM (JACM)}, 68(1):1--39, 2021.
\newblock \href {https://doi.org/10.1145/3424305} {\path{doi:10.1145/3424305}}.

\bibitem[CM77]{chandra77}
Ashok~K. Chandra and Philip~M. Merlin.
\newblock Optimal implementation of conjunctive queries in relational data
  bases.
\newblock In {\em Proceedings of the Ninth Annual ACM Symposium on Theory of
  Computing}, STOC '77, pages 77--90, New York, NY, USA, 1977. ACM.
\newblock \href {https://doi.org/10.1145/800105.803397}
  {\path{doi:10.1145/800105.803397}}.

\bibitem[CRVD11]{calders11}
Toon Calders, Jan Ramon, and Dries Van~Dyck.
\newblock All normalized anti-monotonic overlap graph measures are bounded.
\newblock {\em Data Mining and Knowledge Discovery}, 23(3):503--548, 2011.
\newblock \href {https://doi.org/10.1007/s10618-011-0217-y}
  {\path{doi:10.1007/s10618-011-0217-y}}.

\bibitem[DR14]{Dwork2014a}
Cynthia Dwork and Aaron Roth.
\newblock {The Algorithmic Foundations of Differential Privacy}.
\newblock {\em Foundations and Trends in Theoretical Computer Science},
  9(3--4):211--407, 2014.
\newblock \href {https://doi.org/10.1561/0400000042}
  {\path{doi:10.1561/0400000042}}.

\bibitem[FB07]{fiedler07}
Mathias Fiedler and Christian Borgelt.
\newblock Support computation for mining frequent subgraphs in a single graph.
\newblock In {\em MLG}. Citeseer, 2007.

\bibitem[FGK90]{ampl}
Robert Fourer, David~M Gay, and Brian~W Kernighan.
\newblock A modeling language for mathematical programming.
\newblock {\em Management Science}, 36(5):519--554, 1990.
\newblock \href {https://doi.org/10.1287/mnsc.36.5.519}
  {\path{doi:10.1287/mnsc.36.5.519}}.

\bibitem[GJ79]{garey02}
Michael~R Garey and David~S Johnson.
\newblock {\em Computers and intractability, A Guide to the Theory of
  {NP}-Completeness}.
\newblock {W.H. Freeman And Company}, 1979.

\bibitem[GLS99]{gottlob1999tractable}
Georg Gottlob, Nicola Leone, and Francesco Scarcello.
\newblock On tractable queries and constraints.
\newblock In {\em International Conference on Database and Expert Systems
  Applications}, pages 1--15. Springer, 1999.
\newblock \href {https://doi.org/10.1007/3-540-48309-8_1}
  {\path{doi:10.1007/3-540-48309-8_1}}.

\bibitem[GLS02]{gottlob2002}
Georg Gottlob, Nicola Leone, and Francesco Scarcello.
\newblock Hypertree {Decompositions} and {Tractable} {Queries}.
\newblock {\em Journal of Computer and System Sciences}, 64(3):579--627, 2002.
\newblock \href {https://doi.org/10.1145/303976.303979}
  {\path{doi:10.1145/303976.303979}}.

\bibitem[GM14]{grohe2014constraint}
Martin Grohe and D{\'a}niel Marx.
\newblock Constraint solving via fractional edge covers.
\newblock {\em ACM Transactions on Algorithms (TALG)}, 11(1):4, 2014.
\newblock \href {https://doi.org/10.1145/2636918} {\path{doi:10.1145/2636918}}.

\bibitem[Gro06]{grohe2006structure}
Martin Grohe.
\newblock The structure of tractable constraint satisfaction problems.
\newblock In {\em International Symposium on Mathematical Foundations of
  Computer Science}, pages 58--72. Springer, 2006.
\newblock \href {https://doi.org/10.1007/11821069_5}
  {\path{doi:10.1007/11821069_5}}.

\bibitem[Kar84]{DBLP:journals/combinatorica/Karmarkar84}
Narendra Karmarkar.
\newblock A new polynomial-time algorithm for linear programming.
\newblock In {\em Proceedings of the sixteenth annual ACM symposium on Theory
  of computing}, pages 302--311. ACM, 1984.
\newblock \href {https://doi.org/10.1145/800057.808695}
  {\path{doi:10.1145/800057.808695}}.

\bibitem[Klo94]{kloks1994treewidth}
Ton Kloks.
\newblock {\em Treewidth: computations and approximations}, volume 842.
\newblock Springer Science \& Business Media, 1994.

\bibitem[KPT13]{kolaitis_efficient_2013}
Phokion~G. Kolaitis, Enela Pema, and Wang-Chiew Tan.
\newblock Efficient querying of inconsistent databases with binary integer
  programming.
\newblock {\em Proceedings of the VLDB Endowment}, 6(6):397--408, 2013.
\newblock \href {https://doi.org/10.14778/2536336.2536341}
  {\path{doi:10.14778/2536336.2536341}}.

\bibitem[Lib13]{libkin2013elements}
Leonid Libkin.
\newblock {\em Elements of finite model theory}.
\newblock Springer Science \& Business Media, 2013.
\newblock \href {https://doi.org/10.1007/978-3-662-07003-1}
  {\path{doi:10.1007/978-3-662-07003-1}}.

\bibitem[MS12]{tiresias}
Alexandra Meliou and Dan Suciu.
\newblock Tiresias: The database oracle for how-to queries.
\newblock In {\em Proceedings of the 2012 ACM SIGMOD International Conference
  on Management of Data}, SIGMOD '12, pages 337--348, New York, NY, USA, 2012.
  ACM.
\newblock \href {https://doi.org/10.1145/2213836.2213875}
  {\path{doi:10.1145/2213836.2213875}}.

\bibitem[NSB{\etalchar{+}}07]{minizinc}
Nicholas Nethercote, Peter~J Stuckey, Ralph Becket, Sebastian Brand, Gregory~J
  Duck, and Guido Tack.
\newblock Minizinc: Towards a standard {CP} modelling language.
\newblock In {\em International Conference on Principles and Practice of
  Constraint Programming}, pages 529--543. Springer, 2007.
\newblock \href {https://doi.org/10.1007/978-3-540-74970-7_38}
  {\path{doi:10.1007/978-3-540-74970-7_38}}.

\bibitem[OZ12]{olteanu2012factorised}
Dan Olteanu and Jakub Z{\'a}vodn{\`y}.
\newblock Factorised representations of query results: size bounds and
  readability.
\newblock In {\em Proceedings of the 15th International Conference on Database
  Theory}, pages 285--298. ACM, 2012.
\newblock \href {https://doi.org/10.1145/2274576.2274607}
  {\path{doi:10.1145/2274576.2274607}}.

\bibitem[OZ15]{olteanu2015size}
Dan Olteanu and Jakub Z{\'a}vodn{\`y}.
\newblock Size bounds for factorised representations of query results.
\newblock {\em ACM Transactions on Database Systems (TODS)}, 40(1):1--44, 2015.
\newblock \href {https://doi.org/10.1145/2693969} {\path{doi:10.1145/2693969}}.

\bibitem[PS13]{pichler2013}
Reinhard Pichler and Sebastian Skritek.
\newblock Tractable counting of the answers to conjunctive queries.
\newblock {\em Journal of Computer and System Sciences}, 79:984--1001, 2013.
\newblock \href {https://doi.org/10.1016/j.jcss.2013.01.012}
  {\path{doi:10.1016/j.jcss.2013.01.012}}.

\bibitem[{\v{S}}P16]{vsikvsnys16}
Laurynas {\v{S}}ik{\v{s}}nys and Torben~Bach Pedersen.
\newblock {SolveDB}: Integrating optimization problem solvers into {SQL}
  databases.
\newblock In {\em Proceedings of the 28th International Conference on
  Scientific and Statistical Database Management}, page~14. ACM, 2016.
\newblock \href {https://doi.org/10.1145/2949689.2949693}
  {\path{doi:10.1145/2949689.2949693}}.

\bibitem[Vel14]{triejoin}
Todd~L. Veldhuizen.
\newblock Triejoin: {A} simple, worst-case optimal join algorithm.
\newblock In Nicole Schweikardt, Vassilis Christophides, and Vincent Leroy,
  editors, {\em Proc. 17th International Conference on Database Theory (ICDT),
  Athens, Greece, March 24-28, 2014}, pages 96--106. OpenProceedings.org, 2014.
\newblock \href {https://doi.org/10.5441/002/icdt.2014.13}
  {\path{doi:10.5441/002/icdt.2014.13}}.

\bibitem[VGS02]{vanetik02}
Natalia Vanetik, Ehud Gudes, and Solomon~Eyal Shimony.
\newblock Computing frequent graph patterns from semistructured data.
\newblock In {\em Data Mining, 2002. ICDM 2003. Proceedings. 2002 IEEE
  International Conference on}, pages 458--465. IEEE, 2002.
\newblock \href {https://doi.org/10.1109/ICDM.2002.1183988}
  {\path{doi:10.1109/ICDM.2002.1183988}}.

\bibitem[WRF13]{wang_efficiently_2013}
Yuyi Wang, Jan Ramon, and Thomas Fannes.
\newblock An efficiently computable subgraph pattern support measure: counting
  independent observations.
\newblock {\em Data Mining and Knowledge Discovery}, 27(3):444--477, 2013.
\newblock \href {https://doi.org/10.1007/s10618-013-0318-x}
  {\path{doi:10.1007/s10618-013-0318-x}}.

\bibitem[Yan81]{Yannakakis}
Mihalis Yannakakis.
\newblock Algorithms for acyclic database schemes.
\newblock In {\em Proceedings of the Seventh International Conference on Very
  Large Data Bases - Volume 7}, VLDB '81, pages 82--94. VLDB Endowment, 1981.
\newblock \href {https://doi.org/10.5555/1286831.1286840}
  {\path{doi:10.5555/1286831.1286840}}.

\end{thebibliography}

\appendix
\section{Proofs of Section~\ref{sec:constructweightings}}

In this section, we provide missing proofs of some easy lemmas stated in Section~\ref{sec:constructweightings}.

\subsection{Proof of Lemma~\ref{lem:projdisjoint}}

If $\alpha_1\not=\alpha_2 \in \projs{A}{X'}$, then there exists $x' \in X'$ such that $\alpha_1(x') \not= \alpha_2(x')$, 
so if $\gamma_1 \in \extset{\alpha_1}{A}$ and $\gamma_2 \in \extset{\alpha_2}{A}$ 
then $\gamma_1(x') = \alpha_1(x') \not= \alpha_2(x')= \gamma_2(x')$.

\subsection{Proof of Lemma~\ref{lem:projproj}}

First note that the union on the right is disjoint by Lemma~\ref{lem:projdisjoint}. 

For \ltri{},  let $\alpha \in \extset{\alpha''}{A}$ and $\alpha' = \projt{\alpha}{X'}$. 
By definition, $\alpha' \in \projs{A}{X'}$ so $\alpha \in \extset{\alpha'}{A}$.
Furthermore, 
$\alpha' \in \extseta{\alpha''}{X'}$ so 
$\alpha \in \biguplus_{\tilde\alpha' \in \extseta{\alpha''}{X'}} \extset{\tilde\alpha'}A$.

For \rtli{}, let
$\alpha \in \biguplus_{\alpha' \in \extseta{\alpha''}{X'}}
\extset{\alpha'}{A}$ and let $\alpha' \in \extseta{\alpha''}{X'}$ be
such that $\alpha \in \extset{\alpha'}{A}$.  By definition,
$\projt{\alpha}{X'} = \alpha'$ and $\projt{\alpha'}{X''} =
\alpha''$. Since $X'' \subseteq X'$,
$\projt{\alpha}{X''} = \projt{\alpha'}{X''} = \alpha''$.  Thus
$\alpha \in \extset{\alpha''}{A}$.

\subsection{Proof of Lemma~\ref{prop:projproj}}

Let $\alpha'' \in
  \projs{A}{X''}$. We have: 

  \begin{align*}
    \projw{\omega}{X''}(\alpha'') 
        & = \sum_{\alpha \in \extset{\alpha''}{A}} \omega(\alpha) & \text{ by definition}\\
        & = \sum_{\alpha' \in \extseta{\alpha''}{X'}} \sum_{\alpha \in \extset{\alpha'}{A}} \omega(\alpha) 
            & \text{ by Lemma~\ref{lem:projproj}} \\
        & = \sum_{\alpha' \in \extseta{\alpha''}{X'}} \projw{\omega}{X'}(\alpha') 
            & \text{ by definition of } \projw{\omega}{X'} \\
        & = \projw{\projw{\omega}{X'}}{X''}(\alpha'') 
            & \text{ by definition of } \projw{\projw{\omega}{X'}}{X''}.
  \end{align*}
  The last equality is well defined since $\alpha'' \in \projt{A}{X''} = \projt{(\projt{A}{X'})}{X''}$.

\end{document}